\documentclass[]{aastex631}

\usepackage{amsmath}
\usepackage{graphicx}
\usepackage{txfonts}
\usepackage{subfigure}

\usepackage{cleveref}
\begin{document}

\title{Photometric and Spectroscopic analysis of eight totally eclipsing contact binaries with small mass ratios}

\correspondingauthor{Kai Li}
\email{kaili@sdu.edu.cn}

\author{Li-Heng Wang}
\affiliation{Shandong Key Laboratory of Optical Astronomy and Solar-Terrestrial Environment, School of Space Science and Physics, Institute of Space Sciences, Shandong University, Weihai, Shandong 264209, China}

\author{Kai Li}
\affiliation{Shandong Key Laboratory of Optical Astronomy and Solar-Terrestrial Environment, School of Space Science and Physics, Institute of Space Sciences, Shandong University, Weihai, Shandong 264209, China}

\author{Ya-Ni Guo}
\affiliation{Shandong Key Laboratory of Optical Astronomy and Solar-Terrestrial Environment, School of Space Science and Physics, Institute of Space Sciences, Shandong University, Weihai, Shandong 264209, China}

\author{Jing-Yi Wang}
\affiliation{Shandong Key Laboratory of Optical Astronomy and Solar-Terrestrial Environment, School of Space Science and Physics, Institute of Space Sciences, Shandong University, Weihai, Shandong 264209, China}

\author{Xiang Gao}
\affiliation{Shandong Key Laboratory of Optical Astronomy and Solar-Terrestrial Environment, School of Space Science and Physics, Institute of Space Sciences, Shandong University, Weihai, Shandong 264209, China}

\author{Xing Gao}
\affiliation{Xinjiang Astronomical Observatory, 150 Science 1-Street, Urumqi 830011, China}

\author{Guo-You Sun}
\affiliation{Xingming Observatory, Urumqi, Xinjiang, China }

%% Note that the \and command from previous versions of AASTeX is now
%% depreciated in this version as it is no longer necessary. AASTeX 
%% automatically takes care of all commas and "and"s between authors names.

%% AASTeX 6.31 has the new \collaboration and \nocollaboration commands to
%% provide the collaboration status of a group of authors. These commands 
%% can be used either before or after the list of corresponding authors. The
%% argument for \collaboration is the collaboration identifier. Authors are
%% encouraged to surround collaboration identifiers with ()s. The 
%% \nocollaboration command takes no argument and exists to indicate that
%% the nearby authors are not part of surrounding collaborations.

%% Mark off the abstract in the ``abstract'' environment. 
\begin{abstract}

This paper selected eight totally eclipsing contact binaries for photometric and spectroscopic studies, spectral data were analyzed by ULySS, and photometric data were analyzed using PHOEBE through MCMC sampling. We used two methods to calculate the initial values for running MCMC: one method is a new approach proposed by ourselves to model light curves without spots, while the other method is the genetic algorithm (GA) which can determine physical parameters with spot.  Due to the results, these eight targets are all small mass ratio contact binary stars with a mass ratio below 0.25. There are four systems exhibiting O’Connell effect. By adding a dark spot on the primary component, the ideal fitting can be obtained. Meanwhile, it was found that two systems are shallow contact binaries, while the remaining six are moderate contact binaries. An O-C analysis of the eight eclipsing binary stars revealed that seven of them exhibit long-term changes. Four of them display a long-term decreasing trend, while the other three show a long-term increasing trend, and two targets exhibit periodic variations. The decrease in period may be caused by the transfer of matter from the more massive component to the less massive component, while the increase in period may be caused by the transfer of matter from the less massive component to the more massive component. The absolute physical parameters, orbital angular momentum, initial masses, and ages of these eight systems were calculated. Additionally, their mass-luminosity and mass-radius distributions were analyzed. 

\end{abstract}

%% Keywords should appear after the \end{abstract} command. 
%% The AAS Journals now uses Unified Astronomy Thesaurus concepts:
%% https://astrothesaurus.org
%% You will be asked to selected these concepts during the submission process
%% but this old "keyword" functionality is maintained in case authors want
%% to include these concepts in their preprints.

%% From the front matter, we move on to the body of the paper.
%% Sections are demarcated by \section and \subsection, respectively.
%% Observe the use of the LaTeX \label
%% command after the \subsection to give a symbolic KEY to the
%% subsection for cross-referencing in a \ref command.
%% You can use LaTeX's \ref and \label commands to keep track of
%% cross-references to sections, equations, tables, and figures.
%% That way, if you change the order of any elements, LaTeX will
%% automatically renumber them.
%%
%% We recommend that authors also use the natbib \citep
%% and \citet commands to identify citations.  The citations are
%% tied to the reference list via symbolic KEYs. The KEY corresponds
%% to the KEY in the \bibitem in the reference list below. 

\section{Introduction}

   Binary stars are extremely common star systems in the universe. A thorough understanding of the evolution of binary stars will be of great help in studying the evolution of the universe and the existence of interesting astronomical phenomena. For example, a binary star merger event: two short-period binary stars merge into a rapidly rotating single star due to loss of angular momentum \citep{qian2017,bbb1994, mercy2011}. According to the filling factor of the two Roche lobes of binary stars, they are divided into detached binary (not filled), semidetached binary (only one Roche lobe is filled), and contact binary (two Roche lobe overfilling). Contact binaries are in the late stage of binary star evolution and serve as a connection between single stars and binary stars, making them of great research value. 
There are still many unresolved issues in the current research on contact binary, such as short period end \citep{jiang2012111,li,zhangxiaobin2023}, the O’Connell effect \citep[e.g.,][] {Connell,liu,qian}, and the minimum mass  ratio limit \citep{rasio,arbutina,jiang2010}, etc.  Physical parameters of a large number of contact binaries are expected as a basis to better solve related issues.

\cite{xin1996} first noticed a short period cutoff of 0.22 days by using contact binary data in the General Catalogue of Variable Stars (GCVS).
Recently, \cite{li} combined with the Variable Star Index (VSX), GCVS and photometric surveys from around the world to determine the period distribution of contact binary stars and found that there is still a significant decline around 0.22 days. Many researchers have tried to solve this problem in history. \cite{xin1996} claimed that the short-period cutoff could possibly be explained by  the fully convective limit.
\cite{step2006} suggested that the time scale of the angular momentum loss (AML) is too long, and even at the age of the universe, short period contact binaries below the short period end should not be discovered. \cite{jiang2012111} suggested that the main reason for the short period limit may be the unstable mass transfer when the initial low mass primary component fills its inner Roche lobe.

Contact binary often exhibits unequal heights at the two maxima of the light curve. This phenomenon is called the O’Connell effect \citep{Connell}. \cite{WB2009} outlined three possible explanations for the O’Connell effect: spots on one or both components, the material accretion between two components \citep{shaw1994}, or the circumstellar material surrounding the binary  \citep{Liu_2003}. The most common explanation is star-spot due to magnetic activity \citep{occ1}. 

In past studies, the theoretically determined lower limit of the mass ratio of contact binary was about 0.05-0.09 \citep[e.g., ][]{rasio,liandzhang,Wadhwa2021}, but observations in recent years have continuously broken this lower limit, such as: V857 Her \cite[$q$$\sim $0.065;][]{qian2005b}, ASAS J083241+2332.4 \cite[$q$$\sim $0.068;][]{sriram2016}, V1187 Her \cite[$q$$\sim $0.044;][]{v1187}, VSX J082700.8+462850 \cite[$q$$\sim $0.055;][]{li2021a}, TYC 4002-2628-1 \cite[$q$$\sim$0.0482;][]{guo2022}, CRTS J224827.6+341351 \cite[$q$$\sim $0.0791;][]{liuxinyi} and WISE J185503.7+592234 \cite[$q$$\sim $0.0514;][]{guo2023}.
Therefore, physical parameters of a large number low-mass ratio contact binaries are needed to determine the true lower limit of the mass ratio and the evolutionary fate of the contact binary. 

Due to the statistical work on contact binaries with both spectroscopic and photometric observations by \cite{Apeople} and \cite{li2021a} and the numerical simulations by \cite{Bpeople}, it was found that the mass ratios of total eclipsing contact bianries should be reliable even without spectroscopic observations.  In this paper,  we selected eight contact binary stars with total eclipses from All-Sky Automated Survey for SuperNovae \cite[ASAS-SN;][]{shappee2014,jayasinghe2018}. The relevant information of the selected targets is shown in  \Cref{tab:label1}. We analyzed the light curves and orbital period changes of these eight targets.

\section{Observations} \label{section:section2}

\subsection{Photometric Observations }\label{2.1}

From 2019 to 2021, we used the 60 cm Ningbo Bureau of Education and Xinjiang Observatory Telescope (NEXT) in China to observe the eight binary stars. This telescope is equipped with an FLI PL23042 CCD  and has a field of view of $ 22{'}$×$ 22{'}$.
During the observations,  standard Johnson–Cousins filter and Sloan  filter were used. The observation details are shown in \Cref{tab:label2}  , and the data reduction is using the IRAF\footnote{IRAF  (\href{}{\url{http://iraf.noao.edu/}}) is distributed by the National Optical Astronomy Observatories, which are operated by the Association of Universities for Research in Astronomy, Inc., under cooperative agreement with the National Science Foundation.}  package. First, bias subtraction and flat correction were performed, followed by aperture photometry to extract the instrumental magnitudes.  Stars with constant brightness were selected as comparison and check stars, and differential photometry method was used to obtain the light curve of the target. The data of the light curves can be found in ChinaVO (\url{https://nadc.china-vo.org/res/r101411/}). 

\subsection{Spectroscopic Observations}

\begin{figure*}
    \centering
    \includegraphics[width=\linewidth]{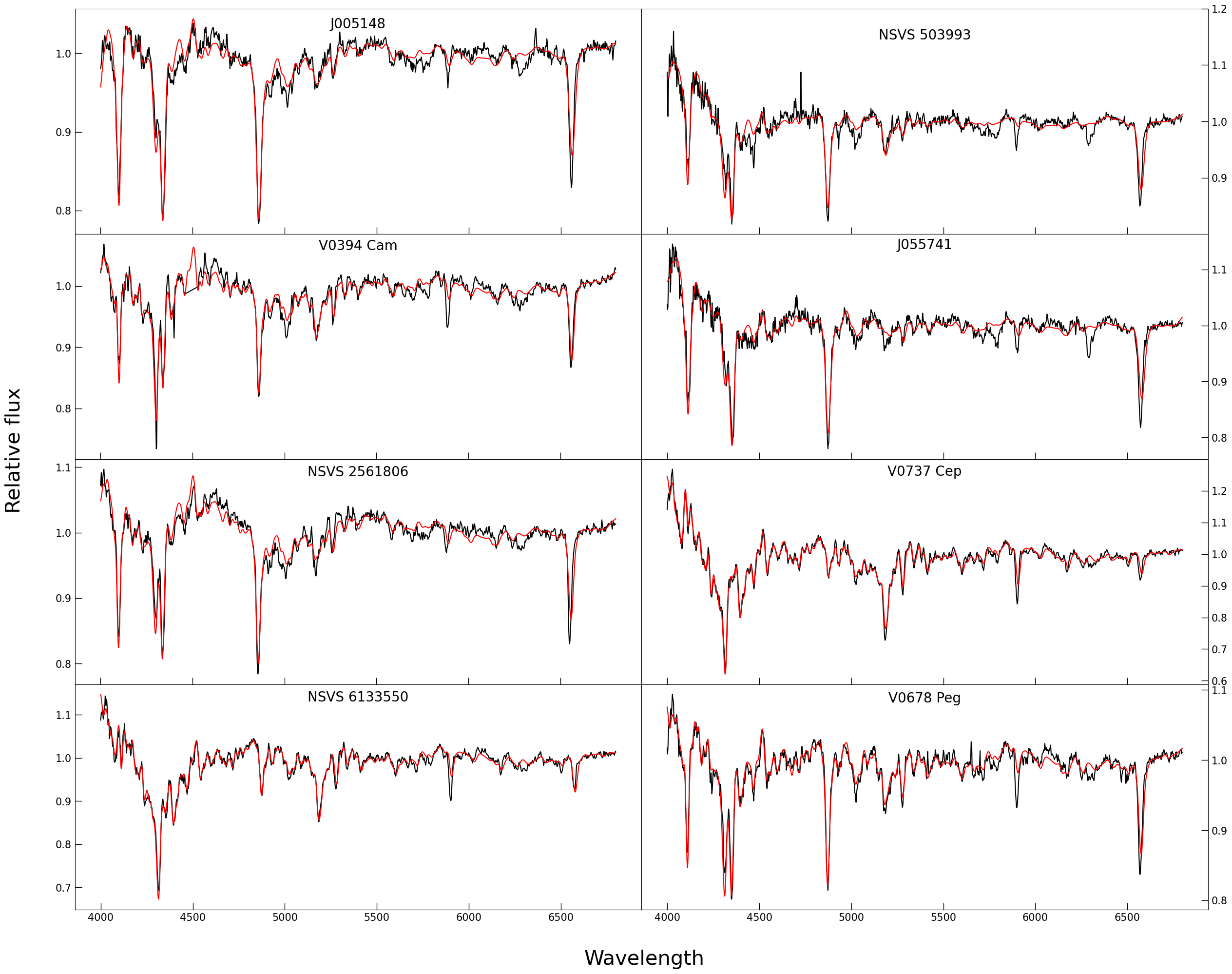}
    \caption{Using ULySS to fit the spectral data. The black line represents the original data and the red line represents the fitted data.}
    \label{fig:fig1}
\end{figure*}

From 2021 to 2022, we conducted spectroscopic observations of these targets using the Xinglong Observatory 2.16 m telescope. During our observation, we used Beijing Faint Object Spectrograph and Camera (BFOSC) and employed G4 mode, with a single pixel spectral resolution of 2.97  ($Å$) and a wavelength coverage of 3850-7000  ($Å$). The observation time, exposure time, and signal-to-noise ratio of these eight targets are shown in  \Cref{tab:label3}. These spectral data were processed using IRAF. Bias subtraction and flat correction were performed, followed by the elimination of cosmic rays.
Subsequently, the processed spectra were obtained through spectral extraction, wavelength calibration, and flux normalization.
Then, the University of Lyon Spectroscopic analysis Software \cite[ULySS;][]{ULySS} was used to fit the normalized spectral data, and the atmospheric parameters were obtained, such as the temperature, [Fe/H], and log g, the corresponding parameters are recorded in  \Cref{tab:label3}. The fitted image of the spectrum is shown in \Cref{fig:fig1}.

\renewcommand\arraystretch{1.2}%设置行高
\begin{table*}[t]
\caption{\centering The information of the eight Targets }
\label{tab:label1}
\centering
\small
\begin{tabular}{cccccccc}

\hline
Star&Other name& R.A.&   Decl.&Period (d)&   $HJD_{0}$ &V(mag)&Amplitude\\

\hline
ASASSN-V J005148.89+203951.3 &J005148&00 51 48.89&  +20 39 51.3&0.403534&   2459491.17611&12.36&0.43\\ 
ASASSN-V J032300.09+723548.2 &NSVS 503993& 03 23 00.09&   +72 35 48.2&0.363766&   2459528.18458&12.33&0.44\\ 
ASASSN-V J053450.34+722645.7 &V0394 Cam& 05 34 50.34&   +72 26 45.7&0.381499&   2458558.14560&13.22&0.38\\ 
ASASSN-V J055741.40+374319.0 &J055741&05 57 41.40&  +37 43 19.0&0.414000&   2459206.25179&12.31&0.42\\

 ASASSN-V J104331.45+632301.8 &NSVS 2561806& 10 43 31.45& +63 23 01.8& 0.420700&   2459299.19981&11.69&0.41\\ 
 ASASSN-V J212348.30+633329.2 &V0737 Cep& 21 23 48.30& +63 33 29.2& 0.298763&   2459107.23350&12.04&0.43\\ 
 ASASSN-V J225517.42+421633.8 &NSVS 6133550& 22 55 17.42& +42 16 33.8& 0.275135&   2459114.37803&11.48&0.42\\ 
 ASASSN-V J231929.22+180135.0 &V0678 Peg& 23 19 29.22& +18 01 35.0& 0.408399&   2459128.40184&12.19&0.46\\ 
 
\hline
\end{tabular}

\end{table*}

\renewcommand\arraystretch{1.2}%设置行高
\begin{table*}[t]
\caption{\centering Photometric observation Details of the eight Targets}
\label{tab:label2}
\centering
\small
\begin{tabular}{clcccc} 

\hline
Star &Observing Date& Exposure Time(s)&   Observational Errors(mag)&Comparison Star&   Check Star \\ 
\hline

J005148&2021 Sep 25,26,27,29,Oct 3&g23 r20 i25&   g0.009 r0.009 i0.010 
& J005144+204237&   J005140+204028\\ 
NSVS 503993&2021 Nov 9,16,21& g23 r20 i25&   g0.012 r0.010 i0.015&J032051+723822&   J032429+722646\\ 
V0394 Cam&2019 Feb 25,27,Mar,15,16& B100 V60 R40 I50&   B0.005 V0.006 R0.005 I0.007&J053432+722607&   J053413+722620\\ 
J055741&2020 Dec 21,22&V25 R18 I25&  V0.005 R0.005 I0.005&J055742+374239&   J055745+374147\\ 

 NSVS 2561806&2021 Mar 25,26,Apr 4,5,14& g12 r9 i15& g0.013 r0.011 i0.012& J104430+631928&   J104317+631952\\ 
 V0737 Cep&2020 Sep 11,14& V25 R18 I23& V0.005 R0.008 I0.008& J212316+633559&   J212352+633610\\ 
 NSVS 6133550&2020 Sep 21,22& V13 R9 I12& V0.006 R0.006 I0.006& J225554+421858&   J225542+422124\\ 
 V0678 Peg&2020 Sep 23,Oct 5,8& V25 R18 I23& V0.009 R0.011 I0.010& J231940+180014&   J231951+180306\\ 
 
\hline
\end{tabular}

\end{table*}

\renewcommand\arraystretch{1.2}%设置行高
\begin{table*}[t]
\caption{\centering Spectral observation information and the atmospheric parameters}
\label{tab:label3}

\begin{tabular}{ccccccc}

\hline
Star 
 & Observing Date&Exposure Time(s) &SNR&   Teff(K)&[Fe/H](dex)&   log g(dex)\\

\hline
J005148
 & 2022 Jan 5&600 &141&  6546($\pm$16)& -0.269($\pm$0.046)&   3.975($\pm$0.054)\\ 
NSVS 503993
 & 2023 Jan 15& 750&84&   6294($\pm$24)&-0.717($\pm$0.056)&   4.434($\pm$0.062)\\ 
V0394 Cam
 & 2022 Jan 27& 1500&179&   5994($\pm$21)&-0.427($\pm$0.044)&   4.031($\pm$0.064)\\ 
J055741& 2023 Jan 15& 750&68&  6529($\pm$20)&-0.198($\pm$0.049)&   3.582($\pm$0.064)\\

 NSVS 2561806
 & 2022 Jan 27& 500&182& 6311($\pm$18)& -0.291($\pm$0.043)&   3.831($\pm$0.059)\\ 
 V0737 Cep
 & 2022 Nov 24& 600&76& 5126($\pm$19)& -0.148($\pm$0.028)&   4.224($\pm$0.032)\\ 
 NSVS 6133550
 & 2022 Nov 24& 400&101& 5465($\pm$17)& -0.404($\pm$0.030)&   4.174($\pm$0.037)\\ 
 V0678 Peg & 2022 Nov 24& 600&95& 6215($\pm$16)& -0.028($\pm$0.028)&   4.012($\pm$0.042)\\ 
 
\hline
\end{tabular}

\end{table*}

\section{Analysis of the light curves}

\subsection{data preparation }

First, the  method of \cite{kw1956} was used to calculate the eclipsing times of the light curves of these eight contact binaries. 
Then we used the primary eclipsing times as the zero point and the equation: $HJD=HJD_0+P\times E$, to convert the data from time to phase. Next, obvious outliers were manually removed to ensure data quality. Finally, the magnitudes were converted into flux, which were then normalized to obtain the final light curves.

Version 2.4 of PHOEBE \citep{Prša2005,Prša2016,Horvat2018,Conroy2020,Jones2020} was used to determine the physical parameters of the eight contact binaries. The contact model  was chosen for calculations. 
First, we set the effective temperature obtained from the spectral analysis as the temperature of the primary component.  
The gravity darkening and bolometric albedo coefficients of all stars are set to $g_{1,2}$ = 0.32 and $A_{1,2}$=0.5 respectively due to \cite{Lucy1967} and \cite{Rucinski1969}. We choose the atmospheric model of \cite{ck2004}, and limb-darkening coefficients were derived by a logarithmic law.

 \subsection{Calculation of initial values of PHOEBE }
According to PHOEBE's introduction, an accurate prior can help the model converge faster. Therefore, we would like to use accurate values as priors for Markov Chain Monte Carlo (MCMC) sampling. The physical parameters that need to be calculated for contact binary stars are mass ratio ($q$), the effective temperature of the secondary component ($T_{2}$), orbital inclination ($i$), luminosity of the primary component ($l_1$), and degree of contact ($f$). 
If  the target exhibits the O'Connell effect, we add a spot on the primary component. To quickly obtain the spot parameters, $\theta$ (the latitude of the spot) is fixed at $90^{\circ}$, and other parameters, such as $\lambda$ (the longitude of the spot), $r_s$ (the radius of the spot) and $T_s$ (the relative temperature of the spot), are calculated accordingly. 
When calculating the initial values of these parameters, we set PHOEBE's $pblum\_mode$ to $dataset-scaled$ to eliminate the need to calculate the luminosity of the primary component.

Two methods are used to obtain the initial values of the parameters for PHOEBE MCMC sampling. One method offers faster calculation speed but cannot model parameters with spots, while the other method can model parameters with spots but takes longer calculation time. 
Since there is a possibility that $q$ is greater than 1, we need to set the parameter phaseshift to 0.5 and recalculate the parameters using the same method. By comparing with the Residual Sum of Squares (RSS) of the fitting residuals, the parameters with the smallest RSS  are selected as the final results.

The first method is a new algorithm proposed by ourselves. For the light curve without O'Connell effect, where only four parameters ($q$, $T_2$, $i$, $f$) need to be determined, the parameter space can be divided into equally space intervals. Subsequently, computational traversals can be executed to pinpoint the optimal fitting parameters. 
To enhance efficiency, a two-step calculation approach is employed. Initially, exploration occurs within a parameter space featuring wide intervals. For example, in this step, the parameter interval for $q$ is 0.1, and for $i$ is 2$^\circ$.
Subsequently, a narrow interval is applied for parameter search in proximity to the optimal solution from the previous step, thereby enhancing both speed and accuracy. And in this step, the parameter interval for $q$ is 0.02, and for $i$ is 1$^\circ$. 
Of course, we can include other parameters, such as the parameters of the spot, in this calculation if we want, but this may cause the computation time to grow exponentially, which is undesirable for us. 
Utilizing a computer equipped with 80 cores to process 400 data points yields the desired initial values within a mere 10 minutes. Compared to the second method, this method demonstrates superior speed and result stability. 
This method's precision is listed in  \Cref{tab:label4}. However, it is unsuitable for light curve with O'Connell effect, presenting a notable limitation in its applicability. 

Therefore, we employed the second method to compute the light curve with  O'Connell effect. The second method is Genetic algorithm \cite[GA;][]{ga}. 
GA is a computational method inspired by natural evolution, rooted in Darwin's theory. It can efficiently explores parameter spaces, seeking optimal solutions without predefined rules, and autonomously adapts and refines search strategies, exhibiting robust exploratory capabilities. 
The procedure of a GA involves evaluating the fitness of each individual based on predefined criteria, selecting the fittest individuals from the population, and applying genetic operations like crossover and mutation to generate a new population. This process iterates until termination criteria are satisfied. GA excels in exploring parameter spaces devoid of directional cues, enabling the search for globally optimal solutions.
In the actual operation, python package scikit-opt\footnote{\url{https://github.com/guofei9987/scikit-opt}} was used to implement GA, setting the population size $size\_pop=500$, the mutation rate $prob\_mut=0.01$, and the number of iterations $max\_iter=100$.
Since it is necessary to obtain an initial value close to the real value, the  space range of each parameter is divided into a finite number of equidistant data points, and this addresses the issue of potentially excessive computation time caused by overly high precision. For example, we set the minimum interval for $q$ to 0.02, \Cref{tab:label4} gives the precision for each parameter.

After obtaining the values of  $q$, $i$, $T_2$, $f$, $\lambda$, $r_s$ and $T_s$, the value of primary luminosity can be automatically determined, and these initial values are input as the priors of the MCMC calculation.

It should be noted that we additionally obtain a set of parameters which contributes to the third light ($l_3$) in our calculation. Therefore, we end up with two distinct sets of parameters; one includes $l_3$, whereas the other does not. We have incorporated the third light into both algorithms mentioned above, with specific search ranges and interval settings as detailed in \Cref{tab:label4}.
\renewcommand\arraystretch{1.2}%设置行高
\begin{table*}[t]
\caption{\centering Search range and precision for the physical parameters}
\label{tab:label4}
\begin{tabular}{ccccc} 

\hline
Parameters&Range& Minimum precision of Method 1&   Minimum precision of Method 2&$\sigma$\\ 
\hline

$q$&[0, 1]&0.02&   0.02 &0.01\\ 
$i(deg)$&[60, 90]& 1&   1 &5\\ 
$T_{2}(K)$&[$T_{1}$-1000, $T_{1}$+500]& 100&   100 &50\\ 
$f$&[0, 1]&0.1&  0.1 &0.05\\ 

 $\lambda(deg)$&[0, 360]& /& 10 &5\\ 
 $r_s(deg)$&[6, 60]& /& 2 &1\\ 
 $T_s$&[0.65, 0.95]& /& 0.05 &0.05\\
 
 $L_1$& /& /& /&0.1\\
 $l_3$& [0, 0.6]& 0.1& 0.1&0.05\\
 \hline
\end{tabular}
\end{table*}

 \subsection{MCMC calculation}\label{mcmc}
 Before performing MCMC, the prior distribution of the parameters was chosen to follow a Gaussian distribution, with the previously calculated parameters used as the mean value of the Gaussian distribution, and set $\sigma$ for different parameters as shown in  \Cref{tab:label4}.
When dealing with parameters containing spots, a higher number of  $nwalkers$  is needed. Consequently, $walker$ is set to 32 for parameters with spots, while the remaining uses 24. The initial iteration number is set as 2000. 
In order to confirm the convergence of the MCMC calculation, it is ensured that the number of iterations for each parameter is 10 times of the autocorrelation time according to \cite{Conroy2020}. During the operation, it can be seen that certain $nwalker$ chains may converge to local values. However, this does not affect the ultimate result. To uphold calculation rigor, we opt to discard convergence chains with lower $lnprobability$ (the cost function) branch and only retain those with higher $lnprobability$ (indicating closer proximity to the real value) branch, and resample all walkers from the higher branch and continue iterations from there. 

\begin{figure*}
    \centering
    \includegraphics[width=\linewidth]{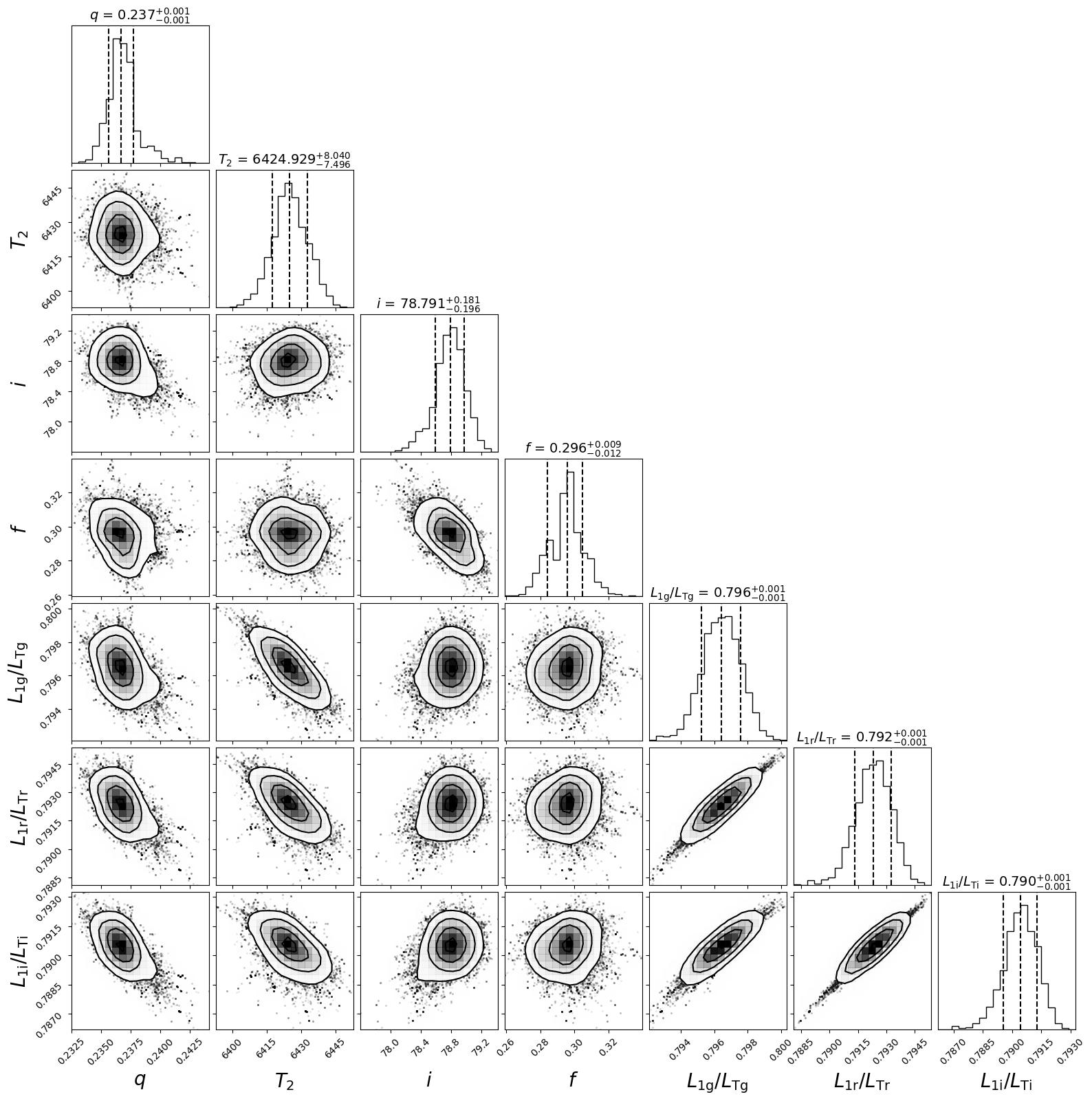}
    \caption{Probability distributions of $q$, $T_{2}$, $i$, $f$, $L_{1g}/L_{Tg}$, $L_{1r}/L_{Tr}$ and $L_{1i}/L_{Ti}$ determined by the $MCMC$ modeling of J005148.}
    \label{fig:fig3}
\end{figure*}

\begin{figure*}
    \centering
    \includegraphics[width=\linewidth]{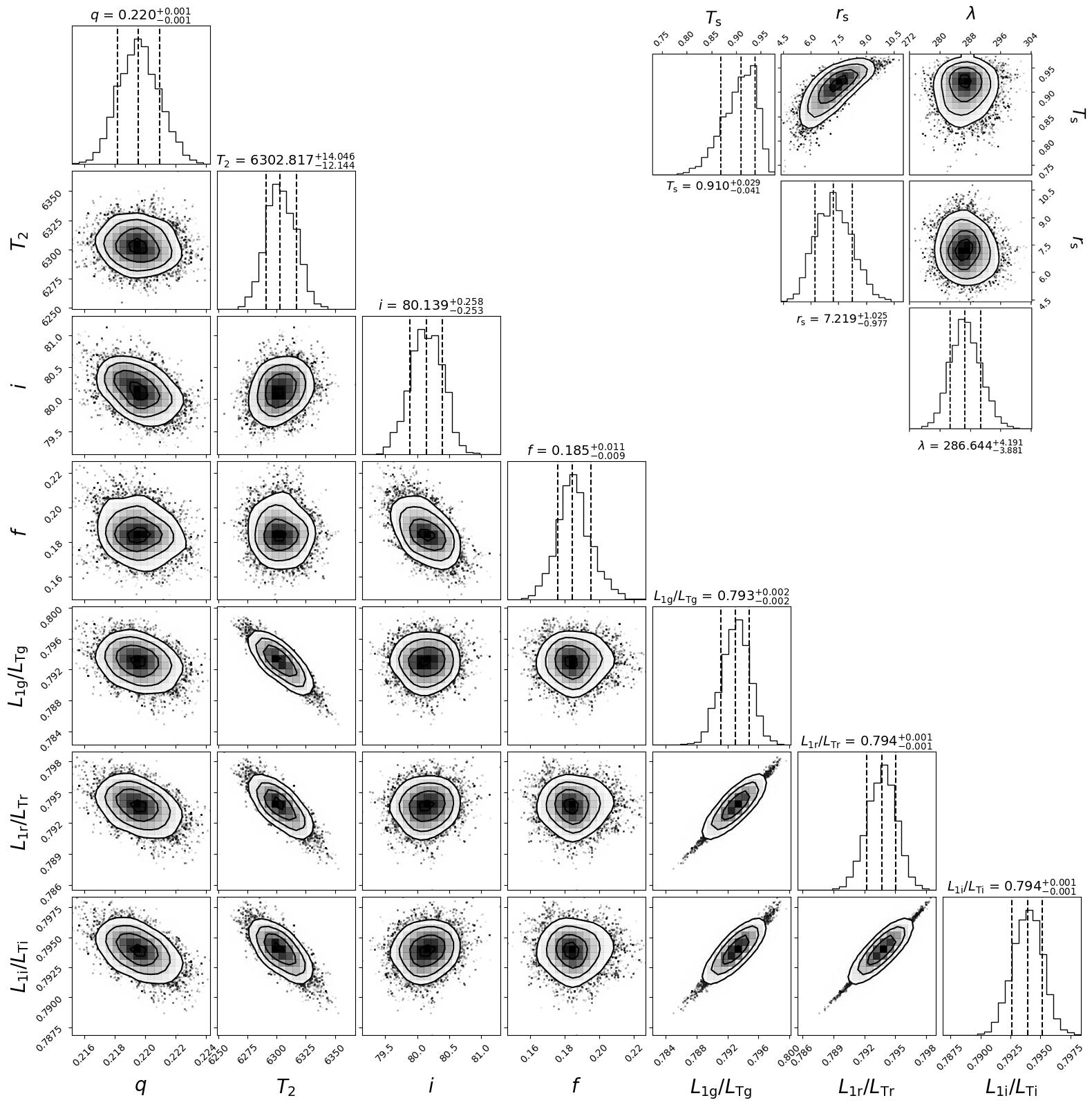}
    \caption{Probability distributions of $q$, $T_{2}$, $i$, $f$, $L_{1g}/L_{Tg}$, $L_{1r}/L_{Tr}$, $L_{1i}/L_{Ti}$, $T_s$, $r_s$ and $\lambda$ determined by the $MCMC$ modeling of NSVS 503993.}
    \label{fig:fig4}
\end{figure*}

\begin{figure*}[htbp]

\subfigure{
\includegraphics[width=9cm,height = 5cm]{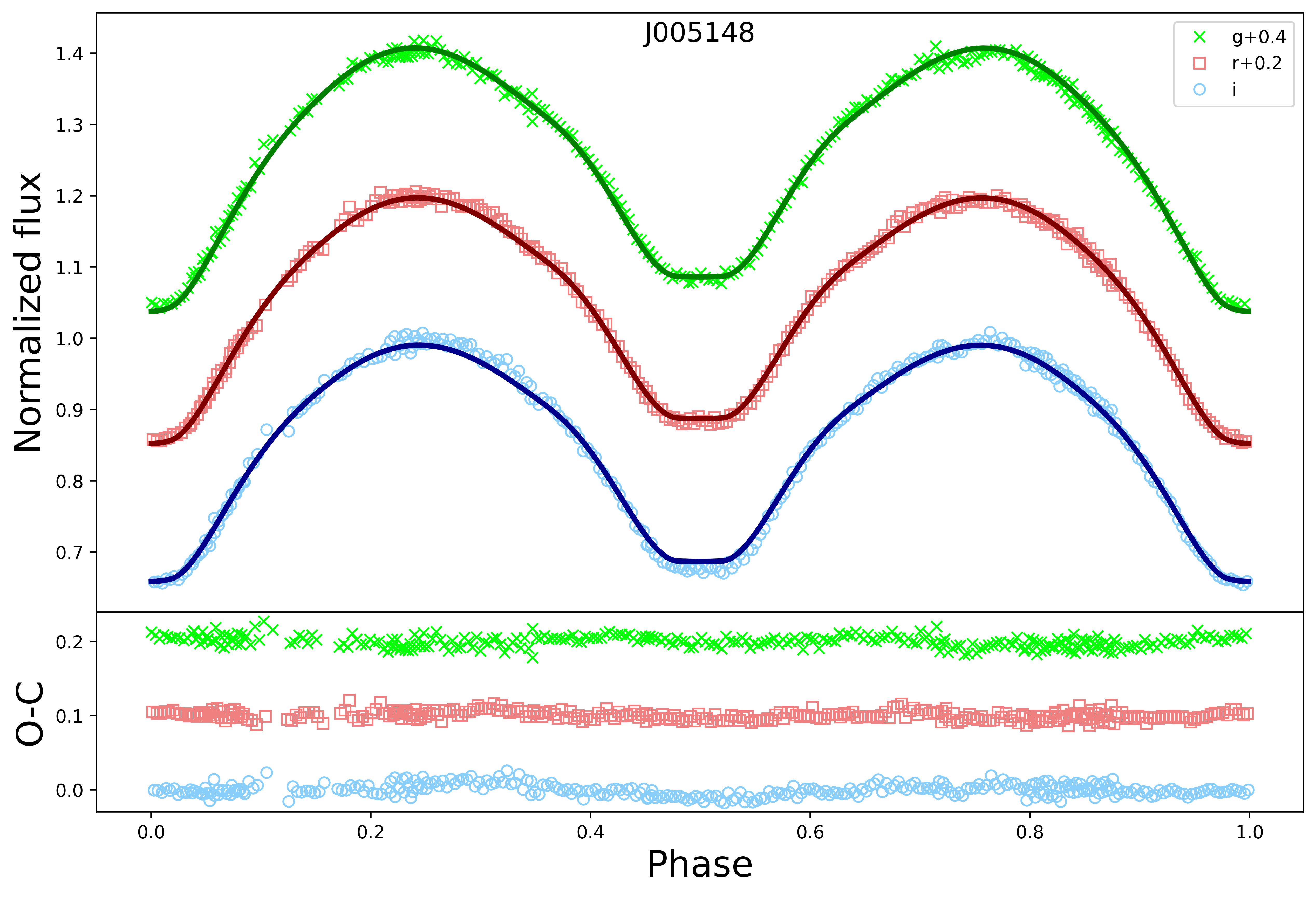}}
\subfigure{
\includegraphics[width=9cm,height = 5cm]{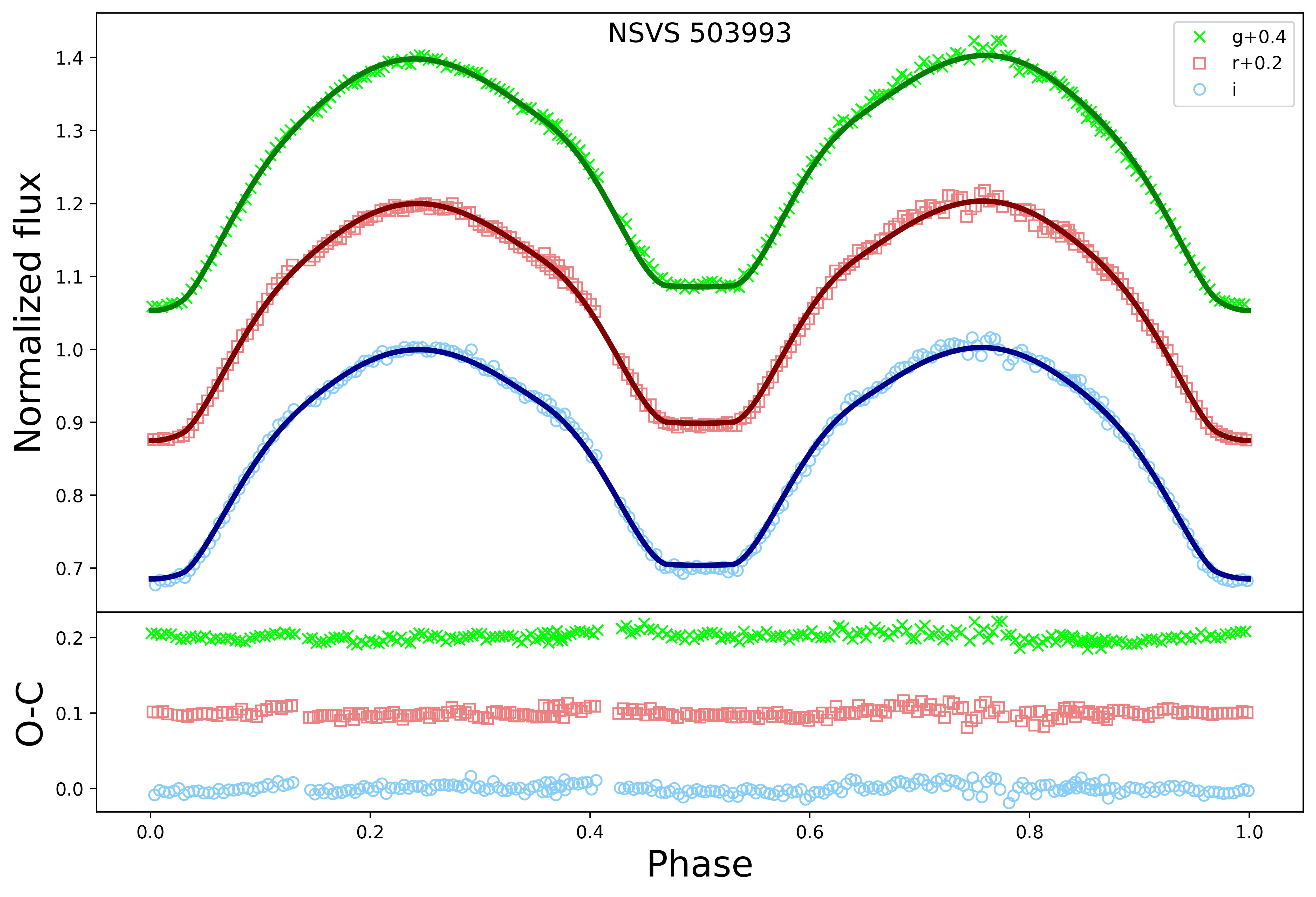}}
\subfigure{
\includegraphics[width=9cm,height = 5cm]{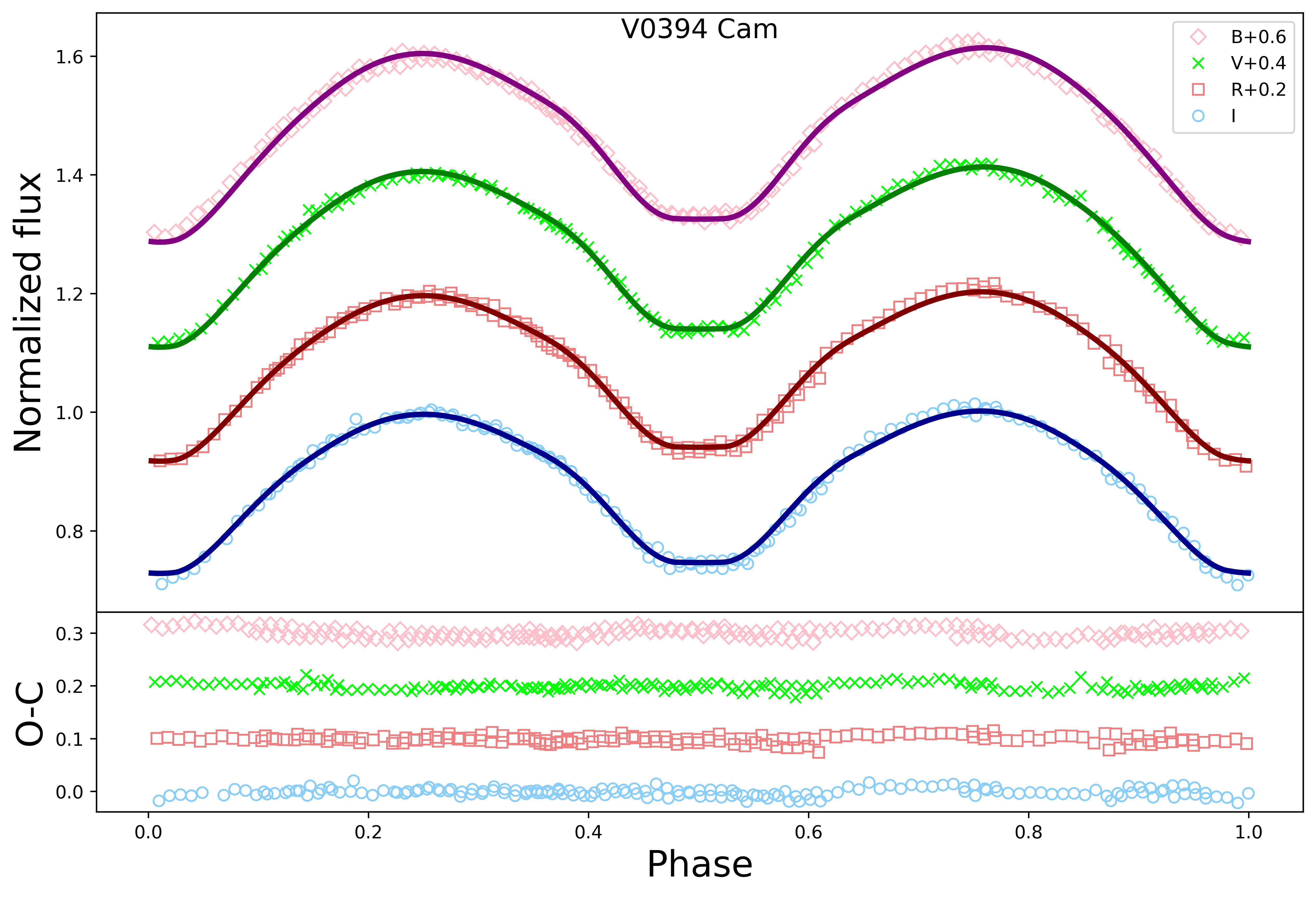}}
\subfigure{
\includegraphics[width=9cm,height = 5cm]{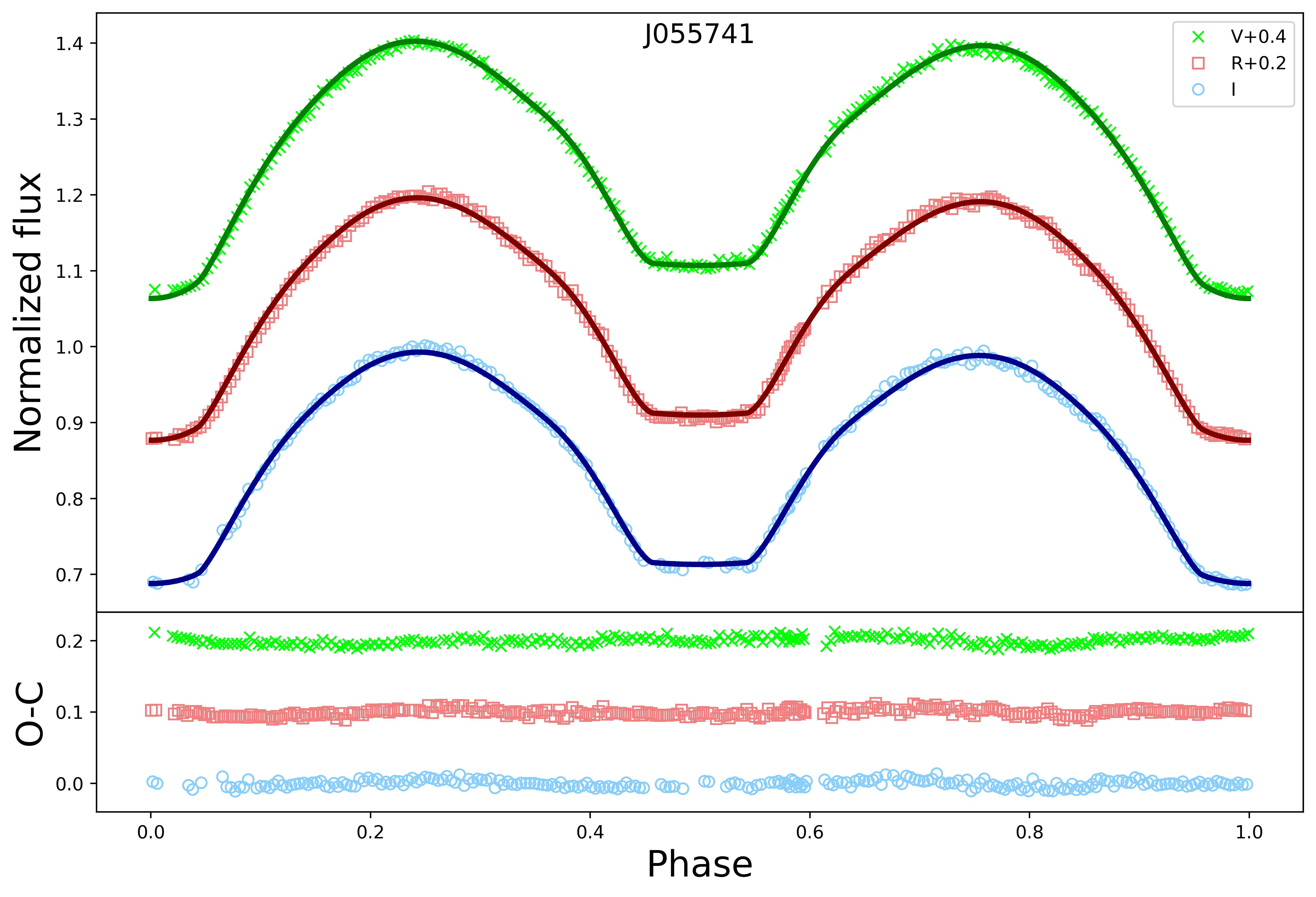}}
\subfigure{
\includegraphics[width=9cm,height = 5cm]{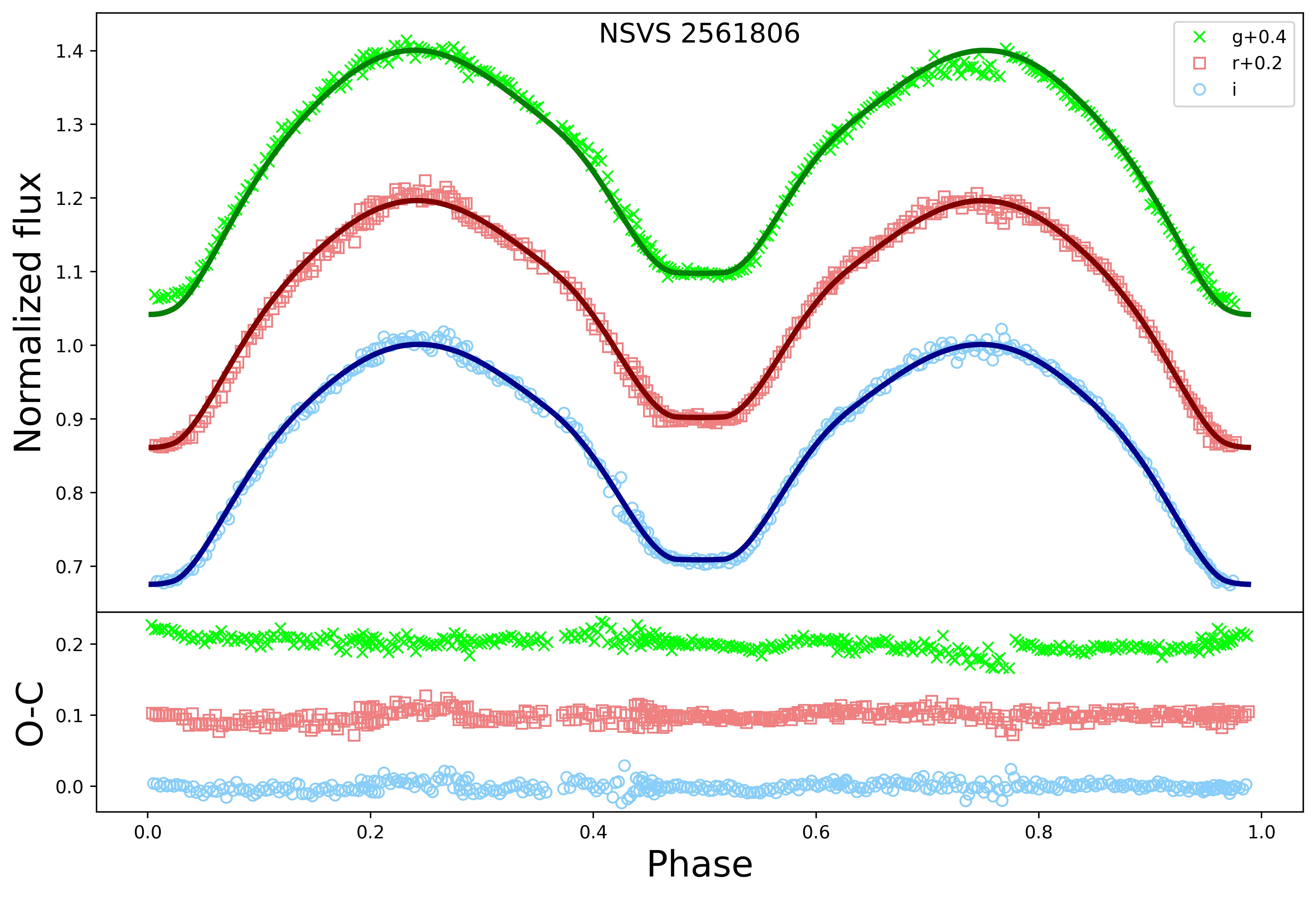}}
\subfigure{
\includegraphics[width=9cm,height = 5cm]{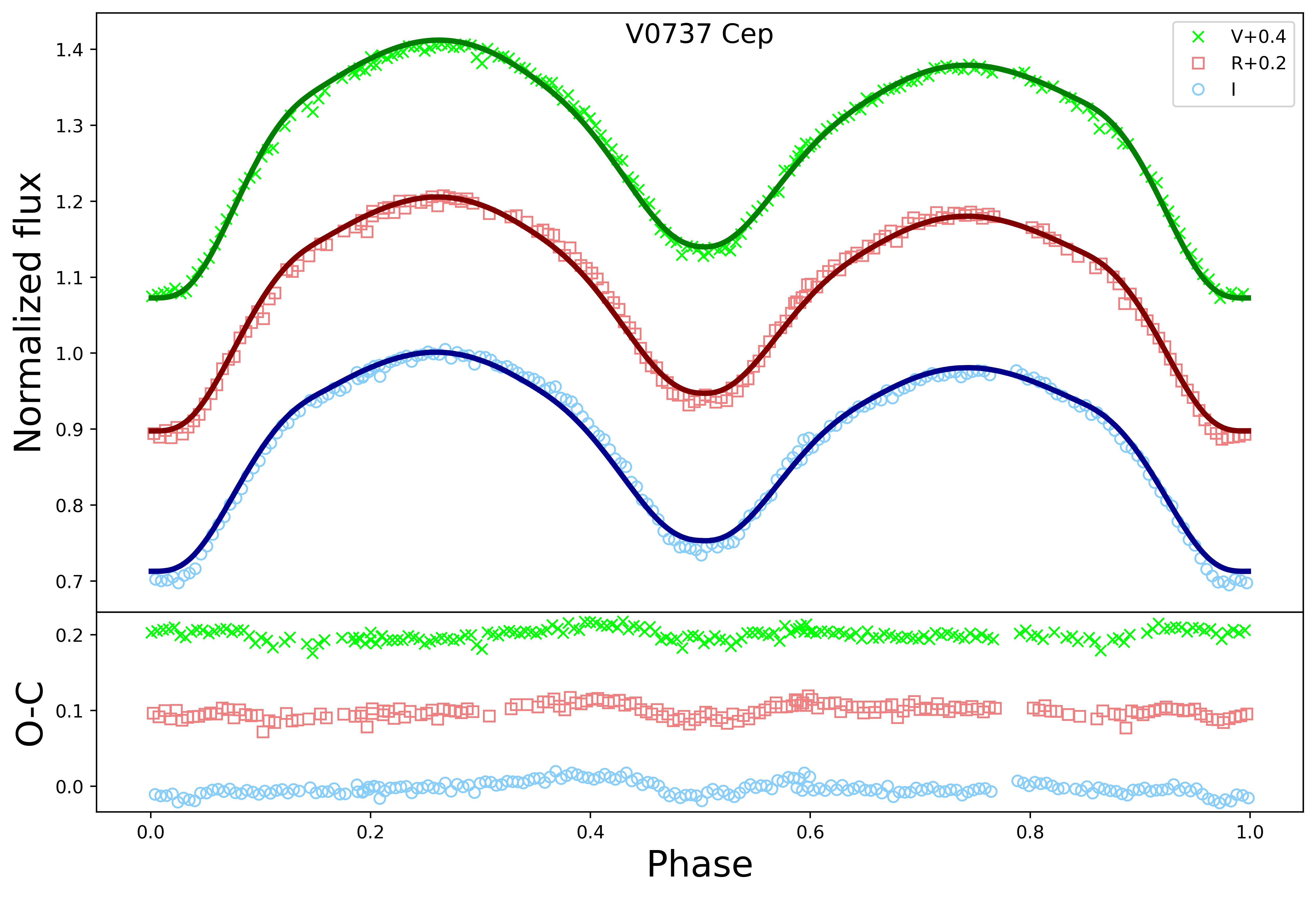}}
\subfigure{
\includegraphics[width=9cm,height = 5cm]{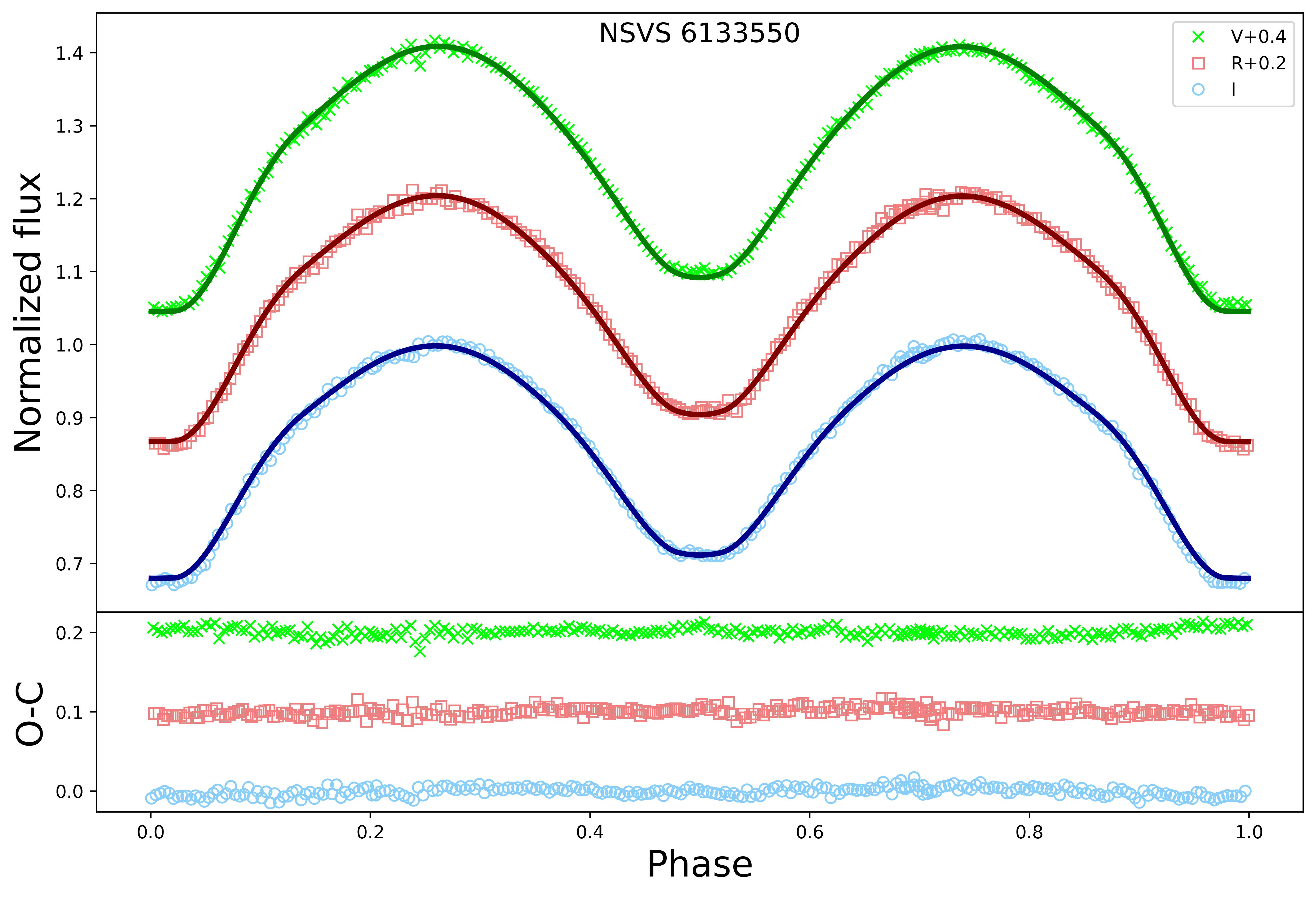}}
\subfigure{
\includegraphics[width=9cm,height = 5cm]{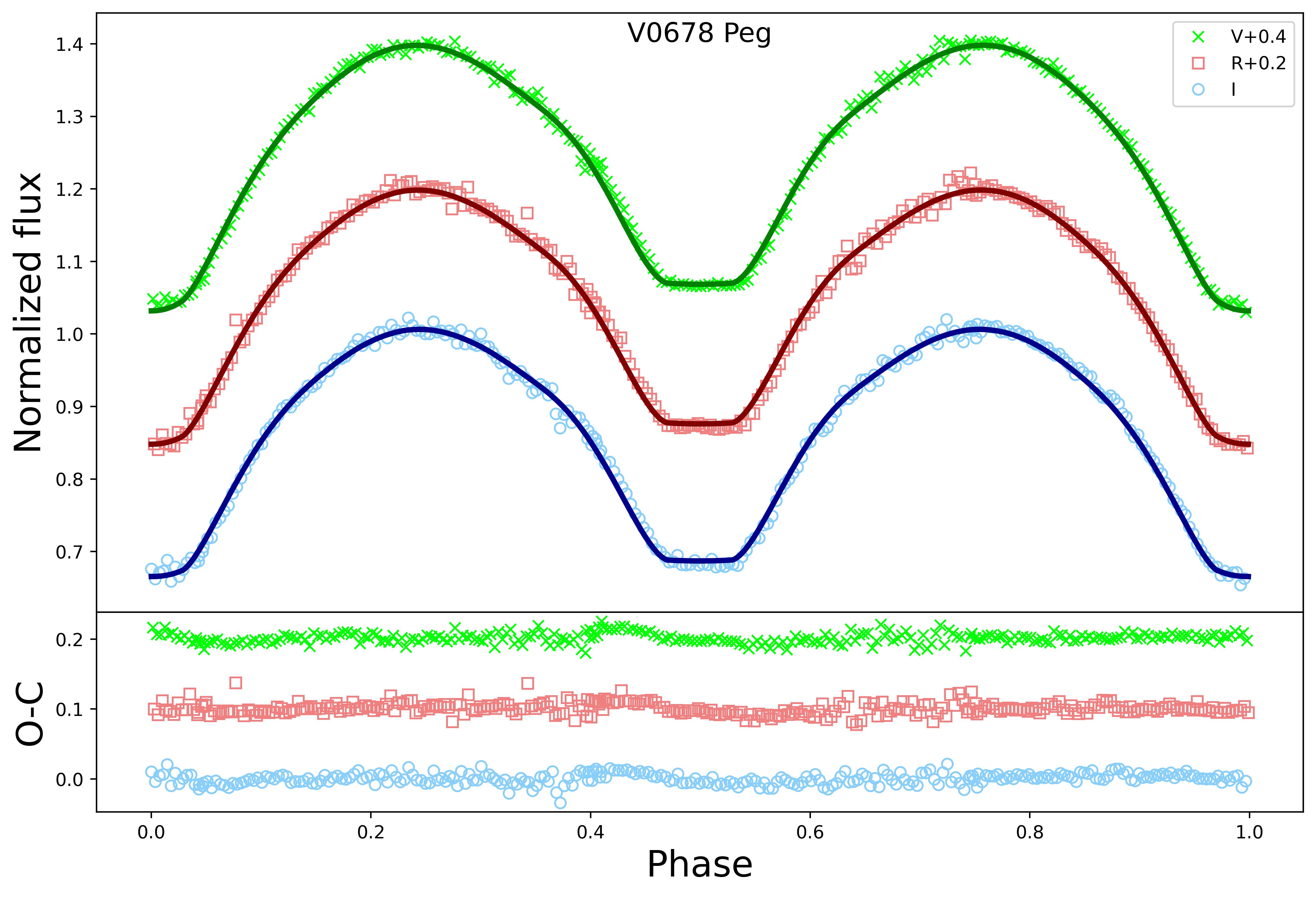}}
\caption{These figures show the fitting results of each target. The scatter points are the real observation data, the curves are the fitting results, and the fitting residuals are shown in the lower panel.}
\label{fig:fig2}
\end{figure*}

\begin{figure*}
    \centering
    \includegraphics[width=\linewidth]{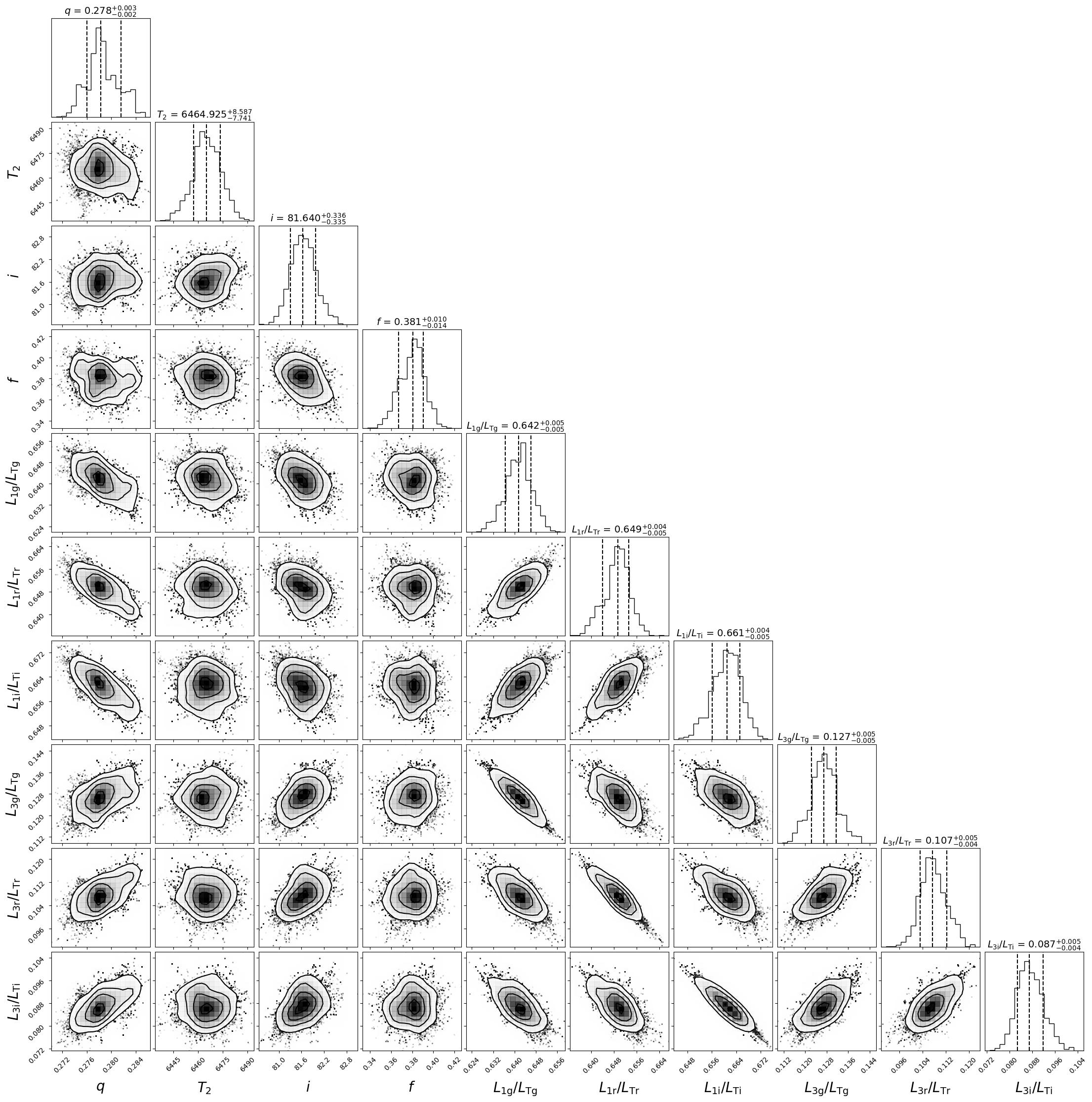}
    \caption{Probability distributions of $q$, $T_{2}$, $i$, $f$, $L_{1g}/L_{Tg}$, $L_{1r}/L_{Tr}$ , $L_{1i}/L_{Ti}$, $L_{3g}/L_{Tg}$, $L_{3r}/L_{Tr}$ and $L_{3i}/L_{Ti}$ determined by the $MCMC$ modeling of J005148.}
    \label{fig:fig3_mcmc}
\end{figure*}

\begin{figure*}[htbp]

\subfigure{
\includegraphics[width=9cm,height = 5cm]{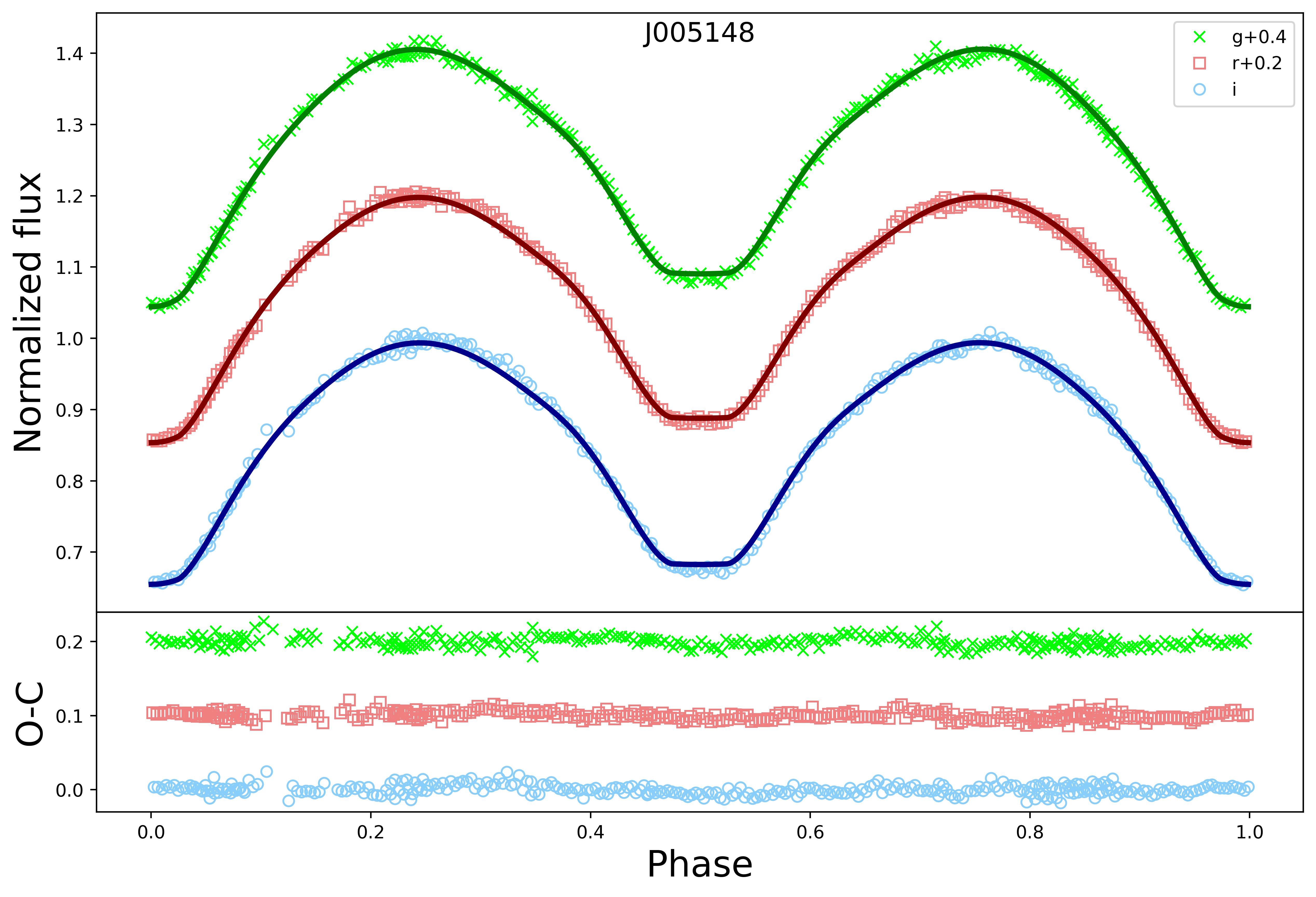}}
\subfigure{
\includegraphics[width=9cm,height = 5cm]{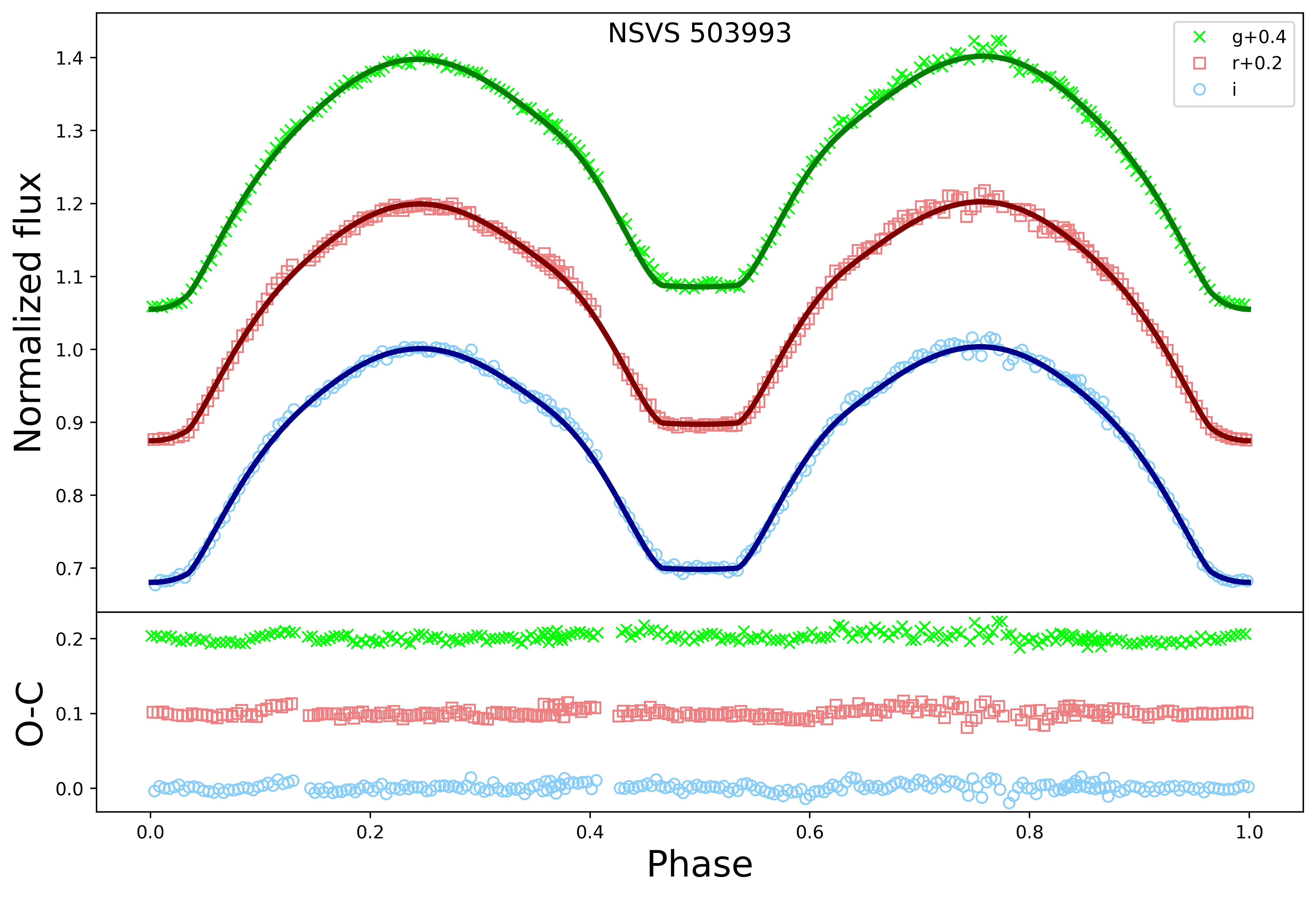}}
\subfigure{
\includegraphics[width=9cm,height = 5cm]{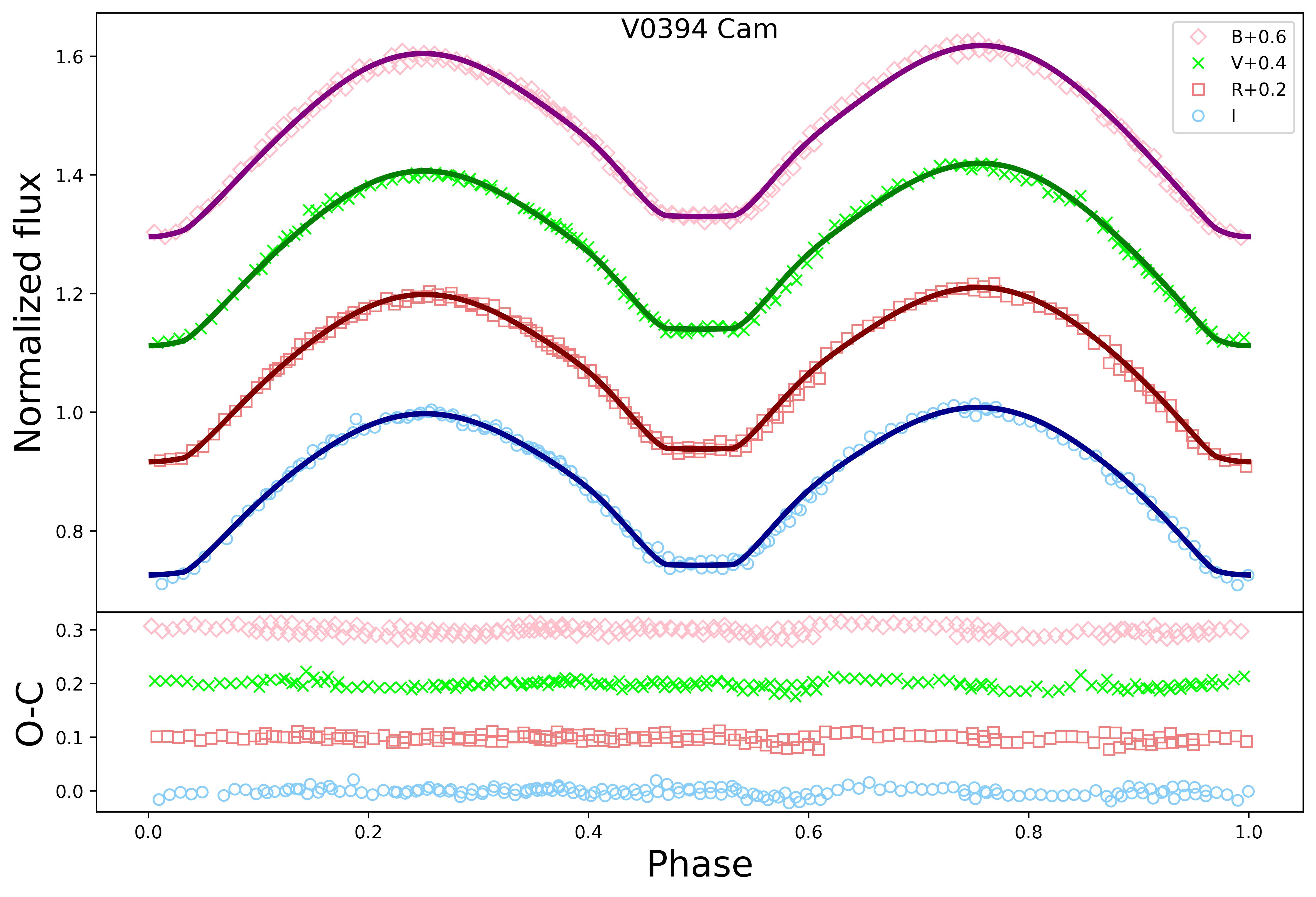}}
\subfigure{
\includegraphics[width=9cm,height = 5cm]{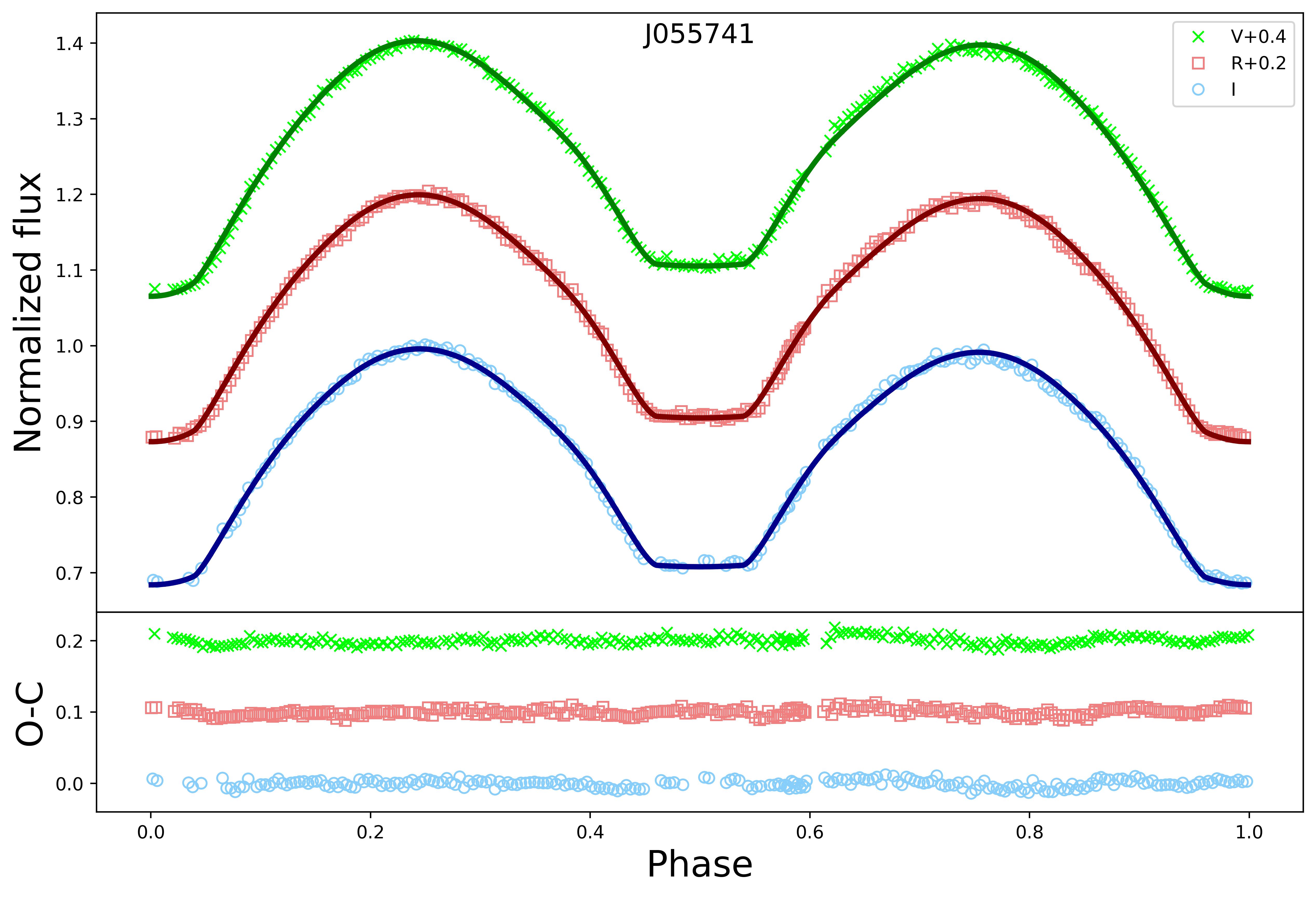}}
\subfigure{
\includegraphics[width=9cm,height = 5cm]{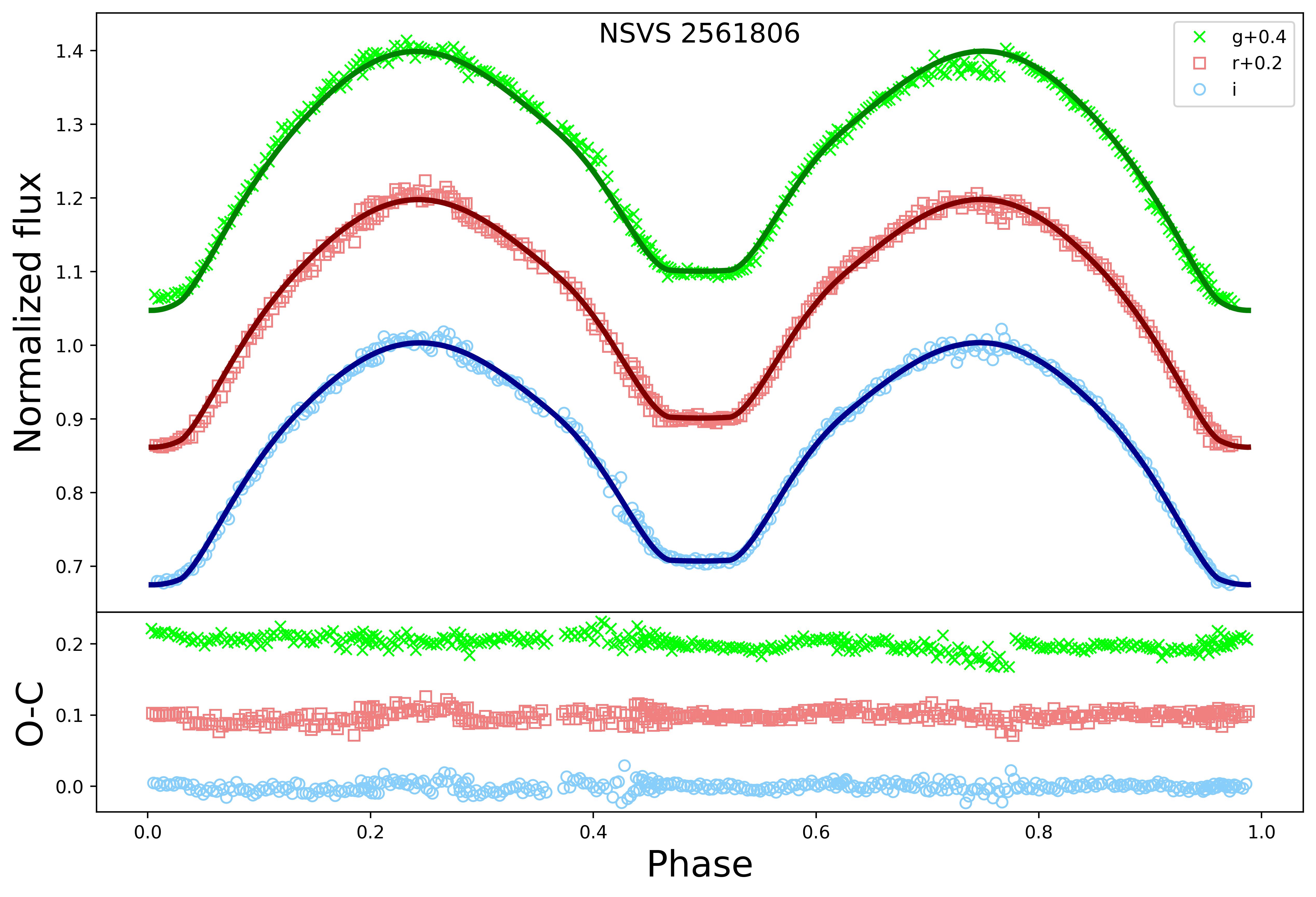}}
\subfigure{
\includegraphics[width=9cm,height = 5cm]{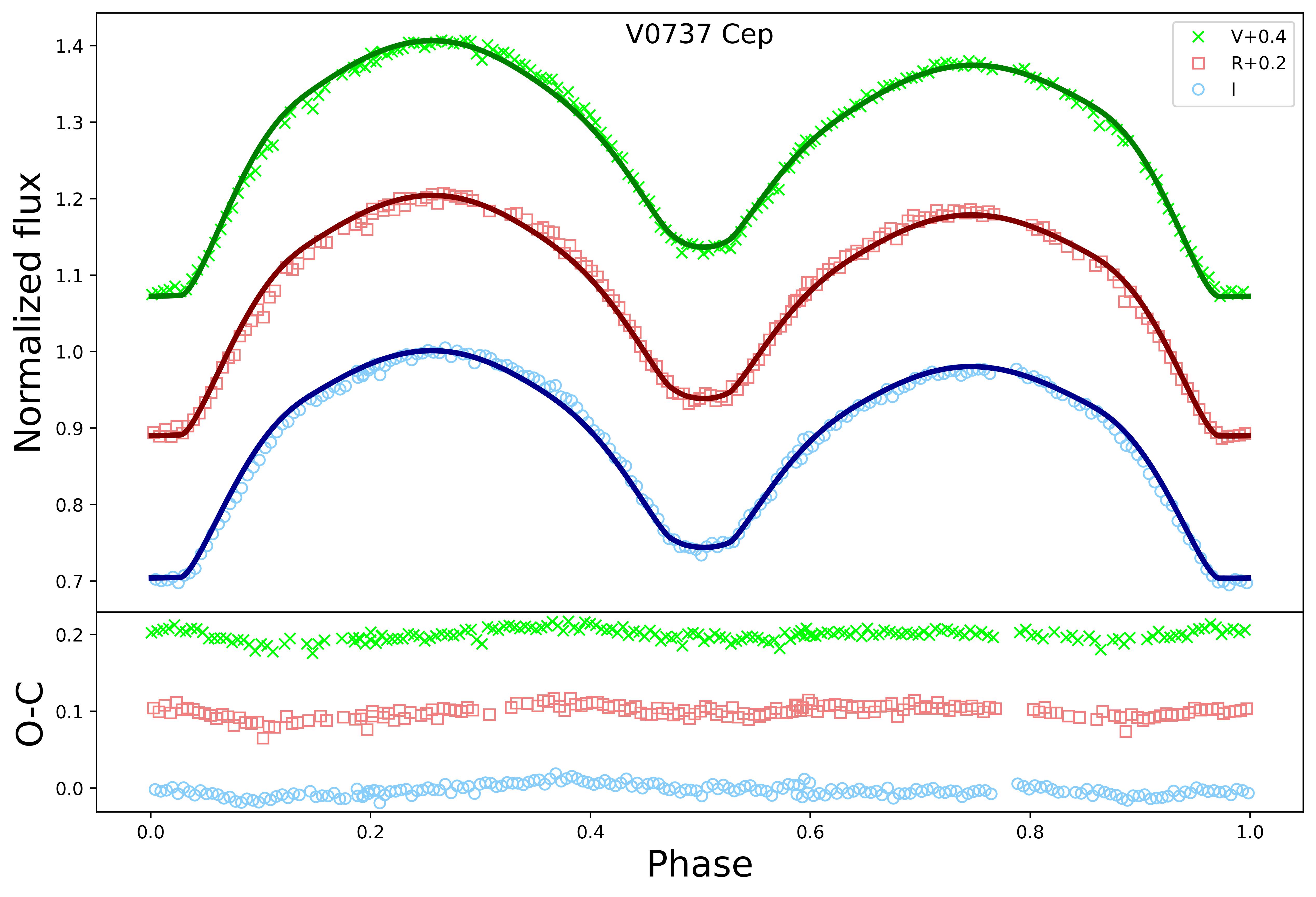}}
\subfigure{
\includegraphics[width=9cm,height = 5cm]{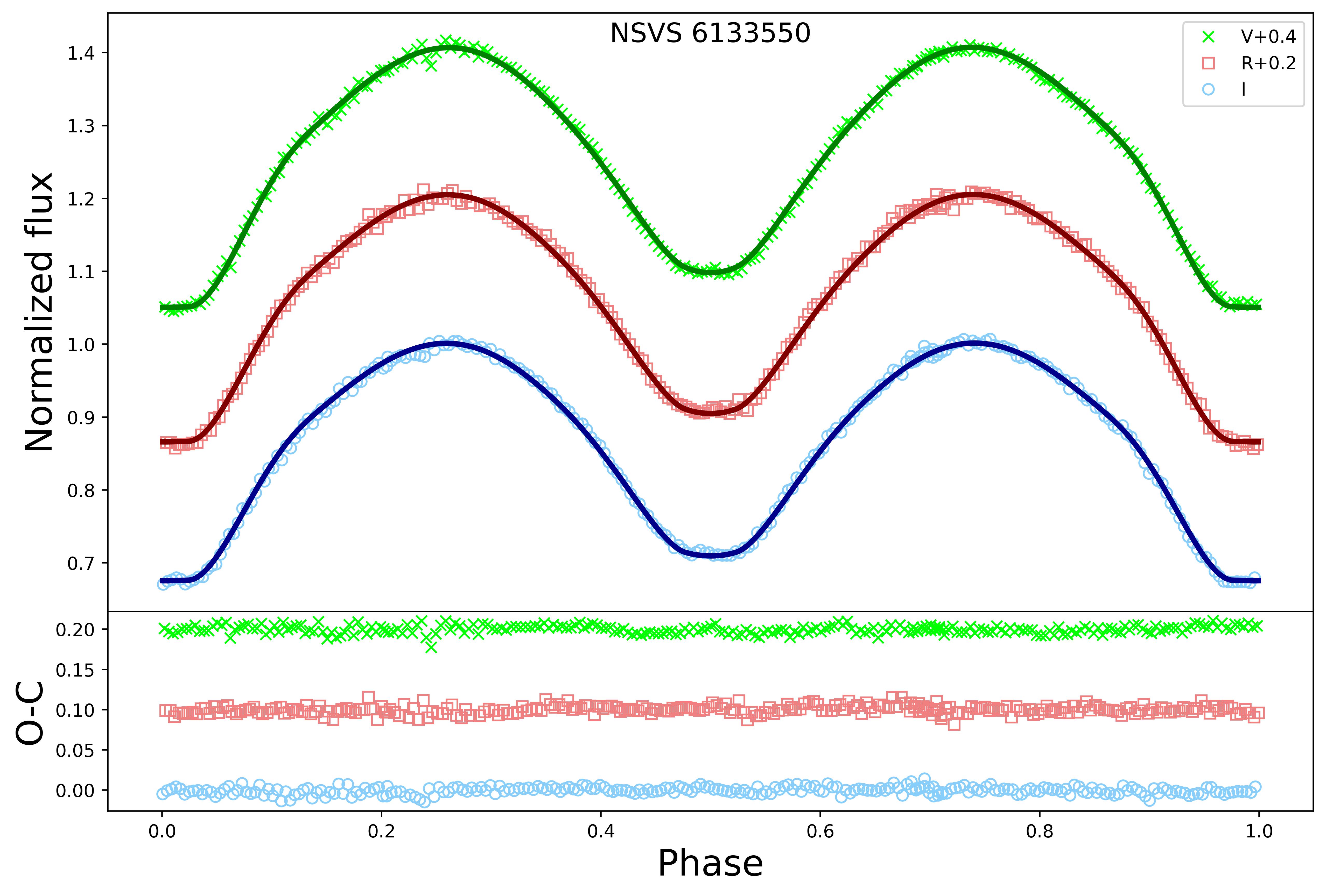}}
\subfigure{
\includegraphics[width=9cm,height = 5cm]{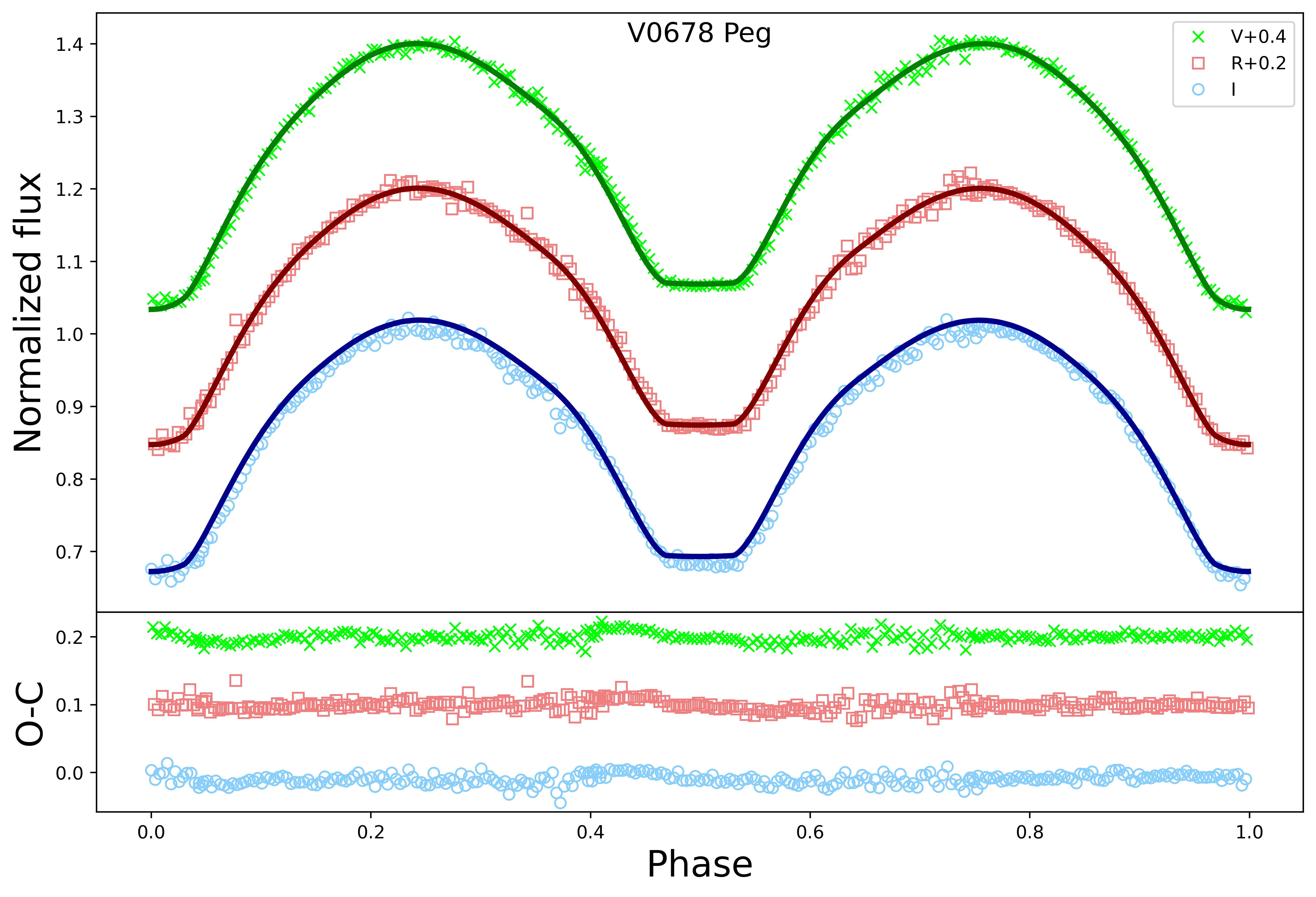}}
\caption{These figures show the fitting results of each target after incorporating $l_3$. The scatter points are the real observation data, the curves are the fitting results, and the fitting residuals are shown in the lower panel.}
\label{fig:fig2_l3}
\end{figure*}

Regarding results that do not include  $l_3$, the posterior distribution  of two targets (with spot and without spot, respectively) are shown in \Cref{fig:fig3,fig:fig4}  for example. \Cref{fig:fig2} shows the observed data, fitted curves, and O-C residuals. 
And physical parameters of the eight systems are summarized in \Cref{tab:label5,tab:label6}.  It should be mentioned that, the errors of these parameters are all underestimated \citep{Prša2005}. We combined the error of $T_1$ due to spectral fitting with the error of $T_2$ determined by MCMC  sampling, resulting in the final error for $T_2$. 
And we show the result with $l_3$, the fitted curves and one of the posterior distribution are shown in \Cref{fig:fig3_mcmc,fig:fig2_l3}, and the physical parameters of the eight systems are shown in  \Cref{tab:label5_l3,tab:label7_l3}.

We found that by introducing the $l_3$ parameter, there are significant shifts in the $q$ values for all targets, which we attribute to the strong correlation between $q$ and $l_3$. This high degree of correlation makes it challenging to ascertain the true value of $q$. To investigate whether our observations are influenced by other stars, we cross-matched our targets with  5 arcsec radius in the Gaia catalog \citep{gaia1, gaia2} . The results revealed that, apart from the target V0678 Peg, no additional stars were detected in the vicinity of the targets. Concerning V0678 Peg, we found an adjacent star, however, the brightness of V0678 Peg far surpasses that of this neighboring star by a factor of 489, meaning that our observations were not affected by nearby stars. Thus, we proposed that the results without third-light are more reliable. This viewpoint aligns with the stance presented by \cite{liuliang}, which posits that $l_3$ has a significant impact on $q$, yet is very faint, it is better to set it to zero. In light of these considerations, we have adopted the results excluding third-light contributions for all subsequent analyses and computations. 

\renewcommand\arraystretch{1.2}%设置行高
\begin{table*}[t]
\centering
\caption{\centering The physical parameters of the eight Targets}
\label{tab:label5}
\begin{tabular}{lcccccccccc} 

\hline
 Parameter&$q$& $i (deg)$&   $T_{2} (K)$\footnote{The error of $T_2$ is corrected by $T_1$}&$\Omega_1$=$\Omega_2$
&$f(\%)$& $\lambda (deg)$& $r_s (deg)$&$T_s$&$r_1$ & $r_2$\\ 
\hline

 J005148&0.237$^{+0.001}_{-0.001}$&78.8$^{+0.2}_{-0.2}$&   6425$^{+24}_{-24}$&2.278$^{+0.003}_{-0.003}$
&29.6$^{+0.9}_{-1.8}$& /& /& / & 0.523$^{+0.001}_{-0.001}$& 0.279$^{+0.001}_{-0.001}$\\

NSVS 503993&0.220$^{+0.001}_{-0.001}$& 80.1$^{+0.3}_{-0.3}$&   6303$^{+38}_{-36}$&2.255$^{+0.004}_{-0.003}$
&18.5$^{+1.1}_{-0.9}$& 287$^{+4}_{-4}$& 7$^{+1}_{-1}$& 0.91$^{+0.03}_{-0.04}$ & 0.524$^{+0.001}_{-0.001}$& 0.267$^{+0.001}_{-0.001}$\\ 
 V0394 Cam&0.157$^{+0.001}_{-0.001}$& 75.0$^{+0.2}_{-0.2}$&   6050$^{+33}_{-41}$&2.083$^{+0.005}_{-0.003}$
&38.2$^{+1.5}_{-0.7}$& 351$^{+1}_{-1}$& 25$^{+1}_{-1}$& 0.94$^{+0.01}_{-0.01}$ & 0.559$^{+0.001}_{-0.001}$& 0.251$^{+0.001}_{-0.001}$\\

J055741&0.188$^{+0.001}_{-0.001}$&83.9$^{+0.5}_{-0.4}$&  6470$^{+27}_{-27}$&2.146$^{+0.003}_{-0.005}$
&46.3$^{+2.0}_{-1.3}$& 10$^{+1}_{-1}$& 8$^{+1}_{-1}$& 0.82$^{+0.02}_{-0.03}$ & 0.549$^{+0.001}_{-0.001}$& 0.269$^{+0.001}_{-0.001}$\\ 

  NSVS 2561806&0.219$^{+0.002}_{-0.002}$& 78.2$^{+0.2}_{-0.2}$& 6142$^{+27}_{-25}$&2.234$^{+0.004}_{-0.006}$
&33.3$^{+1.1}_{-1.2}$& /& /& / & 0.531$^{+0.001}_{-0.001}$& 0.275$^{+0.001}_{-0.001}$\\

V0737 Cep&0.160$^{+0.000}_{-0.000}$& 73.6$^{+0.1}_{-0.2}$& 4722$^{+23}_{-22}$&2.125$^{+0.001}_{-0.001}$
&36.0$^{+0.1}_{-0.1}$& 287$^{+2}_{-2}$& 22$^{+1}_{-1}$& 0.96$^{+0.00}_{-0.00}$ & 0.544$^{+0.001}_{-0.001}$& 0.237$^{+0.001}_{-0.001}$\\ 
  NSVS 6133550&0.174$^{+0.002}_{-0.002}$& 76.0$^{+0.2}_{-0.2}$& 5051$^{+21}_{-21}$&2.117$^{+0.004}_{-0.004}$
&43.1$^{+1.3}_{-1.0}$& /& /& / & 0.553$^{+0.001}_{-0.001}$& 0.261$^{+0.001}_{-0.001}$\\
 
  V0678 Peg& 0.250$^{+0.002}_{-0.002}$& 81.5$^{+0.4}_{-0.3}$& 6180$^{+27}_{-29}$&2.315$^{+0.006}_{-0.006}$&23.8$^{+1.3}_{-1.4}$& /& /& / & 0.516$^{+0.001}_{-0.001}$& 0.280$^{+0.001}_{-0.001}$\\
 \hline
\end{tabular}
\end{table*}

\renewcommand\arraystretch{1.2}%设置行高
\begin{table*}[t]
\caption{\centering Luminosity ratio of the primary component  for eight targets}
\label{tab:label6}

\begin{tabular}{lccccccc} 

\hline
 Parameter&$L_{1g}/L_{Tg}$& $L_{1r}/L_{Tr}$& $L_{1i}/L_{Ti}$& $L_{1B}/L_{TB}$& $L_{1V}/L_{TV}$& $L_{1R}/L_{TR}$& $L_{1I}/L_{TI}$\\ 
\hline

 J005148& 0.796$^{+0.001}_{-0.001}$& 0.792$^{+0.001}_{-0.001}$& 0.790$^{+0.001}_{-0.001}$& /& /& /& /\\

NSVS 503993& 0.793$^{+0.002}_{-0.002}$& 0.794$^{+0.001}_{-0.001}$& 0.794$^{+0.001}_{-0.001}$& /& /& /& /\\ 
 V0394 Cam& /& /& /& 0.825$^{+0.001}_{-0.001}$& 0.828$^{+0.001}_{-0.001}$& 0.829$^{+0.001}_{-0.001}$& 0.830$^{+0.001}_{-0.001}$\\

J055741& /& /& /& /& 0.815$^{+0.001}_{-0.001}$& 0.814$^{+0.001}_{-0.001}$& 0.813$^{+0.001}_{-0.001}$\\ 

  NSVS 2561806& 0.814$^{+0.001}_{-0.001}$& 0.808$^{+0.001}_{-0.001}$& 0.805$^{+0.001}_{-0.001}$& /& /& /& /\\

V0737 Cep& /& /& /& /& 0.761$^{+0.001}_{-0.001}$& 0.781$^{+0.001}_{-0.001}$& 0.793$^{+0.001}_{-0.001}$\\ 
  NSVS 6133550& /& /& /& /& 0.749$^{+0.001}_{-0.001}$& 0.765$^{+0.001}_{-0.001}$& 0.775$^{+0.001}_{-0.001}$\\
 
  V0678 Peg& /& /& /& /& 0.778$^{+0.002}_{-0.002}$& 0.777$^{+0.002}_{-0.002}$& 0.777$^{+0.002}_{-0.002}$\\
 \hline
\end{tabular}
\end{table*}

\renewcommand\arraystretch{1.2}%设置行高
\begin{table*}[t]
\centering
\caption{\centering The physical parameters of the eight Targets after incorporating $l_3 $}
\label{tab:label5_l3}
\begin{tabular}{lcccccccccc} 

\hline
 Parameter&$q$& $i(deg)$&   $T_{2}(K)$\footnote{The error of $T_2$ is corrected by $T_1$}&$\Omega_1$=$\Omega_2$
&$f(\%)$& $\lambda(deg)$& $r_s(deg)$&$T_s$  &$r_1$ & $r_2$\\ 
\hline

 J005148&0.278$^{+0.003}_{-0.002}$&81.6$^{+0.3}_{-0.3}$&   6465$^{+24}_{-24}$&2.352$^{+0.007}_{-0.005}$&38.1$^{+1.0}_{-1.4}$& /& /& / & 0.515$^{+0.001}_{-0.001}$& 0.297$^{+0.001}_{-0.001}$\\

NSVS 503993&0.307$^{+0.002}_{-0.001}$& 89.9$^{+0.7}_{-0.8}$&   6365$^{+37}_{-37}$&2.422$^{+0.004}_{-0.003}$&31.7$^{+0.9}_{-1.0}$& 304$^{+5}_{-6}$& 10$^{+1}_{-1}$& 0.96$^{+0.01}_{-0.02}$& 0.504$^{+0.001}_{-0.001}$& 0.302$^{+0.001}_{-0.001}$\\ 
 V0394 Cam&0.347$^{+0.004}_{-0.004}$& 89.9$^{+0.4}_{-0.6}$&   6071$^{+32}_{-36}$&2.389$^{+0.011}_{-0.008}$&83.5$^{+0.7}_{-2.3}$& 355$^{+1}_{-1}$& 24$^{+1}_{-1}$& 0.79$^{+0.03}_{-0.03}$& 0.528$^{+0.001}_{-0.002}$& 0.350$^{+0.001}_{-0.001}$\\

J055741&0.246$^{+0.001}_{-0.002}$&89.9$^{+0.4}_{-0.4}$&  6538$^{+26}_{-26}$&2.225$^{+0.002}_{-0.002}$&76.1$^{+0.6}_{-0.6}$& 10$^{+1}_{-1}$& 11$^{+1}_{-1}$& 0.78$^{+0.04}_{-0.04}$& 0.545$^{+0.001}_{-0.001}$& 0.311$^{+0.001}_{-0.001}$\\ 

  NSVS 2561806&0.276$^{+0.003}_{-0.002}$& 82.2$^{+0.4}_{-0.3}$& 6196$^{+26}_{-26}$&2.334$^{+0.006}_{-0.006}$&45.5$^{+1.6}_{-1.3}$& /& /& / & 0.520$^{+0.001}_{-0.001}$& 0.301$^{+0.001}_{-0.001}$\\

V0737 Cep&0.385$^{+0.004}_{-0.002}$& 89.9$^{+0.8}_{-0.8}$& 4673$^{+24}_{-24}$&2.614$^{+0.005}_{-0.005}$&14.4$^{+1.2}_{-0.9}$& 283$^{+2}_{-2}$& 33$^{+1}_{-1}$& 0.96$^{+0.00}_{-0.00}$ & 0.474$^{+0.001}_{-0.001}$& 0.309$^{+0.001}_{-0.001}$\\ 
  NSVS 6133550&0.200$^{+0.002}_{-0.002}$& 78.0$^{+0.3}_{-0.3}$& 5017$^{+21}_{-21}$&2.166$^{+0.004}_{-0.005}$&52.0$^{+1.1}_{-1.2}$& /& /& / & 0.546$^{+0.001}_{-0.001}$& 0.277$^{+0.001}_{-0.001}$\\
 
  V0678 Peg& 0.276$^{+0.004}_{-0.003}$& 83.4$^{+0.6}_{-0.5}$& 6205$^{+28}_{-28}$&2.365$^{+0.010}_{-0.007}$&26.7$^{+1.6}_{-1.9}$& /& /& / & 0.509$^{+0.001}_{-0.002}$& 0.290$^{+0.001}_{-0.001}$\\
 \hline
\end{tabular}
\end{table*}

\renewcommand\arraystretch{1.2}%设置行高
\begin{table*}[t]
\centering
\caption{\centering The luminosity ratio of the primary component and the third light ratio for eight targets after incorporating $l_3$}
\label{tab:label7_l3}

\begin{tabular}{lcccccccc} 

\hline
 Parameter&J005148& NSVS 503993& V0394 Cam& J055741& NSVS 2561806& V0737 Cep&NSVS 6133550 &V0678 Peg\\ 
\hline

 $L_{1g}/L_{Tg}$& 0.642$^{+0.005}_{-0.005}$& 0.553$^{+0.003}_{-0.003}$& /& /& 0.613$^{+0.004}_{-0.005}$& /
&/
&/
\\

$L_{1r}/L_{Tr}$& 0.649$^{+0.004}_{-0.005}$& 0.562$^{+0.003}_{-0.003}$& /& /& 0.624$^{+0.005}_{-0.005}$& /
&/
&/
\\ 
 $L_{1i}/L_{Ti}$& 0.661$^{+0.004}_{-0.005}$& 0.576$^{+0.003}_{-0.004}$& /& /& 0.631$^{+0.004}_{-0.005}$& /
&/
&/
\\

$L_{1B}/L_{TB}$& /
& /
& 0.387$^{+0.003}_{-0.004}$& /& /& /&/&/\\ 

  $L_{1V}/L_{TV}$& /
& /
& 0.401$^{+0.003}_{-0.003}$& 0.585$^{+0.004}_{-0.003}$& /& 0.336$^{+0.003}_{-0.003}$&0.604$^{+0.005}_{-0.005}$&0.686$^{+0.005}_{-0.005}$\\

$L_{1R}/L_{TR}$& /
& /
& 0.406$^{+0.002}_{-0.003}$& 0.598$^{+0.005}_{-0.003}$& /& 0.361$^{+0.003}_{-0.003}$&0.637$^{+0.005}_{-0.005}$&0.692$^{+0.005}_{-0.006}$\\ 
  $L_{1I}/L_{TI}$& /& /& 0.409$^{+0.003}_{-0.003}$
& 0.599$^{+0.005}_{-0.003}$
& /& 0.373$^{+0.003}_{-0.003}$
&0.657$^{+0.005}_{-0.005}$
&0.697$^{+0.005}_{-0.006}$\\
 
  $L_{3g}/L_{Tg}$& 0.127$^{+0.005}_{-0.005}$& 0.204$^{+0.003}_{-0.004}$& /
& /
& 0.166$^{+0.005}_{-0.004}$& /
&/
&/
\\
 
 $L_{3r}/L_{Tr}$& 0.107$^{+0.005}_{-0.004}$& 0.198$^{+0.003}_{-0.003}$& /
& /
& 0.145$^{+0.005}_{-0.005}$& /
& /
&/
\\
 $L_{3i}/L_{Ti}$& 0.087$^{+0.005}_{-0.004}$
& 0.182$^{+0.004}_{-0.003}$& /& /& 0.140$^{+0.004}_{-0.004}$& /
& /
&/
\\
 $L_{3B}/L_{TB}$& /
& /
& 0.397$^{+0.003}_{-0.003}$& /& /
& /& /&/\\
 $L_{3V}/L_{TV}$& /
& /
& 0.397$^{+0.003}_{-0.003}$& 0.195$^{+0.002}_{-0.003}$& /
& 0.383$^{+0.003}_{-0.003}$& 0.120$^{+0.005}_{-0.005}$&0.060$^{+0.005}_{-0.005}$\\
 $L_{3R}/L_{TR}$& /
& /
& 0.385$^{+0.002}_{-0.002}$& 0.174$^{+0.002}_{-0.005}$& /
& 0.373$^{+0.004}_{-0.003}$& 0.098$^{+0.005}_{-0.005}$&0.053$^{+0.003}_{-0.003}$\\
 $L_{3I}/L_{TI}$& /& /& 0.382$^{+0.003}_{-0.003}$
& 0.171$^{+0.003}_{-0.005}$& /& 0.373$^{+0.003}_{-0.003}$
& 0.083$^{+0.005}_{-0.005}$
&0.055$^{+0.006}_{-0.005}$\\

 \hline
\end{tabular}
\end{table*}

\section{O-C analysis  }
\begin{figure*}[htbp]
\subfigure{
\includegraphics[width=9cm,height = 5cm]{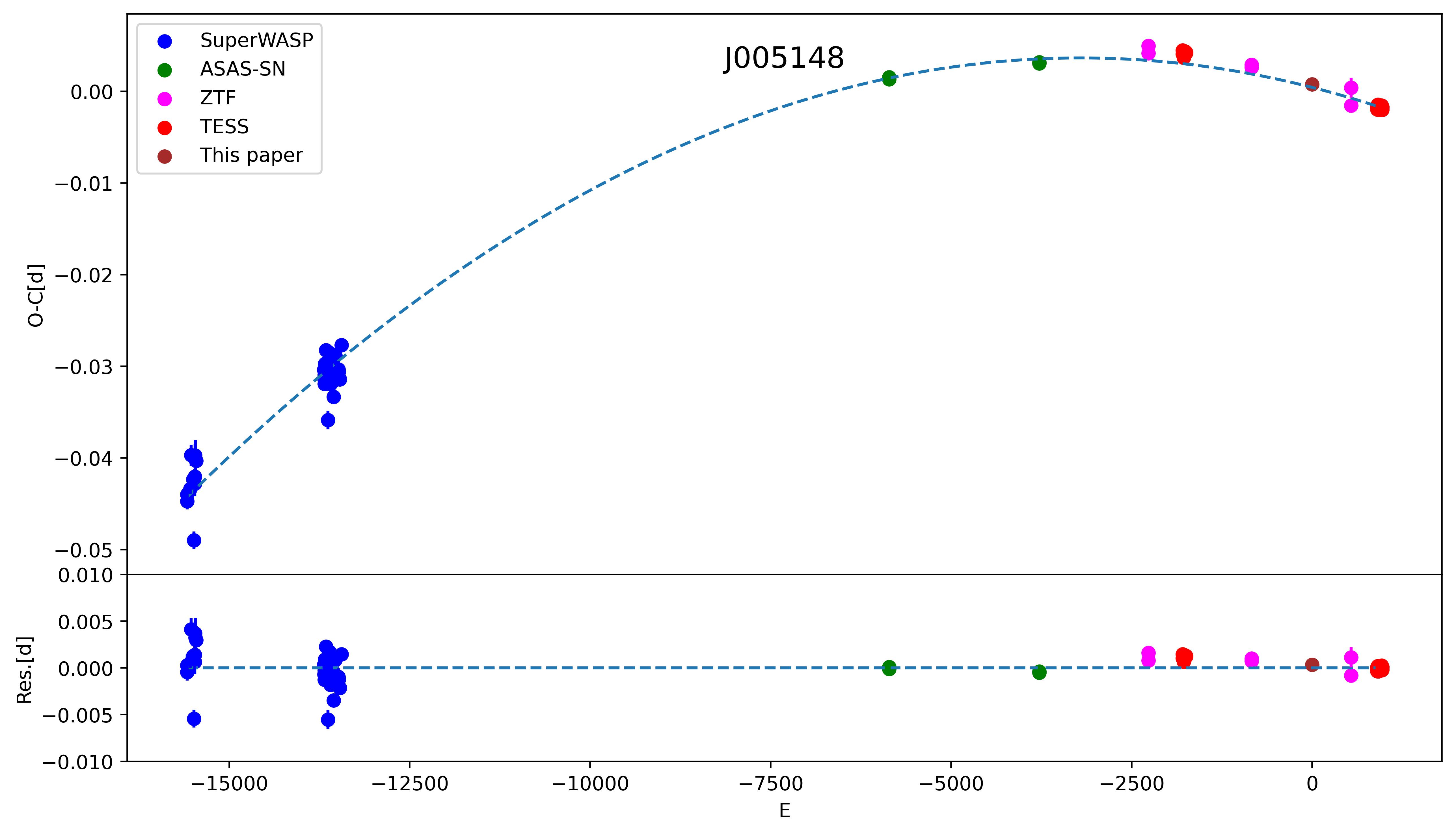}}
\subfigure{
\includegraphics[width=9cm,height = 5cm]{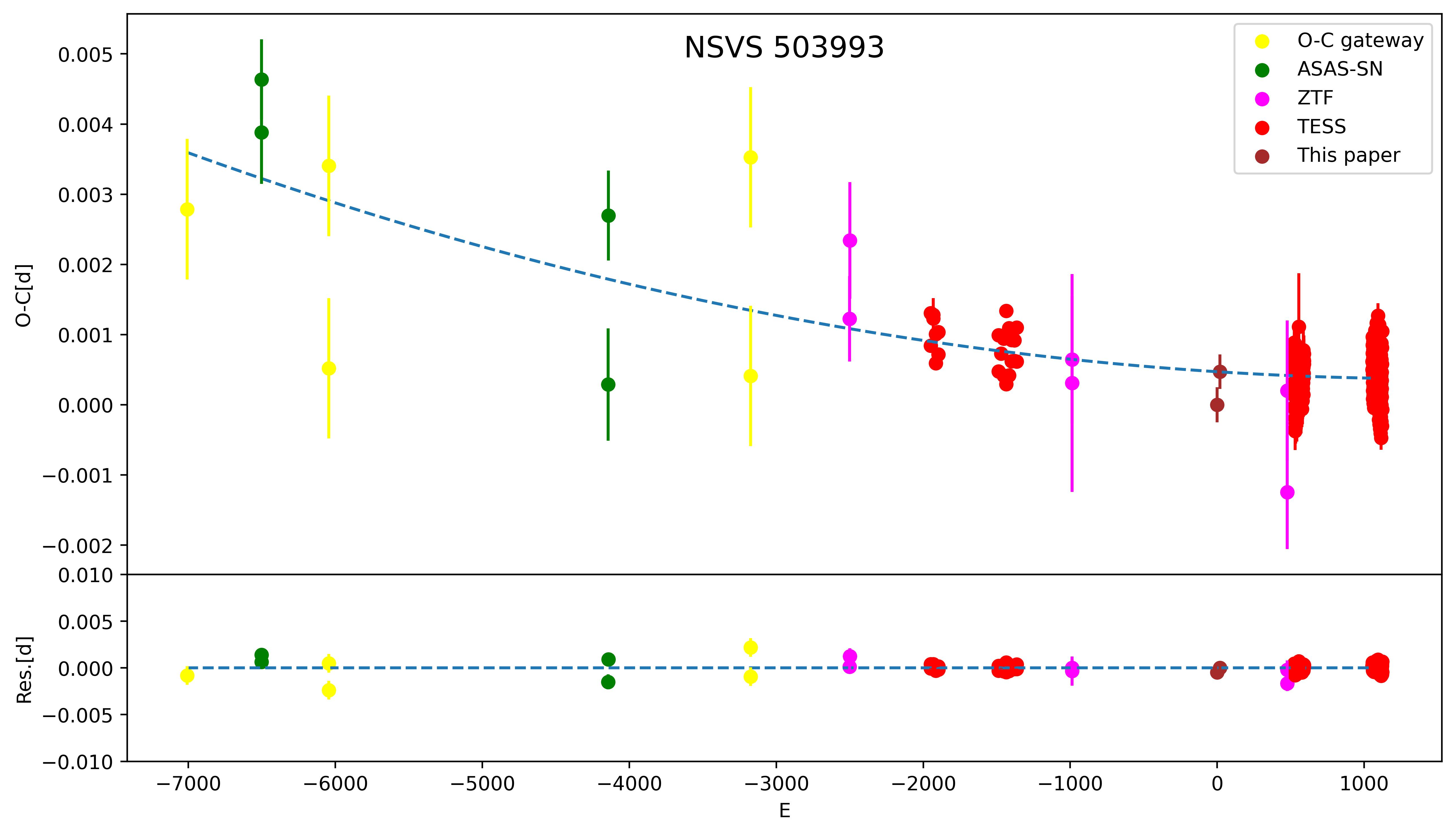}}
\subfigure{
\includegraphics[width=9cm,height = 5cm]{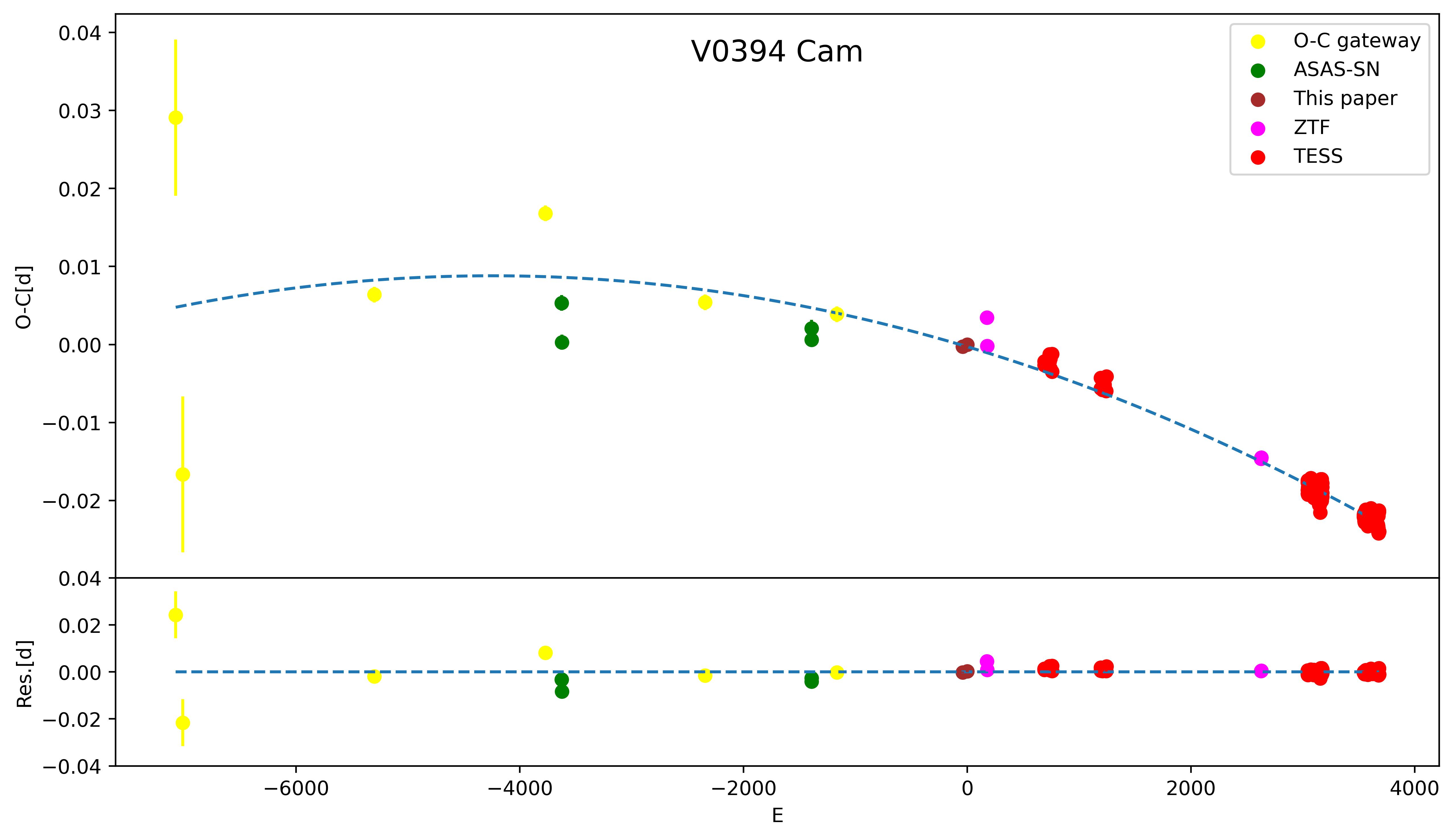}}
\subfigure{
\includegraphics[width=9cm,height = 5cm]{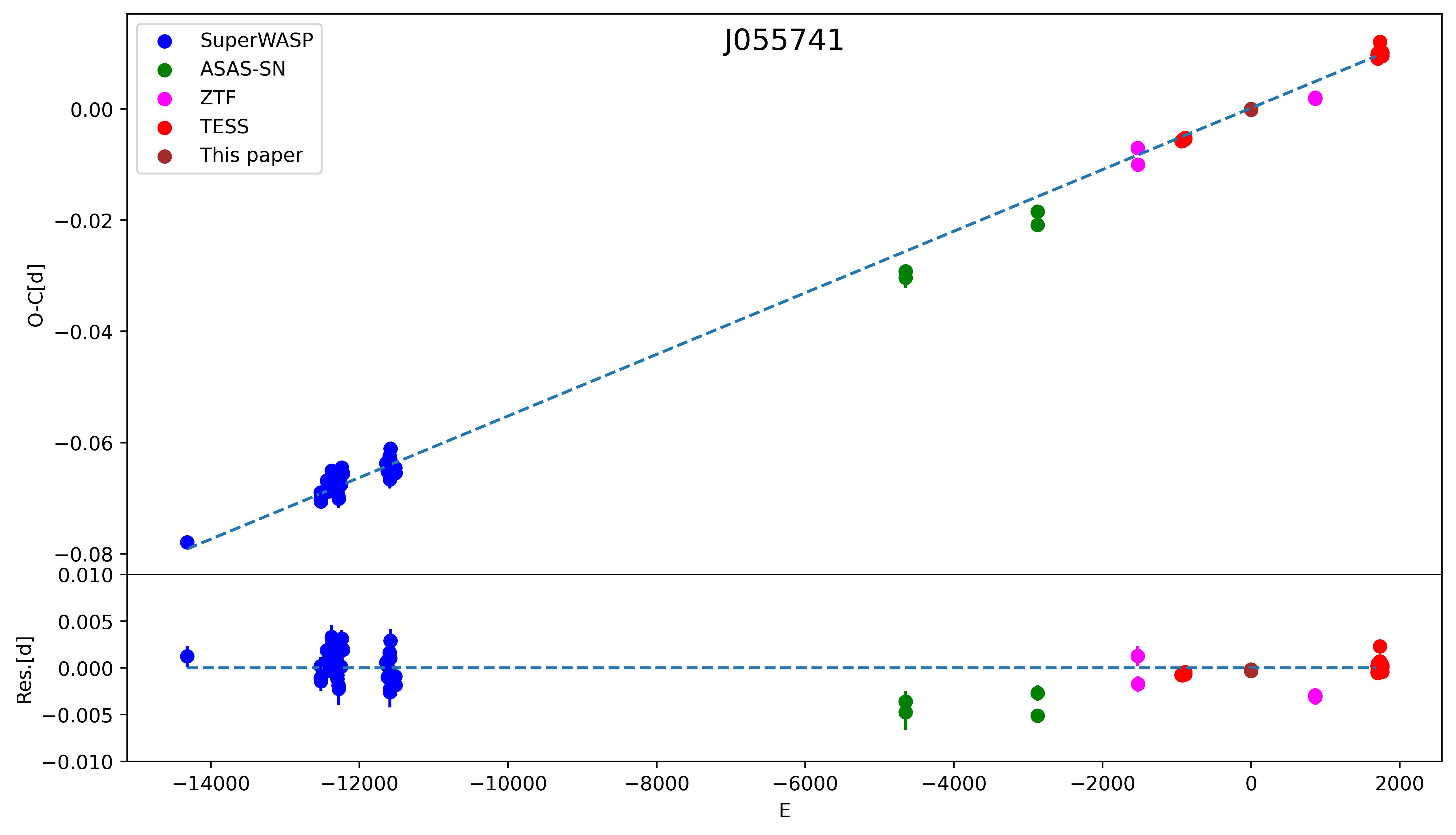}}
\subfigure{
\includegraphics[width=9cm,height = 5cm]{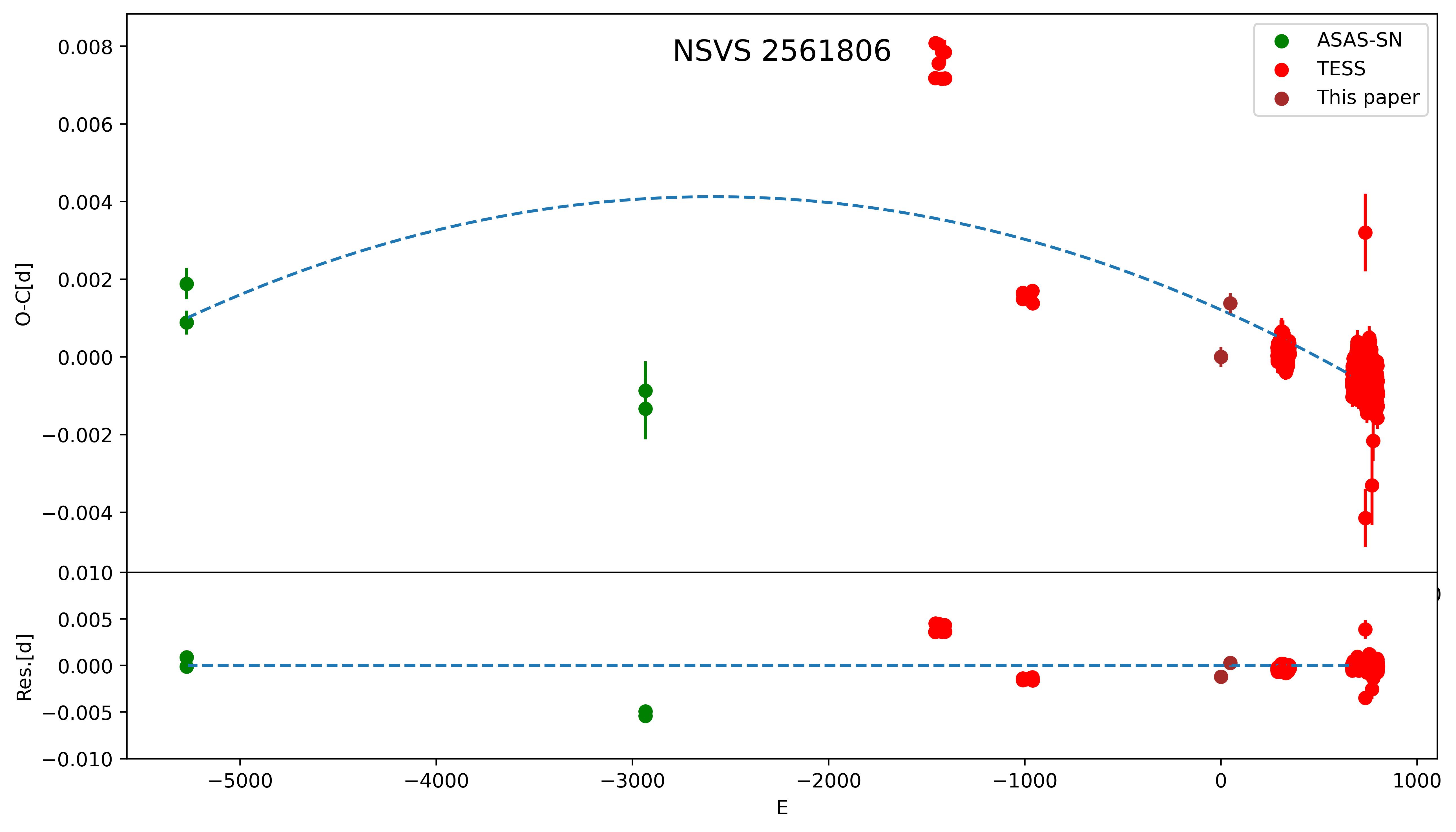}}
\subfigure{
\includegraphics[width=9cm,height = 5cm]{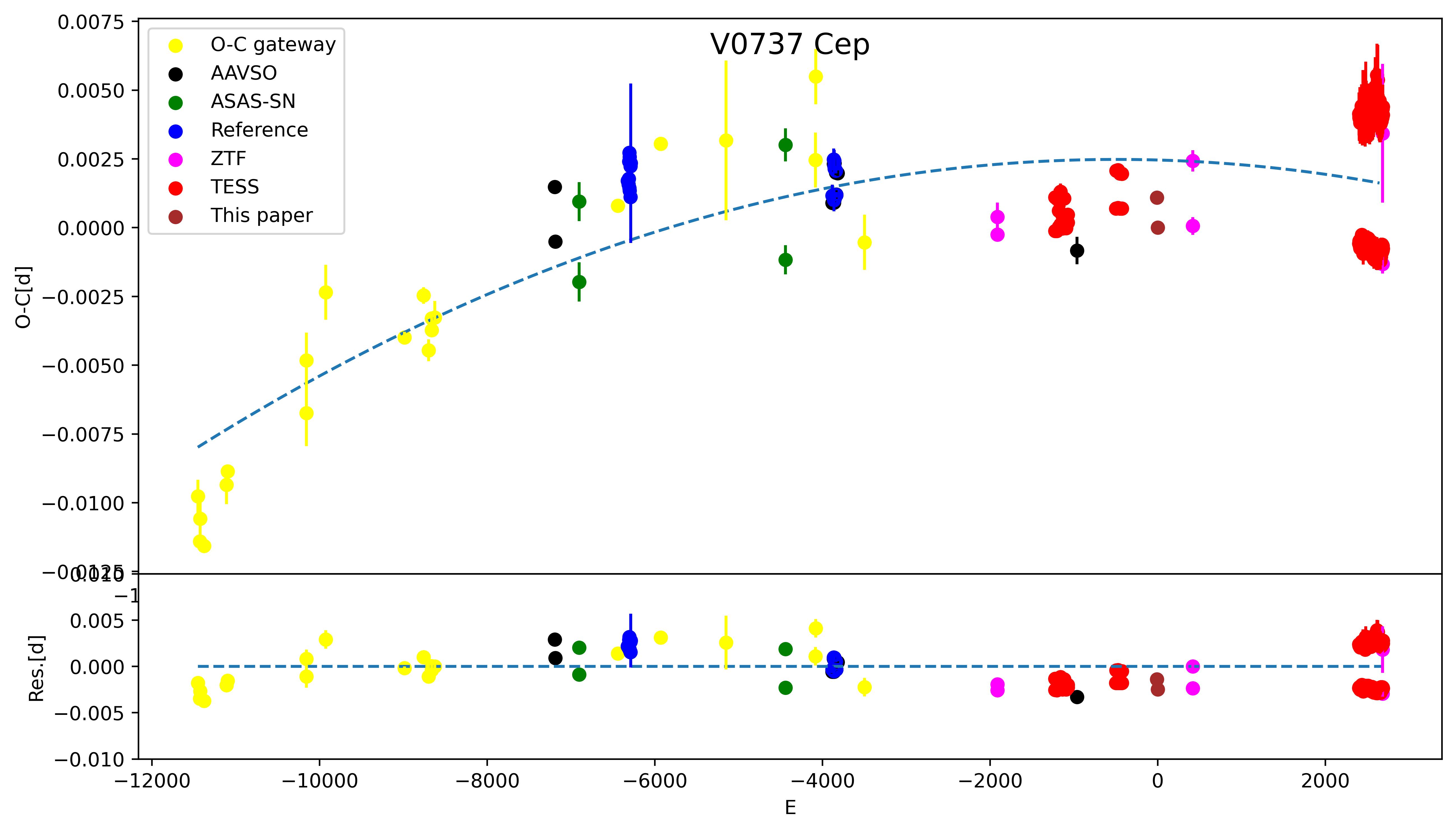}}
\subfigure{
\includegraphics[width=9cm,height = 5cm]{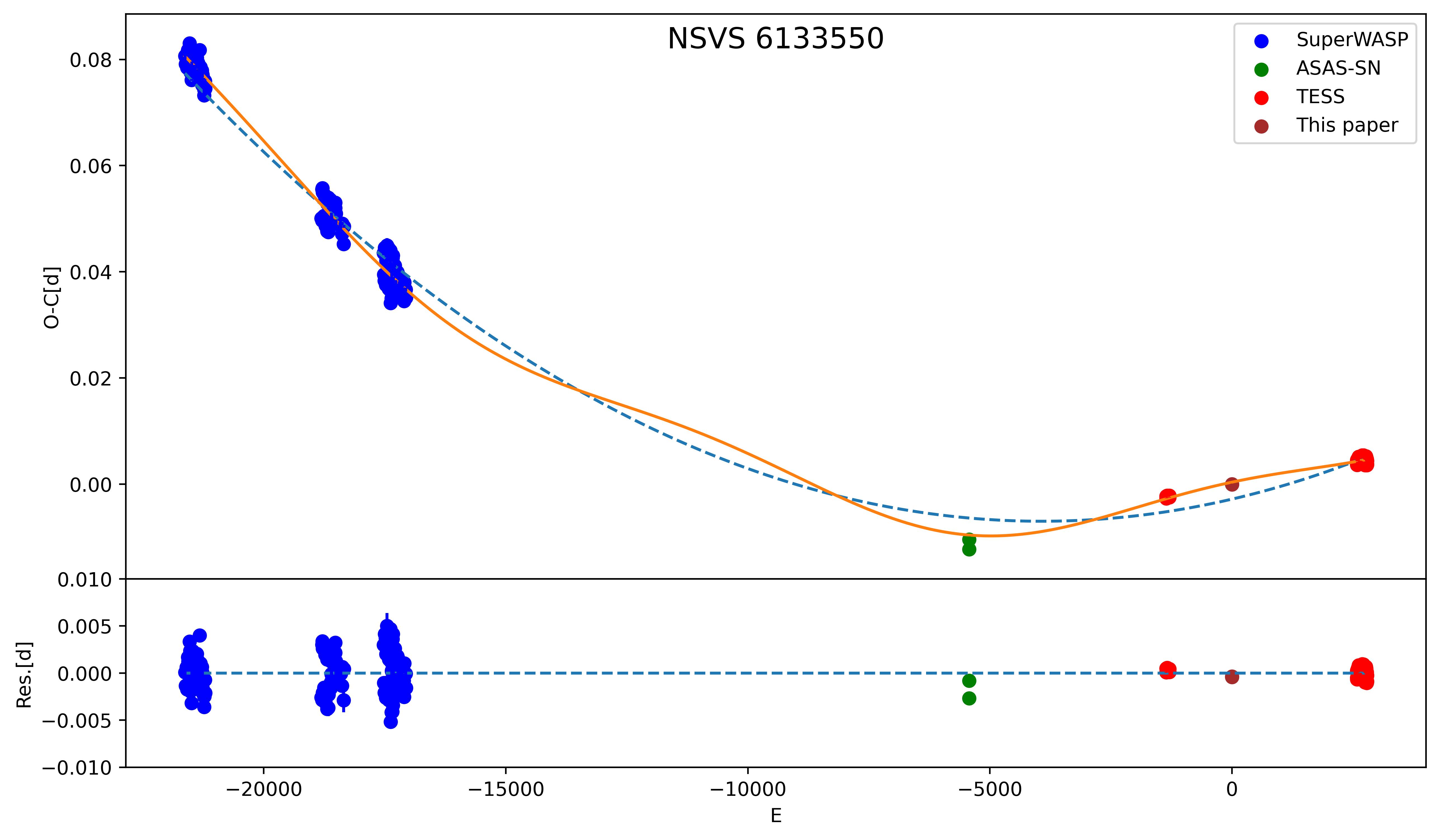}}
\subfigure{
\includegraphics[width=9cm,height = 5cm]{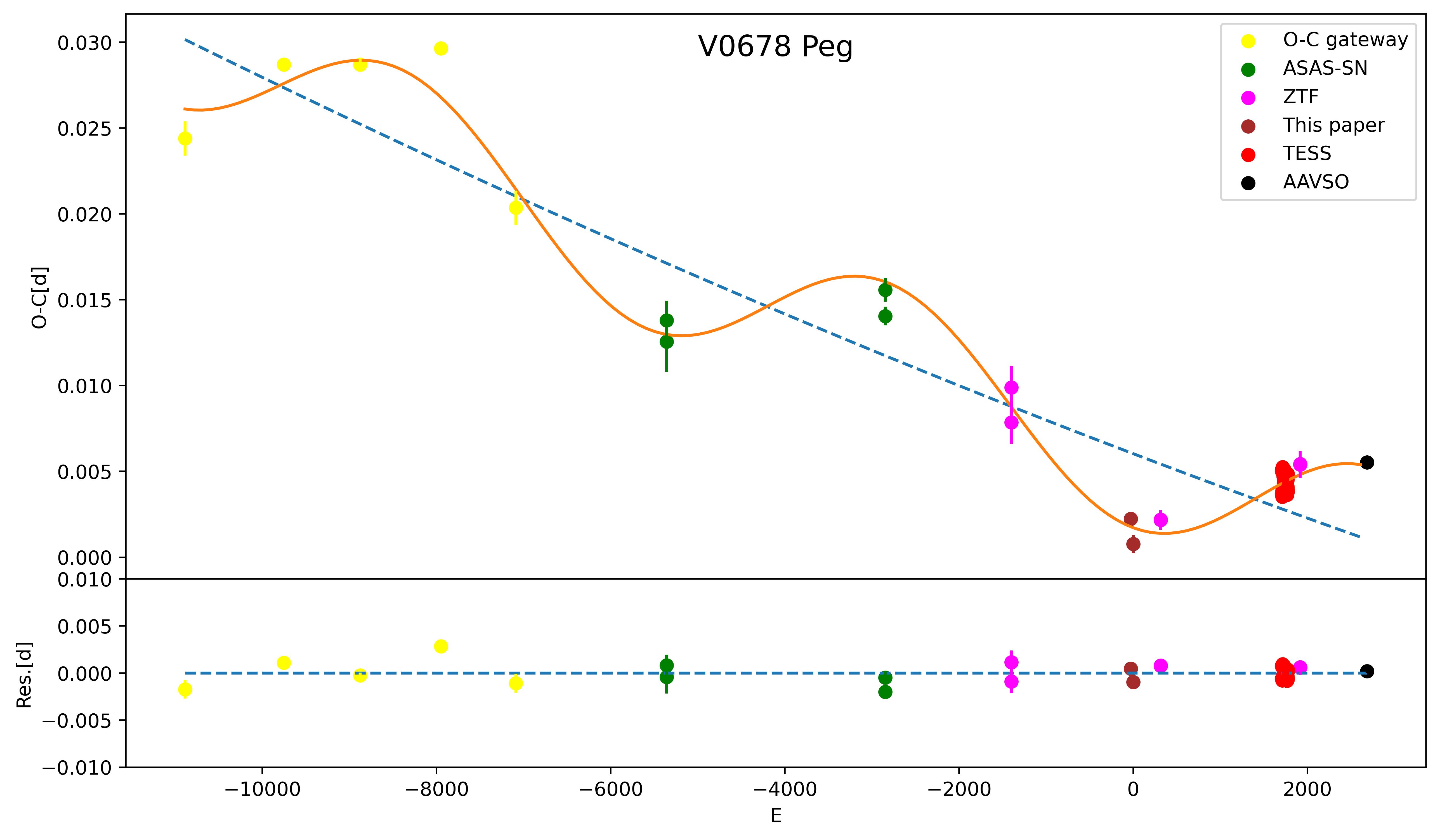}}
\caption{The O-C diagrams  for our eight targets.}
\label{fig:fig5}
\end{figure*}

The O-C analysis is a powerful tool to study the dynamical evolution of contact binaries and search for additional companions \citep{li_o-c_2014, li_o-c_2016}. Therefore, we seek to gather as many eclipsing times as possible from global photometric surveys to analyze the O-C diagrams of the eight targets, the sources for computing eclipsing minima include: ASAS-SN,the Zwicky Transient Facility (ZTF) survey \citep{ztf1,ztf2}, O-C gateway\footnote{O-C gateway (\url{http://var.astro.cz/ocgate/}) }, AAVSO\footnote{AAVSO(\url{http://var.astro.cz/ocgate/})}, the Transiting Exoplanet Survey Satellite \cite[TESS;][]{tess}, and Wide Angle Search for Planets \cite[SuperWASP;][]{wasp}. 
Because the observations, such as ASAS-SN and ZTF, were very dispersed,  we use the period shift method proposed by \cite{Li_2020} to obtain  the eclipsing times. We need to divide the data into groups and use the equation: $HJD=HJD_0+P\times E$, where $HJD$ is the observation time, $HJD_0$ is the reference time, and $P$ is the period, to shift the data into one period. Then, we can calculate the eclipsing times using the \cite{kw1956} method. The eclipsing times for other observations, such as TESS-2min, 10min, and SuperWASP, can be calculated directly. Due to the time of TESS data is BJD, and that of the others is HJD, so we converted HJD to BJD using an online tool\footnote{\url {https://astroutils.astronomy.osu.edu/time/hjd2bjd.html }}. Then, the O-C values of each system are calculated through the following equation,
\begin{equation}
\begin{aligned}
&BJD=BJD_0+P\times E ,
\end{aligned}
\end{equation}   
where $BJD$ is the observational eclipsing times, $BJD_0$ is mentioned in \Cref{tab:label1}, and $P$ is the orbital period obtained  from ASAS-SN. The O-C values of these eight systems are displayed in \Cref{tab:o-c}, and the corresponding O-C diagrams are shown in  \Cref{fig:fig5}. It is evident from this figure that the periods of most targets undergo long-term changes. 

\renewcommand\arraystretch{1.2}%设置行高
\begin{table*}[t]
\caption{\centering Eclipse Timings for the Eight Targets}
\label{tab:o-c}
\footnotesize 
\setlength{\tabcolsep}{1mm}{
\begin{tabular}{cccccccccccccc}

\hline\hline
 Star& BJD& Error& E& O-C& Residual&Ref.& Star& BJD& Error& E& O-C& Residual&Ref.\\

 & 2450000+& & & & & & & 2450000+& & & & &\\ 
 \hline

 J005148&7126.06700&0.00066 & -5861.0 &0.00132 & -0.00010 
& (1)& NSVS 2561806 & 7080.84833& 0.00031 & -5273.0 & 0.00089 & -0.00011 
&(1)\\

 &  
7126.26900&0.00063 & -5860.5 &0.00155 & 0.00013 
&(1)& & 

7081.05968& 0.00040 & -5272.5 & 0.00189 & 0.00089 
&(1)\\

&  7966.02404&0.00045 & -3779.5 &0.00317 & -0.00035 
&(1)& & 8064.86481& 0.00075 & -2934.0 & -0.00087 & -0.00494 
&(1)\\ 
 &  
7966.22568&0.00076 & -3779.0 &0.00304 & -0.00048 
&(1)& & 

8065.07469& 0.00079 & -2933.5 & -0.00134 & -0.00541 
&(1)\\

&  9491.17688&0.00011 & 0.0 &0.00000 & -0.00042 
&(2)& & 9299.20065& 0.00026 & 0.0 & 0.00000 & -0.00121 
&(2)\\ 

  &  
9881.18964&0.00016 & 966.5 &-0.00169 & 0.00011 
&(3)& & 

9319.18530& 0.00026 & 47.5 & 0.00138 & 0.00027 
&(2)\\

&  9881.39108&0.00015 & 967.0 &-0.00202 & -0.00021 
&(3)& & 9635.12900& 0.00031 & 798.5 & -0.00092 & -0.00006 
&(3)\\ 
  &  
9881.59314&0.00020 & 967.5 &-0.00173 & 0.00008 
&(3)& & 

9635.33900& 0.00022 & 799.0 & -0.00127 & -0.00041 
&(3)\\
 
  &  9881.79462&0.00013 & 968.0 &-0.00201 & -0.00020 
&(3)& & 9635.55000& 0.00030 & 799.5 & -0.00062 & 0.00024 
&(3)\\
 & 9881.99670& 0.00015 & 968.5 & -0.00170 & 0.00011 
& (3)& & 

9635.76000& 0.00026 & 800.0 & -0.00097 & -0.00011 
&(3)\\
 NSVS 503993& 6978.36900& 0.00100 & -7009.5 & 0.00279 & -0.00081 
& (4)& V0737 Cep& 7568.45710& 0.00290& -5150.5
& 0.00317& 0.00346&(6)\\
 & 7329.58300& 0.00100 & -6044.0 & 0.00052 & -0.00239 
& (4)& & 7887.38621& 0.00100& -4083& 0.00246& 0.00195&(6)\\
 & 8373.41000& 0.00100 & -3174.5 & 0.00041 & -0.00093 
& (4)& & 7888.43491& 0.00100& -4079.5& 0.00549& 0.00498&(6)\\
 & 8373.59500& 0.00100 & -3174.0 & 0.00353 & 0.00218 
& (4)& & 8062.45851& 0.00100& -3497& -0.00053& -0.00141&(6)\\
 & 8618.58923& 0.00061 & -2500.5 & 0.00123 & 0.00014 
& (5)& & 6958.38074& 0.00021 & -7192.5 
& 0.00148 & 0.00346 
&(7)
\\
 & 
8618.77223& 0.00083 & -2500.0 & 0.00234 & 0.00126 
& (5)& & 6960.32072& 0.00023 & -7186.0 
& -0.00051 & 0.00146 
&(7)
\\
 & 9169.14880& 0.00091 & -987.0 & 0.00064 & 0.00000 
& (5)& & 7946.83855& 0.00020 & -3884.0 & 0.00090 & -0.00006 
&(7)
\\
 & 9169.33035& 0.00155 & -986.5 & 0.00031 & -0.00034 
& (5)& & 7952.66585& 0.00029 & -3864.5 & 0.00232 & 0.00134&(7)\\
 & 9528.18539& 0.00025 & 0.0 & 0.00000 & -0.00047 
& (2)& & 7952.81382& 0.00020 & -3864.0 & 0.00092 & -0.00006&(7)\\
 & 9535.27931& 0.00025 & 19.5 & 0.00047 & 0.00000 
& (2)& & 7953.71015& 0.00021 & -3861.0 & 0.00096 & -0.00002 
&(7)\\

\hline
\end{tabular}}
\centering Note. (1) ASAS-SN; (2) This paper; (3) TESS; (4) O-C gateway; (5) ZTF; (6) SuperWASP; (7) AAVSO
\centering \\Here we only show a portion of the table, the table is available in its entirety from ChinaVO (\url{https://nadc.china-vo.org/res/r101411/}).
\end{table*}

For  J055741, only linear correction was used. For the other stars (except NSVS 6133550 and V0678 Peg), the following equation, 
\begin{equation}
\begin{aligned}
&O-C=\Delta{T_0}+\Delta{P_0}\times E+\frac{\beta}{2}\times {E^2},
\end{aligned}
\end{equation}   
was used to fit the O-C diagram.
For NSVS 6133550 and V0678 Peg, there is a periodic variation except the secular change, we use the following equation to fit their O-C diagrams.
\begin{equation}
\begin{aligned}
&O-C=\Delta{T_0}+\Delta{P_0}\times E+\frac{\beta}{2}\times{E^2}+A\times sin(\frac{2\pi}{P_3}\times E+\varphi),
\end{aligned}
\end{equation}   
The periodic variation in the equation may be caused by the magnetic activity cycle or the light travel time effect of a third body, and we will discuss this in detail in Section \ref{dis}. The results are shown in \Cref{tab:label7}.
We can see that there are some outlier points in the O-C curves . According to our analysis, this may be due to insufficient accuracy when calculating eclipsing times or magnetic activity in our targets. In order to honor the integrity of the original data, we have chosen to retain the existence of this portion.  

 The results illustrate that three stars exhibit an upward trend, indicating a long-term increase orbital periods. In contrast, the other four stars display a downward trend, suggesting a long-term decrease in their orbital periods. 
 It is worth noting that due to the limitations imposed by the observational time span and the density of observations across different phases.   Therefore, we are currently restricted to perform some qualitative analyses, and it will still require more observations in the future to assist us in making quantitative calculations.  

\section{Discussions and Conclusions }\label{dis}

Utilizing PHOEBE, we derived the physical parameters of eight totally eclipsing contact binaries. Because they are all totally eclipsing contact binaries, the photometric physical parameters are reliable \citep{Apeople, Bpeople}. Our analysis revealed that all of them are low mass ratio (q$<=$0.25) contact binaries. The light curves of NSVS 503993, V0394 Cam, J055741, and V0737 Cep exhibit noticeable asymmetry, which is attributed to dark spot on the primary component. Additionally, two systems display shallow contact configuration (f$<=$0.25), while the remaining six systems feature moderate contact configuration. O-C analysis of all available eclipsing times revealed a long-term decrease in the orbital periods of three systems, whereas four systems show a long-term increase in their orbital periods. 

\renewcommand\arraystretch{1.2}%设置行高
\begin{table*}[t]
\centering
\small
\setlength{\tabcolsep}{1mm}{
\caption{\centering O-C fitting coefficients and errors}
\label{tab:label7}

\begin{tabular}{lcccccccccccccc}
\hline
 Parameter& $\Delta{T_0}$& Error& $\Delta{P_0}$& Error& $\beta$&Error & $dM_1/dt$&Error & $A$&Error & $P_3$& Error & $\varphi$&Error \\

 &$(\times {10^{-4}} d)$&& $(\times {10^{-7}} d)$&& $(\times {10^{-7}} d$ $ {yr^{-1}})$& & $(\times {10^{-7}} M\odot$ $ {yr^{-1}})$& & $(\times {10^{-3}} d)$& & $(\times {10^{3}} d)$& & ($\circ$)&\\ 
\hline

 J005148&  4.29&2.01& -20.06&1.60& -5.67&0.22 & -2.05&0.08 & /& /& /& /& /&/\\

NSVS 503993&  4.71&0.72& -1.34&0.68& 0.89&0.32 & 0.32&0.11 & /& /& /& /& /&/\\ 
 V0394 Cam&  -2.88&6.53& -42.81&1.35& -0.97&0.79 & -0.23&0.19 & /& /& /& /& /&/\\

J055741&  1.67&1.64& 55.39&0.28& 0&0 & 0&0 & /& /& /& /& /&/\\ 

  NSVS 2561806&  12.14&1.63& -22.50&2.17& -7.54&1.15& -2.45&0.37& /& /& /& /& /&/\\

V0737 Cep&  24.60&4.29& -0.85&1.24& -2.13&0.51 & -0.61&0.15 & /& /& /& /& /&/\\ 
  NSVS 6133550&  -28.32&4.53& 21.18&1.40& 7.15&0.22& 2.36&0.07& 3.21& 0.52& 2.94& 0& 95.55&9.38\\
 
  V0678 Peg&  60.33&3.75& -19.29&1.99& 0.47&0.41& 0.18&0.16& 4.31& 0.36& 2.29& 0.53& 270.35&7.27\\
 \hline
\end{tabular}}
\end{table*}

\renewcommand\arraystretch{1.2}%设置行高
\begin{table*}[t]
\caption{\centering Absolute parameters of the eight Targets}
\label{tab:label8}
\centering
\small
\setlength{\tabcolsep}{1mm}{
\begin{tabular}{lcccccccccccc} 

\hline
 Parameter& $a(R_{\odot})$& $M_1(M_{\odot})$& $M_2(M_{\odot})$& $R_1(R_{\odot})$& $R_2(R_{\odot})$ & $L_1(L_{\odot})$&$L_2(L_{\odot})$ &$J_{orb}$ & $M_{1i}(M_{\odot})$& $M_{2i}(M_{\odot})$& $M_{lost}(M_{\odot})$&$\tau(Gyr)$
\\ 
\hline

 J005148& $2.77{\pm 0.12}$& $1.41{\pm 0.18}$& $0.33{\pm 0.04}$& $1.45{\pm 0.07}$& $0.77{\pm 0.04}$& $3.45{\pm 0.34}$&$0.91{\pm 0.10}$&51.56 & 0.88
& 1.91& 1.04&4.20
\\

NSVS 503993& $2.55{\pm 0.11}$& $1.37{\pm 0.18}$& $0.30{\pm 0.04}$& $1.33{\pm 0.06}$& $0.68{\pm 0.03}$& $2.50{\pm 0.27}$&$0.65{\pm 0.08}$&51.49 & 0.86
& 1.80& 1.00&5.13
\\ 
 V0394 Cam& $2.65{\pm 0.12}$& $1.47{\pm 0.20}$& $0.23{\pm 0.03}$& $1.48{\pm 0.07}$& $0.66{\pm 0.03}$& $2.53{\pm 0.27}$&$0.53{\pm 0.06}$ &51.41 & 0.95
& 1.80& 1.04&5.38
\\

J055741& $2.83{\pm 0.12}$& $1.49{\pm 0.19}$& $0.28{\pm 0.04}$& $1.55{\pm 0.07}$& $0.76{\pm 0.04}$& $3.92{\pm 0.40}$&$0.91{\pm 0.10}$&51.50 & 0.92
& 1.96& 1.12&3.92
\\ 

  NSVS 2561806& $2.87{\pm 0.12}$& $1.46{\pm 0.19}$& $0.32{\pm 0.04}$& $1.52{\pm 0.07}$& $0.79{\pm 0.04}$& $3.29{\pm 0.33}$&$0.79{\pm 0.09}$&51.55 & 0.94
& 1.86& 1.02&4.57
\\

V0737 Cep& $2.18{\pm 0.10}$& $1.34{\pm 0.19}$& $0.21{\pm 0.03}$& $1.19{\pm 0.06}$& $0.52{\pm 0.03}$& $0.87{\pm 0.10}$&$0.12{\pm 0.01}$ &51.31 & 0.98
& 1.28& 0.71&16.90
\\ 
  NSVS 6133550& $2.05{\pm 0.10}$& $1.29{\pm 0.19}$& $0.23{\pm 0.03}$& $1.13{\pm 0.06}$& $0.53{\pm 0.03}$& $1.02{\pm 0.12}$&$0.17{\pm 0.02}$ &51.31 & 0.91
& 1.38& 0.77&13.04
\\
 
  V0678 Peg& $2.80{\pm 0.12}$& $1.41{\pm 0.18}$& $0.35{\pm 0.05}$& $1.44{\pm 0.06}$& $0.78{\pm 0.04}$& $2.78{\pm 0.28}$&$0.80{\pm 0.09}$&51.58 & 0.91& 1.83& 0.98&4.75\\
 \hline
 
\end{tabular}}
\end{table*}

To investigate the reasons for the changes in orbital periods, as well as the evolutionary status, initial masses, and ages of the eight contact binaries, we need to derive their absolute parameters. Due to the lack of radial velocity curve, it is difficult to obtain them. Nevertheless, for contact binary stars with a temperature less than 10000K, there is a relation between the orbital period ($P$) and the semi-major axis ($a$) \citep{li2022fo}, 
\begin{equation}
\begin{aligned}
&a = 0.501(\pm 0.063)+5.621(\pm 0.138)\times P.
\end{aligned}
\end{equation}   
Combining this equation with the period of the eight binaries, their absolute physical parameters can be obtained as shown in \Cref{tab:label8}.

The long-term orbital period increase is caused by the transfer of matter from the less massive component to the more massive component. And the long-term orbital period decrease generally caused by angular momentum loss (AML) or the transfer of matter from the more massive component to the less massive component.
Then, the following equation \citep{k1958}, 
\begin{equation}
\begin{aligned}
&\frac{\dot{P}}{P}=-3\dot{M}(\frac{1}{M_1}-\frac{1}{M_2}),
\end{aligned}
\end{equation}   
was used to calculate material transfer rates of these stars, the results are shown in \Cref{tab:label7}. For the three targets with long-term decreasing period, the negative sign indicates that the more massive primary  is losing mass. We use the equation $\tau _{th}=\frac{GM^2_1}{R_1L_1}$ to calculate the thermal timescale, and $\tau_{th}=1.26\times 10^7yr$ for J005148, $\tau_{th}=1.82\times 10^7yr$ for V0394 Cam, $\tau_{th}=1.34\times 10^7yr$ for NSVS 2561806 and $\tau_{th}=5.48\times 10^7yr$ for V0737 Cep. The thermal timescale mass transfer rate is determined to be $M_1/\tau_{th}=1.13\times 10^{-7}M_\odot$ $ {yr^{-1}}$ for  J005148, $M_1/\tau_{th}=8.08\times 10^{-8}M_\odot$ $ {yr^{-1}}$ for V0394 Cam, $M_1/\tau_{th}=1.09\times 10^{-7}M_\odot$ $ {yr^{-1}}$ for NSVS 2561806 and $M_1/\tau_{th}=2.45\times 10^{-8}M_\odot$ $ {yr^{-1}}$ for V0737 Cep. For all of them, the thermal mass transfer rates are similar to those listed in  \Cref{tab:label7}, suggesting that the long-term decrease in orbital period may be caused by mass transfer. For the two targets with periodic variations, the periodic changes may be due to the magnetic activity cycle or the light travel time effect of a third body. We do not provide detailed discussions, mainly because the time span of the eclipsing times is short and the coverage is insufficient, so these results can only be considered preliminary. 

NSVS 6133550 and V0678 Peg showed periodic variations in their O-C diagrams, such variations could potentially be explained by the Applegate mechanism or the light travel time effect caused by a third body \citep{2016ApJ_new}.

The Applegate mechanism \citep{ap1} suggested that variations in the quadrupole moment of solar-type stars, driven by magnetic activity, can account for cyclic orbital period oscillations. Using the equations \citep{L2002,R2000},
\begin{equation}
\begin{aligned}
&\frac{\Delta{P}}{P}=-9\frac{\Delta{Q}}{Ma^2},
\end{aligned}
\end{equation}   
\begin{equation}
\begin{aligned}
&\Delta{P}=\sqrt{[1-cos(2\pi P/P_3)]}\times A,
\end{aligned}
\end{equation}   
we derived the required quadrupole moments. For NSVS 6133550, the primary component has $\Delta Q_1 = 1.1\times10^{47}g$ $cm^2$  and the secondary component has $\Delta Q_2 = 1.9\times10^{46}g$ $cm^2$. For V0678 Peg,  the primary component has $\Delta Q_1 = 3.8\times10^{47}g$ $cm^2$ and the secondary component has  $\Delta Q_2 = 9.5\times10^{46}g$ $cm^2$. However, the typical values of the required quadrupole moments for active close binaries range from $10^{51}$ to $10^{52}$ $g$·$cm^2$.  Therefore, it seems unsuitable to explain the cyclic variations using the Applegate mechanism  for the two targets. 

Another explanation is the light travel time effect due to the presence of a third body. Using the mass function,
\begin{equation}
\begin{aligned}
&f(m)=\frac{(m_3\mathrm{sin}i)^3}{(m_1+m_2+m_3)^2}=\frac{4\pi}{GP^2_3}\times (a_{12}\mathrm{sin}i)^3,
\end{aligned}
\end{equation}   
where $a_{12}\mathrm{sin}i$ $=$$A \times c$ (where $c$ is the speed of light). We obtained $f(m)=2.67(\pm 1.31)\times 10^{-3}M_\odot$ for NSVS 6133550 and $f(m)=10.63(\pm 2.67)\times 10^{-3}M_\odot$ for V0678 Peg.
And we assumed the orbital inclination of the third body is $i=90^\circ$,  the minimum mass of the tertiary companion of NSVS 6133550 was determined to be $m_3=1.99(\pm 0.59)\times 10^{-1}M_\odot$ with a separation of $5.05(\pm 1.71) AU$, while the minimum mass of the tertiary companion V0678 Peg was determined to be  $m_3=3.63(\pm 0.58)\times 10^{-1}M_\odot$ with a separation of  $4.30(\pm 0.77) AU$.
 We assumed the third body to be a main-sequence star and estimated the luminosity by interpolating  from the Table 5 provided by  \cite{tab5}. We obtained the luminosity of the third body for the two targets to be $0.004L_\odot$ and $0.016L_\odot$. The third light contribution for these two targets was calculated to be $0.36\%$ and $0.43\%$, respectively. Since the contribution of the third light was very small, we did not consider its effect in the analysis of the light curve \citep{liuliang}.

\begin{figure*}[htbp]
\subfigure{
\includegraphics[width=9cm,height = 6cm]{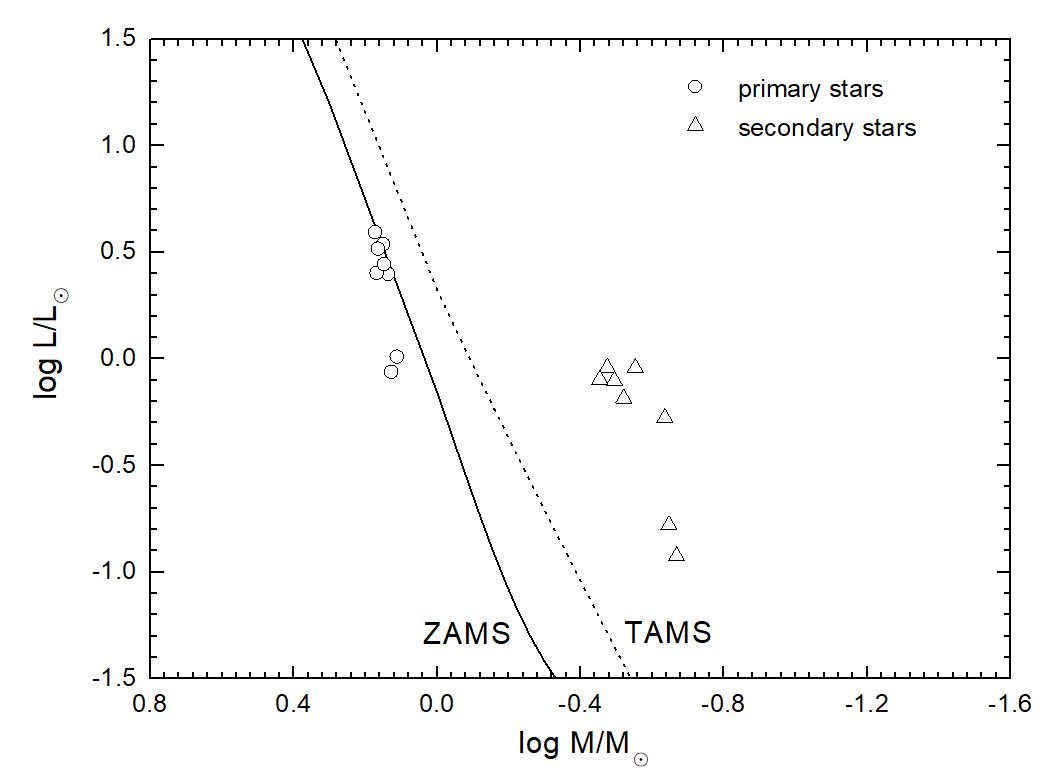}}
\subfigure{
\includegraphics[width=9cm,height = 6cm]{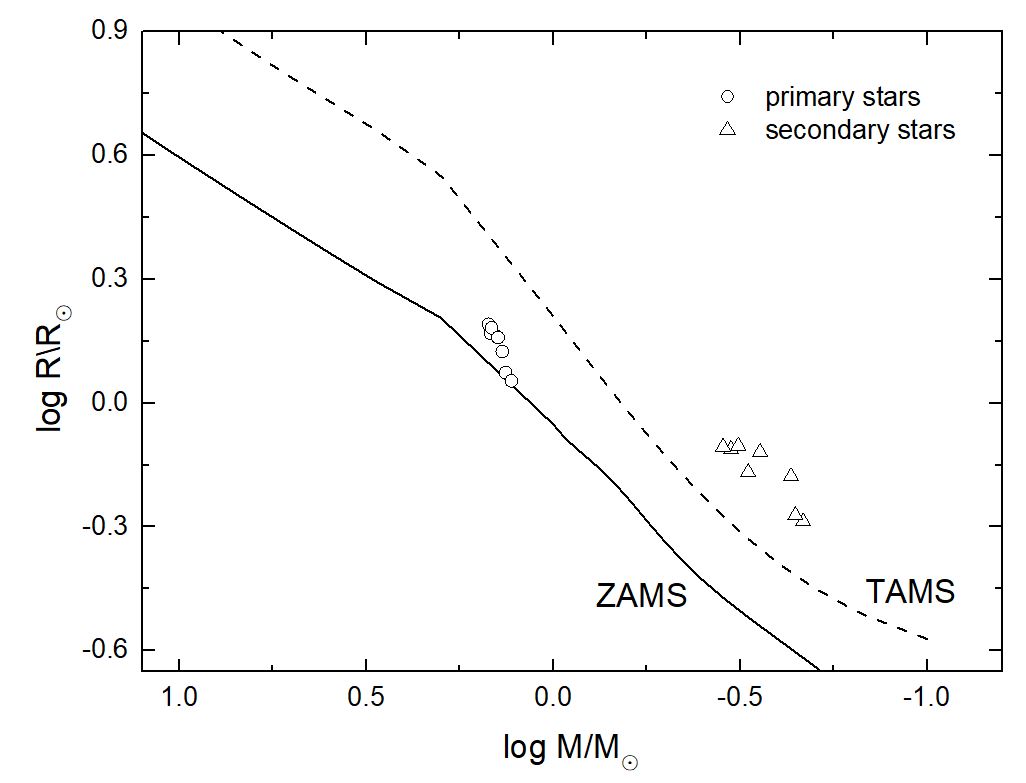}}

\caption{M-L and M-R diagrams. the solid line is the ZAMS and the dotted line is the TAMS. The circles represent the primary components and the triangles represent the secondary components.}
\label{fig:fig12}
\end{figure*}

 Based on the absolute physical parameters, we plotted the mass-luminosity ($M-L$) and the mass-radius ($M-R$) distributions of the eight binaries as shown in \Cref{fig:fig12}. The solid and dotted lines correspond to the zero-age main-sequence (ZAMS) and terminal-age main-sequence (TAMS) provided by \cite{Hurley}, respectively. Circles and triangles denote the more massive and less massive components of the contact binary stars.
 It is obvious that the more massive primary components are around the ZAMS, indicating that they are either non-evolved or only little-evolved main-sequence stars. The less massive secondary components, on the other hand, are above the TAMS, indicating that they have evolved away from the main sequence. This phenomenon may be attributed to the transfer of mass and energy from the more massive star to the less massive star. 
 
\begin{figure}
    \centering
    \includegraphics[scale=0.5]{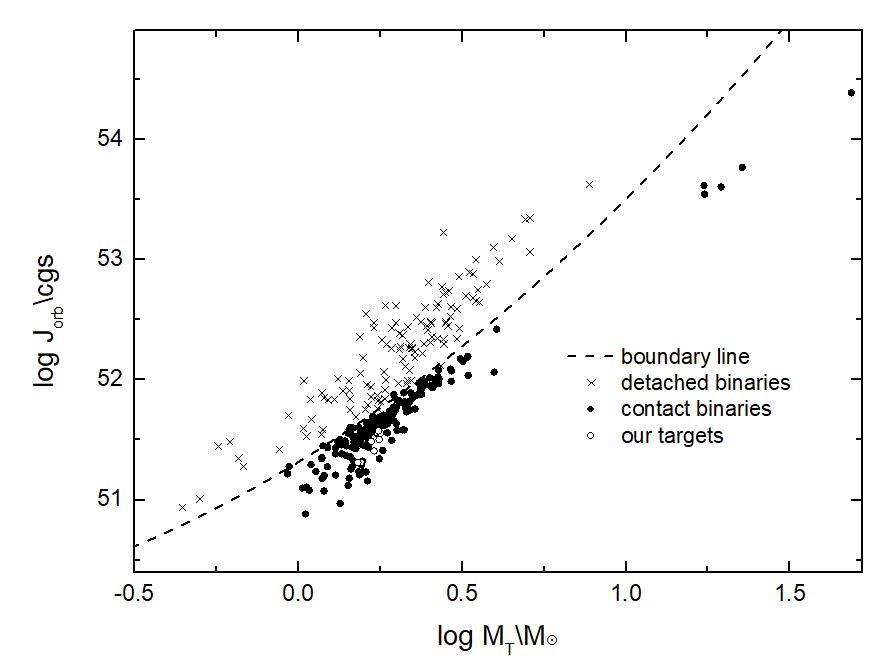}
    \caption{The relationship between log$M_T$ and log$J_{orb}$. The crosses represent detached binaries, the solid circles display contact binaries, and the open circles denote our ten targets. The dashed line refers to the boundary between detached and contact binaries derived by \cite{Eker}. }
    \label{fig:fig13}
\end{figure}

 The orbital angular momentum of the contact binary stars can be calculated by the following equation proposed by \cite{Eker},
\begin{equation}
\begin{aligned}
&J_{orb}=1.24\times 10^{52}\times M^{5/3}_{T}\times P^{1/3}\times q\times (1+q)^{-2} ,
\end{aligned}
\end{equation}   
where $M_T$ is the total mass of the binaries, $P$ is the period, and $q$ is the mass ratio. The relationship between $logJ_{orb}$ and $logM_{T}$ is shown in  \Cref{fig:fig13}. The detached binaries from \cite{Eker} and the contact binaries from \cite{li2021a} are displayed in \Cref{fig:fig13} for comparison. The boundary line between detached and contact binary stars obtained by \cite{Eker} is also marked in \Cref{fig:fig13}. It was found that the orbital angular momentum of our eight targets is within the contact binary region. For detached binaries and contact binaries with the same total mass, contact binaries have smaller orbital angular momentum. This may be due to the loss of orbital angular momentum during the formation and evolution of the contact binary stars.

For contact binary stars, their initial masses are crucial to their evolution. Therefore, we used the method suggested by \cite{yd2013} to calculate the initial masses of the two components. First, the following equation is used to calculate the initial mass of the secondary component:
\begin{equation}
\begin{aligned}
&M_{2i}=M_{2}+\Delta M=M_{2}+2.50(M_{L}-M_{2}-0.07)^{0.64},
\end{aligned}
\end{equation}   
where $M_{2i}$ is the initial mass of the secondary component, $M_2$ is the mass of the secondary component at present, and we can calculate $M_L$ by  mass–luminosity relation:
$M_L = (L_2/1.49)^{1/4.216}$. Then we can calculate the initial mass of the primary component using the following equation,
\begin{equation}
\begin{aligned}
&M_{1i}=M_{1}-(\Delta M-M_{lost})=M_{1}-\Delta M(1-\gamma),
\end{aligned}
\end{equation}   
where $M_{1i}$ is the initial mass of the primary component, $M_1$ is the mass of the primary component at present, $M_{lost}$ is the mass lost by the system during the evolution, $\gamma$ is the ratio of $M_{lost}$ to $\Delta M$. 
We set it to 0.664, the same value as used in  \cite{yd2013}, and the results are shown in   \Cref{tab:label8}. At last, we calculated the age of the targets using the equation proposed by \cite{yd2014},

\begin{equation}
\begin{aligned}
&t\approx t_{MS}(M_{2i})+t_{MS}(\overline{M_2}),
\end{aligned}
\end{equation}   
\begin{equation}
\begin{aligned}
&t_{MS}=\frac{10}{(M/M_{\odot})^{4.05}}\times(5.60\times 10^{-3}(\frac{M}{M_{\odot}}+3.993)^{3.16}+0.042) ,
\end{aligned}
\end{equation}
where $\overline{M_2}=(M_{2i}+M_L)/2$. We tabulated the results in  \Cref{tab:label8}  and we can see that six of them, except  V0737 Cep and NSVS 6133550, are similar to the results obtained by \cite{yd2013} and \cite{yd2014}. The masses of V0737 Cep and NSVS 6133550, derived from the empirical relationship, may be inaccurate, and radial velocity curves are required to calculate more accurate absolute parameters in the future.

These eight stars have all been studied by \cite{lixuzhi}. In addition, V0737 Cep has been studied by \cite{LAST2}, and V0678 Peg has been studied by \cite{sunwei}. In order to compare the physical parameters of these binaries, we listed our determined parameters alongside those from \cite{lixuzhi} in \Cref{tab:label9}. Notable differences between the results can be observed. 
For V0737 Cep, \cite{LAST2} use Wilson–Devinney (W-D) code \citep{w1971,w1979,w1990,w1994} get $q=2.5$. To find out the true paraments, we use PHOEBE and W-D to model the data from \cite{LAST2}. We obtained  $q=0.18$ and $q=0.16$ respectively using PHOEBE and W-D, with setting the phaseshift to 0.5. For V0678 Peg, \cite{sunwei} get $q=0.25$, which is equal to our result. Regarding other targets, due to our higher precision of our observational data and the fact that all were observed in three bands, we consider our results to be more reliable. 

\renewcommand\arraystretch{1.2}%设置行高
\begin{table*}[t] 
\centering
\caption{Comparison with the physical parameters obtained by \cite{lixuzhi}}
\label{tab:label9}
\begin{tabular}{lcccccccccc} 

\hline
 Parameter&\multicolumn{2}{c}{$q$}& \multicolumn{2}{c}{$T_{1}$}& \multicolumn{2}{c}{$T_{2}$}&  \multicolumn{2}{c}{$L_s/L_p$}&\multicolumn{2}{c}{$f$}\\
 &  Ours&Li et al.&  Ours&Li et al.&  Ours&Li et al.&  Ours&Li et al.& Ours&Li et al.\\ 
\hline

 J005148& 0.237&0.200& 6546&6538& 6425&6604&  0.256&0.251 
&0.296&0.227
\\

NSVS 503993& 0.220&0.910& 6294&6389& 6303&6516&  0.261&1.025 
&0.185& 0.162
\\ 
 V0394 Cam& 0.157&0.370& 5994&5705& 6050&5534 
&  0.208&0.369 
&0.382& 0.251
\\

J055741& 0.188&0.160& 6529&6259 
& 6470&6322&  0.227&0.220 
&0.463& 0.523
\\ 

  NSVS 2561806& 0.219&0.200& 6311&6493& 6142&6428&  0.229&0.234 
&0.333& 0.346
\\

V0737 Cep& 0.160&0.290& 5126&5103& 4722&4899&  0.314&0.263 
&0.360& 0.066
\\ 
  NSVS 6133550& 0.174&0.150& 5465&5857& 5051&6320 
&  0.335&0.304 
&0.431& 0.532
\\
 
  V0678 Peg& 0.250&0.490& 6215&6073& 6180&6137&  0.285&0.566 &0.238& 0.301
\\
 \hline
 
\end{tabular}
\end{table*}

In conclusion, we have observed eight totally eclipsing contact binaries, and  their light curves are analyzed using PHOEBE. O-C investigation was performed with all available eclipsing times.
We studied their evolutionary states, initial masses, and ages. Future observations of radial velocities are required to obtain more accurate physical parameters, which will help in understanding their formation and evolution. 
Due to the short time span of  the observations, further observations are required to confirm the results of orbital period variations.

\section*{ACKNOWLEDGEMENTS}
We extend our heartfelt gratitude to the anonymous reviewer for the insightful comments and constructive suggestions, which have significantly enhanced the quality and clarity of our manuscript. This work was supported by National Natural Science Foundation of China (NSFC) (No. 12273018), and the Joint Research Fund in Astronomy (No. U1931103) under cooperative agreement between NSFC and Chinese Academy of Sciences (CAS), and by the Qilu Young Researcher Project of Shandong University, and by Young Data Scientist Project of the National Astronomical Data Center and by the Cultivation Project for LAMOST Scientific Payoff and Research Achievement of CAMSCAS. The calculations in this work were carried out at Supercomputing Center of Shandong University, Weihai.

This paper makes use of data from ASAS-SN. ASAS-SN isfunded in part by the Gordon and Betty Moore Foundation through grant numbers GBMF5490 and GBMF10501 to the Ohio State University, and also funded in part by the Alfred P. Sloan Foundation grant number G-2021-14192. 

This work includes data collected by the TESS mission. Funding for the TESS mission is provided by NASA Science Mission Directorate. We acknowledge the TESS team for its support of this work. 

This paper makes use of observation from the Two Micron All Sky Survey (MASS), a joint project of the University of Massachusetts and the Infrared Processing and Analysis Center/California Institute of Technology. Funding of MASS is provided by the National Aeronautics and Space Administration and the National Science Foundation. 

This paper makes use of data from the DR1 of the WASP data (Butters et al 2010) as provided by the WASP consortium, and computational resources supplied by the project ‘e-Infrastruktura CZ’ (e-INFRA CZ LM2018140) supported by the Ministry of Education, Youth and Sports of the Czech Republic. 

%% For this sample we use BibTeX plus aasjournals.bst to generate the
%% the bibliography. The sample631.bib file was populated from ADS. To
%% get the citations to show in the compiled file do the following:
%%
%% pdflatex sample631.tex
%% bibtext sample631
%% pdflatex sample631.tex
%% pdflatex sample631.tex

\bibliographystyle{aasjournal}
\bibliography{sample631}

\begin{thebibliography}{}
\expandafter\ifx\csname natexlab\endcsname\relax\def\natexlab#1{#1}\fi
\providecommand{\url}[1]{\href{#1}{#1}}
\providecommand{\dodoi}[1]{doi:~\href{http://doi.org/#1}{\nolinkurl{#1}}}
\providecommand{\doeprint}[1]{\href{http://ascl.net/#1}{\nolinkurl{http://ascl.net/#1}}}
\providecommand{\doarXiv}[1]{\href{https://arxiv.org/abs/#1}{\nolinkurl{https://arxiv.org/abs/#1}}}

\bibitem[{{Applegate}(1992)}]{ap1}
{Applegate}, J.~H. 1992, \apj, 385, 621, \dodoi{10.1086/170967}

\bibitem[{{Arbutina}(2007)}]{arbutina}
{Arbutina}, B. 2007, \mnras, 377, 1635, \dodoi{10.1111/j.1365-2966.2007.11723.x}

\bibitem[{{Bellm} {et~al.}(2019){Bellm}, {Kulkarni}, {Barlow}, {Feindt}, {Graham}, {Goobar}, {Kupfer}, {Ngeow}, {Nugent}, {Ofek}, {Prince}, {Riddle}, {Walters}, \& {Ye}}]{ztf1}
{Bellm}, E.~C., {Kulkarni}, S.~R., {Barlow}, T., {et~al.} 2019, \pasp, 131, 068003, \dodoi{10.1088/1538-3873/ab0c2a}

\bibitem[{{Bradstreet} \& {Guinan}(1994)}]{bbb1994}
{Bradstreet}, D.~H., \& {Guinan}, E.~F. 1994, in Astronomical Society of the Pacific Conference Series, Vol.~56, Interacting Binary Stars, ed. A.~W. {Shafter}, 228

\bibitem[{{Butters} {et~al.}(2010){Butters}, {West}, {Anderson}, {Collier Cameron}, {Clarkson}, {Enoch}, {Haswell}, {Hellier}, {Horne}, {Joshi}, {Kane}, {Lister}, {Maxted}, {Parley}, {Pollacco}, {Smalley}, {Street}, {Todd}, {Wheatley}, \& {Wilson}}]{wasp}
{Butters}, O.~W., {West}, R.~G., {Anderson}, D.~R., {et~al.} 2010, \aap, 520, L10, \dodoi{10.1051/0004-6361/201015655}

\bibitem[{{Castelli} \& {Kurucz}(2004)}]{ck2004}
{Castelli}, F., \& {Kurucz}, R.~L. 2004, \aap, 419, 725, \dodoi{10.1051/0004-6361:20040079}

\bibitem[{{Caton} {et~al.}(2019){Caton}, {Gentry}, {Samec}, {Chamberlain}, {Robb}, {Faulkner}, \& {Hill}}]{v1187}
{Caton}, D., {Gentry}, D.~R., {Samec}, R.~G., {et~al.} 2019, \pasp, 131, 054203, \dodoi{10.1088/1538-3873/aafb8f}

\bibitem[{{Conroy} {et~al.}(2020){Conroy}, {Kochoska}, {Hey}, {Pablo}, {Hambleton}, {Jones}, {Giammarco}, {Abdul-Masih}, \& {Pr{\v{s}}a}}]{Conroy2020}
{Conroy}, K.~E., {Kochoska}, A., {Hey}, D., {et~al.} 2020, \apjs, 250, 34, \dodoi{10.3847/1538-4365/abb4e2}

\bibitem[{{Eker} {et~al.}(2006){Eker}, {Demircan}, {Bilir}, \& {Karata{\c{s}}}}]{Eker}
{Eker}, Z., {Demircan}, O., {Bilir}, S., \& {Karata{\c{s}}}, Y. 2006, \mnras, 373, 1483, \dodoi{10.1111/j.1365-2966.2006.11073.x}

\bibitem[{{Gaia Collaboration} {et~al.}(2016){Gaia Collaboration}, {Prusti}, {de Bruijne}, {Brown}, {Vallenari}, {Babusiaux}, {Bailer-Jones}, {Bastian}, {Biermann}, {Evans}, {Eyer}, {Jansen}, {Jordi}, {Klioner}, {Lammers}, {Lindegren}, {Luri}, {Mignard}, {Milligan}, {Panem}, {Poinsignon}, {Pourbaix}, {Randich}, {Sarri}, {Sartoretti}, {Siddiqui}, {Soubiran}, {Valette}, {van Leeuwen}, {Walton}, {Aerts}, {Arenou}, {Cropper}, {Drimmel}, {H{\o}g}, {Katz}, {Lattanzi}, {O'Mullane}, {Grebel}, {Holland}, {Huc}, {Passot}, {Bramante}, {Cacciari}, {Casta{\~n}eda}, {Chaoul}, {Cheek}, {De Angeli}, {Fabricius}, {Guerra}, {Hern{\'a}ndez}, {Jean-Antoine-Piccolo}, {Masana}, {Messineo}, {Mowlavi}, {Nienartowicz}, {Ord{\'o}{\~n}ez-Blanco}, {Panuzzo}, {Portell}, {Richards}, {Riello}, {Seabroke}, {Tanga}, {Th{\'e}venin}, {Torra}, {Els}, {Gracia-Abril}, {Comoretto}, {Garcia-Reinaldos}, {Lock}, {Mercier}, {Altmann}, {Andrae}, {Astraatmadja}, {Bellas-Velidis}, {Benson}, {Berthier}, {Blomme}, {Busso}, {Carry}, {Cellino}, {Clementini},
  {Cowell}, {Creevey}, {Cuypers}, {Davidson}, {De Ridder}, {de Torres}, {Delchambre}, {Dell'Oro}, {Ducourant}, {Fr{\'e}mat}, {Garc{\'\i}a-Torres}, {Gosset}, {Halbwachs}, {Hambly}, {Harrison}, {Hauser}, {Hestroffer}, {Hodgkin}, {Huckle}, {Hutton}, {Jasniewicz}, {Jordan}, {Kontizas}, {Korn}, {Lanzafame}, {Manteiga}, {Moitinho}, {Muinonen}, {Osinde}, {Pancino}, {Pauwels}, {Petit}, {Recio-Blanco}, {Robin}, {Sarro}, {Siopis}, {Smith}, {Smith}, {Sozzetti}, {Thuillot}, {van Reeven}, {Viala}, {Abbas}, {Abreu Aramburu}, {Accart}, {Aguado}, {Allan}, {Allasia}, {Altavilla}, {{\'A}lvarez}, {Alves}, {Anderson}, {Andrei}, {Anglada Varela}, {Antiche}, {Antoja}, {Ant{\'o}n}, {Arcay}, {Atzei}, {Ayache}, {Bach}, {Baker}, {Balaguer-N{\'u}{\~n}ez}, {Barache}, {Barata}, {Barbier}, {Barblan}, {Baroni}, {Barrado y Navascu{\'e}s}, {Barros}, {Barstow}, {Becciani}, {Bellazzini}, {Bellei}, {Bello Garc{\'\i}a}, {Belokurov}, {Bendjoya}, {Berihuete}, {Bianchi}, {Bienaym{\'e}}, {Billebaud}, {Blagorodnova}, {Blanco-Cuaresma}, {Boch},
  {Bombrun}, {Borrachero}, {Bouquillon}, {Bourda}, {Bouy}, {Bragaglia}, {Breddels}, {Brouillet}, {Br{\"u}semeister}, {Bucciarelli}, {Budnik}, {Burgess}, {Burgon}, {Burlacu}, {Busonero}, {Buzzi}, {Caffau}, {Cambras}, {Campbell}, {Cancelliere}, {Cantat-Gaudin}, {Carlucci}, {Carrasco}, {Castellani}, {Charlot}, {Charnas}, {Charvet}, {Chassat}, {Chiavassa}, {Clotet}, {Cocozza}, {Collins}, {Collins}, {Costigan}, {Crifo}, {Cross}, {Crosta}, {Crowley}, {Dafonte}, {Damerdji}, {Dapergolas}, {David}, {David}, {De Cat}, {de Felice}, {de Laverny}, {De Luise}, {De March}, {de Martino}, {de Souza}, {Debosscher}, {del Pozo}, {Delbo}, {Delgado}, {Delgado}, {di Marco}, {Di Matteo}, {Diakite}, {Distefano}, {Dolding}, {Dos Anjos}, {Drazinos}, {Dur{\'a}n}, {Dzigan}, {Ecale}, {Edvardsson}, {Enke}, {Erdmann}, {Escolar}, {Espina}, {Evans}, {Eynard Bontemps}, {Fabre}, {Fabrizio}, {Faigler}, {Falc{\~a}o}, {Farr{\`a}s Casas}, {Faye}, {Federici}, {Fedorets}, {Fern{\'a}ndez-Hern{\'a}ndez}, {Fernique}, {Fienga}, {Figueras}, {Filippi},
  {Findeisen}, {Fonti}, {Fouesneau}, {Fraile}, {Fraser}, {Fuchs}, {Furnell}, {Gai}, {Galleti}, {Galluccio}, {Garabato}, {Garc{\'\i}a-Sedano}, {Gar{\'e}}, {Garofalo}, {Garralda}, {Gavras}, {Gerssen}, {Geyer}, {Gilmore}, {Girona}, {Giuffrida}, {Gomes}, {Gonz{\'a}lez-Marcos}, {Gonz{\'a}lez-N{\'u}{\~n}ez}, {Gonz{\'a}lez-Vidal}, {Granvik}, {Guerrier}, {Guillout}, {Guiraud}, {G{\'u}rpide}, {Guti{\'e}rrez-S{\'a}nchez}, {Guy}, {Haigron}, {Hatzidimitriou}, {Haywood}, {Heiter}, {Helmi}, {Hobbs}, {Hofmann}, {Holl}, {Holland}, {Hunt}, {Hypki}, {Icardi}, {Irwin}, {Jevardat de Fombelle}, {Jofr{\'e}}, {Jonker}, {Jorissen}, {Julbe}, {Karampelas}, {Kochoska}, {Kohley}, {Kolenberg}, {Kontizas}, {Koposov}, {Kordopatis}, {Koubsky}, {Kowalczyk}, {Krone-Martins}, {Kudryashova}, {Kull}, {Bachchan}, {Lacoste-Seris}, {Lanza}, {Lavigne}, {Le Poncin-Lafitte}, {Lebreton}, {Lebzelter}, {Leccia}, {Leclerc}, {Lecoeur-Taibi}, {Lemaitre}, {Lenhardt}, {Leroux}, {Liao}, {Licata}, {Lindstr{\o}m}, {Lister}, {Livanou}, {Lobel}, {L{\"o}ffler},
  {L{\'o}pez}, {Lopez-Lozano}, {Lorenz}, {Loureiro}, {MacDonald}, {Magalh{\~a}es Fernandes}, {Managau}, {Mann}, {Mantelet}, {Marchal}, {Marchant}, {Marconi}, {Marie}, {Marinoni}, {Marrese}, {Marschalk{\'o}}, {Marshall}, {Mart{\'\i}n-Fleitas}, {Martino}, {Mary}, {Matijevi{\v{c}}}, {Mazeh}, {McMillan}, {Messina}, {Mestre}, {Michalik}, {Millar}, {Miranda}, {Molina}, {Molinaro}, {Molinaro}, {Moln{\'a}r}, {Moniez}, {Montegriffo}, {Monteiro}, {Mor}, {Mora}, {Morbidelli}, {Morel}, {Morgenthaler}, {Morley}, {Morris}, {Mulone}, {Muraveva}, {Musella}, {Narbonne}, {Nelemans}, {Nicastro}, {Noval}, {Ord{\'e}novic}, {Ordieres-Mer{\'e}}, {Osborne}, {Pagani}, {Pagano}, {Pailler}, {Palacin}, {Palaversa}, {Parsons}, {Paulsen}, {Pecoraro}, {Pedrosa}, {Pentik{\"a}inen}, {Pereira}, {Pichon}, {Piersimoni}, {Pineau}, {Plachy}, {Plum}, {Poujoulet}, {Pr{\v{s}}a}, {Pulone}, {Ragaini}, {Rago}, {Rambaux}, {Ramos-Lerate}, {Ranalli}, {Rauw}, {Read}, {Regibo}, {Renk}, {Reyl{\'e}}, {Ribeiro}, {Rimoldini}, {Ripepi}, {Riva}, {Rixon},
  {Roelens}, {Romero-G{\'o}mez}, {Rowell}, {Royer}, {Rudolph}, {Ruiz-Dern}, {Sadowski}, {Sagrist{\`a} Sell{\'e}s}, {Sahlmann}, {Salgado}, {Salguero}, {Sarasso}, {Savietto}, {Schnorhk}, {Schultheis}, {Sciacca}, {Segol}, {Segovia}, {Segransan}, {Serpell}, {Shih}, {Smareglia}, {Smart}, {Smith}, {Solano}, {Solitro}, {Sordo}, {Soria Nieto}, {Souchay}, {Spagna}, {Spoto}, {Stampa}, {Steele}, {Steidelm{\"u}ller}, {Stephenson}, {Stoev}, {Suess}, {S{\"u}veges}, {Surdej}, {Szabados}, {Szegedi-Elek}, {Tapiador}, {Taris}, {Tauran}, {Taylor}, {Teixeira}, {Terrett}, {Tingley}, {Trager}, {Turon}, {Ulla}, {Utrilla}, {Valentini}, {van Elteren}, {Van Hemelryck}, {van Leeuwen}, {Varadi}, {Vecchiato}, {Veljanoski}, {Via}, {Vicente}, {Vogt}, {Voss}, {Votruba}, {Voutsinas}, {Walmsley}, {Weiler}, {Weingrill}, {Werner}, {Wevers}, {Whitehead}, {Wyrzykowski}, {Yoldas}, {{\v{Z}}erjal}, {Zucker}, {Zurbach}, {Zwitter}, {Alecu}, {Allen}, {Allende Prieto}, {Amorim}, {Anglada-Escud{\'e}}, {Arsenijevic}, {Azaz}, {Balm}, {Beck}, {Bernstein},
  {Bigot}, {Bijaoui}, {Blasco}, {Bonfigli}, {Bono}, {Boudreault}, {Bressan}, {Brown}, {Brunet}, {Bunclark}, {Buonanno}, {Butkevich}, {Carret}, {Carrion}, {Chemin}, {Ch{\'e}reau}, {Corcione}, {Darmigny}, {de Boer}, {de Teodoro}, {de Zeeuw}, {Delle Luche}, {Domingues}, {Dubath}, {Fodor}, {Fr{\'e}zouls}, {Fries}, {Fustes}, {Fyfe}, {Gallardo}, {Gallegos}, {Gardiol}, {Gebran}, {Gomboc}, {G{\'o}mez}, {Grux}, {Gueguen}, {Heyrovsky}, {Hoar}, {Iannicola}, {Isasi Parache}, {Janotto}, {Joliet}, {Jonckheere}, {Keil}, {Kim}, {Klagyivik}, {Klar}, {Knude}, {Kochukhov}, {Kolka}, {Kos}, {Kutka}, {Lainey}, {LeBouquin}, {Liu}, {Loreggia}, {Makarov}, {Marseille}, {Martayan}, {Martinez-Rubi}, {Massart}, {Meynadier}, {Mignot}, {Munari}, {Nguyen}, {Nordlander}, {Ocvirk}, {O'Flaherty}, {Olias Sanz}, {Ortiz}, {Osorio}, {Oszkiewicz}, {Ouzounis}, {Palmer}, {Park}, {Pasquato}, {Peltzer}, {Peralta}, {P{\'e}turaud}, {Pieniluoma}, {Pigozzi}, {Poels}, {Prat}, {Prod'homme}, {Raison}, {Rebordao}, {Risquez}, {Rocca-Volmerange}, {Rosen},
  {Ruiz-Fuertes}, {Russo}, {Sembay}, {Serraller Vizcaino}, {Short}, {Siebert}, {Silva}, {Sinachopoulos}, {Slezak}, {Soffel}, {Sosnowska}, {Strai{\v{z}}ys}, {ter Linden}, {Terrell}, {Theil}, {Tiede}, {Troisi}, {Tsalmantza}, {Tur}, {Vaccari}, {Vachier}, {Valles}, {Van Hamme}, {Veltz}, {Virtanen}, {Wallut}, {Wichmann}, {Wilkinson}, {Ziaeepour}, \& {Zschocke}}]{gaia1}
{Gaia Collaboration}, {Prusti}, T., {de Bruijne}, J.~H.~J., {et~al.} 2016, \aap, 595, A1, \dodoi{10.1051/0004-6361/201629272}

\bibitem[{{Gaia Collaboration} {et~al.}(2021){Gaia Collaboration}, {Brown}, {Vallenari}, {Prusti}, {de Bruijne}, {Babusiaux}, {Biermann}, {Creevey}, {Evans}, {Eyer}, {Hutton}, {Jansen}, {Jordi}, {Klioner}, {Lammers}, {Lindegren}, {Luri}, {Mignard}, {Panem}, {Pourbaix}, {Randich}, {Sartoretti}, {Soubiran}, {Walton}, {Arenou}, {Bailer-Jones}, {Bastian}, {Cropper}, {Drimmel}, {Katz}, {Lattanzi}, {van Leeuwen}, {Bakker}, {Cacciari}, {Casta{\~n}eda}, {De Angeli}, {Ducourant}, {Fabricius}, {Fouesneau}, {Fr{\'e}mat}, {Guerra}, {Guerrier}, {Guiraud}, {Jean-Antoine Piccolo}, {Masana}, {Messineo}, {Mowlavi}, {Nicolas}, {Nienartowicz}, {Pailler}, {Panuzzo}, {Riclet}, {Roux}, {Seabroke}, {Sordo}, {Tanga}, {Th{\'e}venin}, {Gracia-Abril}, {Portell}, {Teyssier}, {Altmann}, {Andrae}, {Bellas-Velidis}, {Benson}, {Berthier}, {Blomme}, {Brugaletta}, {Burgess}, {Busso}, {Carry}, {Cellino}, {Cheek}, {Clementini}, {Damerdji}, {Davidson}, {Delchambre}, {Dell'Oro}, {Fern{\'a}ndez-Hern{\'a}ndez}, {Galluccio}, {Garc{\'\i}a-Lario},
  {Garcia-Reinaldos}, {Gonz{\'a}lez-N{\'u}{\~n}ez}, {Gosset}, {Haigron}, {Halbwachs}, {Hambly}, {Harrison}, {Hatzidimitriou}, {Heiter}, {Hern{\'a}ndez}, {Hestroffer}, {Hodgkin}, {Holl}, {Jan{\ss}en}, {Jevardat de Fombelle}, {Jordan}, {Krone-Martins}, {Lanzafame}, {L{\"o}ffler}, {Lorca}, {Manteiga}, {Marchal}, {Marrese}, {Moitinho}, {Mora}, {Muinonen}, {Osborne}, {Pancino}, {Pauwels}, {Petit}, {Recio-Blanco}, {Richards}, {Riello}, {Rimoldini}, {Robin}, {Roegiers}, {Rybizki}, {Sarro}, {Siopis}, {Smith}, {Sozzetti}, {Ulla}, {Utrilla}, {van Leeuwen}, {van Reeven}, {Abbas}, {Abreu Aramburu}, {Accart}, {Aerts}, {Aguado}, {Ajaj}, {Altavilla}, {{\'A}lvarez}, {{\'A}lvarez Cid-Fuentes}, {Alves}, {Anderson}, {Anglada Varela}, {Antoja}, {Audard}, {Baines}, {Baker}, {Balaguer-N{\'u}{\~n}ez}, {Balbinot}, {Balog}, {Barache}, {Barbato}, {Barros}, {Barstow}, {Bartolom{\'e}}, {Bassilana}, {Bauchet}, {Baudesson-Stella}, {Becciani}, {Bellazzini}, {Bernet}, {Bertone}, {Bianchi}, {Blanco-Cuaresma}, {Boch}, {Bombrun}, {Bossini},
  {Bouquillon}, {Bragaglia}, {Bramante}, {Breedt}, {Bressan}, {Brouillet}, {Bucciarelli}, {Burlacu}, {Busonero}, {Butkevich}, {Buzzi}, {Caffau}, {Cancelliere}, {C{\'a}novas}, {Cantat-Gaudin}, {Carballo}, {Carlucci}, {Carnerero}, {Carrasco}, {Casamiquela}, {Castellani}, {Castro-Ginard}, {Castro Sampol}, {Chaoul}, {Charlot}, {Chemin}, {Chiavassa}, {Cioni}, {Comoretto}, {Cooper}, {Cornez}, {Cowell}, {Crifo}, {Crosta}, {Crowley}, {Dafonte}, {Dapergolas}, {David}, {David}, {de Laverny}, {De Luise}, {De March}, {De Ridder}, {de Souza}, {de Teodoro}, {de Torres}, {del Peloso}, {del Pozo}, {Delbo}, {Delgado}, {Delgado}, {Delisle}, {Di Matteo}, {Diakite}, {Diener}, {Distefano}, {Dolding}, {Eappachen}, {Edvardsson}, {Enke}, {Esquej}, {Fabre}, {Fabrizio}, {Faigler}, {Fedorets}, {Fernique}, {Fienga}, {Figueras}, {Fouron}, {Fragkoudi}, {Fraile}, {Franke}, {Gai}, {Garabato}, {Garcia-Gutierrez}, {Garc{\'\i}a-Torres}, {Garofalo}, {Gavras}, {Gerlach}, {Geyer}, {Giacobbe}, {Gilmore}, {Girona}, {Giuffrida}, {Gomel}, {Gomez},
  {Gonzalez-Santamaria}, {Gonz{\'a}lez-Vidal}, {Granvik}, {Guti{\'e}rrez-S{\'a}nchez}, {Guy}, {Hauser}, {Haywood}, {Helmi}, {Hidalgo}, {Hilger}, {H{\l}adczuk}, {Hobbs}, {Holland}, {Huckle}, {Jasniewicz}, {Jonker}, {Juaristi Campillo}, {Julbe}, {Karbevska}, {Kervella}, {Khanna}, {Kochoska}, {Kontizas}, {Kordopatis}, {Korn}, {Kostrzewa-Rutkowska}, {Kruszy{\'n}ska}, {Lambert}, {Lanza}, {Lasne}, {Le Campion}, {Le Fustec}, {Lebreton}, {Lebzelter}, {Leccia}, {Leclerc}, {Lecoeur-Taibi}, {Liao}, {Licata}, {Lindstr{\o}m}, {Lister}, {Livanou}, {Lobel}, {Madrero Pardo}, {Managau}, {Mann}, {Marchant}, {Marconi}, {Marcos Santos}, {Marinoni}, {Marocco}, {Marshall}, {Martin Polo}, {Mart{\'\i}n-Fleitas}, {Masip}, {Massari}, {Mastrobuono-Battisti}, {Mazeh}, {McMillan}, {Messina}, {Michalik}, {Millar}, {Mints}, {Molina}, {Molinaro}, {Moln{\'a}r}, {Montegriffo}, {Mor}, {Morbidelli}, {Morel}, {Morris}, {Mulone}, {Munoz}, {Muraveva}, {Murphy}, {Musella}, {Noval}, {Ord{\'e}novic}, {Orr{\`u}}, {Osinde}, {Pagani}, {Pagano},
  {Palaversa}, {Palicio}, {Panahi}, {Pawlak}, {Pe{\~n}alosa Esteller}, {Penttil{\"a}}, {Piersimoni}, {Pineau}, {Plachy}, {Plum}, {Poggio}, {Poretti}, {Poujoulet}, {Pr{\v{s}}a}, {Pulone}, {Racero}, {Ragaini}, {Rainer}, {Raiteri}, {Rambaux}, {Ramos}, {Ramos-Lerate}, {Re Fiorentin}, {Regibo}, {Reyl{\'e}}, {Ripepi}, {Riva}, {Rixon}, {Robichon}, {Robin}, {Roelens}, {Rohrbasser}, {Romero-G{\'o}mez}, {Rowell}, {Royer}, {Rybicki}, {Sadowski}, {Sagrist{\`a} Sell{\'e}s}, {Sahlmann}, {Salgado}, {Salguero}, {Samaras}, {Sanchez Gimenez}, {Sanna}, {Santove{\~n}a}, {Sarasso}, {Schultheis}, {Sciacca}, {Segol}, {Segovia}, {S{\'e}gransan}, {Semeux}, {Shahaf}, {Siddiqui}, {Siebert}, {Siltala}, {Slezak}, {Smart}, {Solano}, {Solitro}, {Souami}, {Souchay}, {Spagna}, {Spoto}, {Steele}, {Steidelm{\"u}ller}, {Stephenson}, {S{\"u}veges}, {Szabados}, {Szegedi-Elek}, {Taris}, {Tauran}, {Taylor}, {Teixeira}, {Thuillot}, {Tonello}, {Torra}, {Torra}, {Turon}, {Unger}, {Vaillant}, {van Dillen}, {Vanel}, {Vecchiato}, {Viala}, {Vicente},
  {Voutsinas}, {Weiler}, {Wevers}, {Wyrzykowski}, {Yoldas}, {Yvard}, {Zhao}, {Zorec}, {Zucker}, {Zurbach}, \& {Zwitter}}]{gaia2}
{Gaia Collaboration}, {Brown}, A.~G.~A., {Vallenari}, A., {et~al.} 2021, \aap, 649, A1, \dodoi{10.1051/0004-6361/202039657}

\bibitem[{{Guo} {et~al.}(2023){Guo}, {Li}, {Liu}, {Li}, \& {Liu}}]{guo2023}
{Guo}, D.-F., {Li}, K., {Liu}, F., {Li}, H.-Z., \& {Liu}, X.-Y. 2023, \mnras, 521, 51, \dodoi{10.1093/mnras/stad417}

\bibitem[{{Guo} {et~al.}(2022){Guo}, {Li}, {Liu}, {Li}, {Xia}, {Gao}, {Gao}, {Chen}, {Gao}, \& {Sun}}]{guo2022}
{Guo}, D.-F., {Li}, K., {Liu}, F., {et~al.} 2022, \mnras, 517, 1928, \dodoi{10.1093/mnras/stac2811}

\bibitem[{Holland(1992)}]{ga}
Holland, J.~H. 1992, Scientific American, 267, 66.
\newblock \url{http://www.jstor.org/stable/24939139}

\bibitem[{{Horvat} {et~al.}(2018){Horvat}, {Conroy}, {Pablo}, {Hambleton}, {Kochoska}, {Giammarco}, \& {Pr{\v{s}}a}}]{Horvat2018}
{Horvat}, M., {Conroy}, K.~E., {Pablo}, H., {et~al.} 2018, \apjs, 237, 26, \dodoi{10.3847/1538-4365/aacd0f}

\bibitem[{{Hurley} {et~al.}(2002){Hurley}, {Tout}, \& {Pols}}]{Hurley}
{Hurley}, J.~R., {Tout}, C.~A., \& {Pols}, O.~R. 2002, \mnras, 329, 897, \dodoi{10.1046/j.1365-8711.2002.05038.x}

\bibitem[{{Jayasinghe} {et~al.}(2018){Jayasinghe}, {Kochanek}, {Stanek}, {Shappee}, {Holoien}, {Thompson}, {Prieto}, {Dong}, {Pawlak}, {Shields}, {Pojmanski}, {Otero}, {Britt}, \& {Will}}]{jayasinghe2018}
{Jayasinghe}, T., {Kochanek}, C.~S., {Stanek}, K.~Z., {et~al.} 2018, \mnras, 477, 3145, \dodoi{10.1093/mnras/sty838}

\bibitem[{{Jiang} {et~al.}(2012){Jiang}, {Han}, {Ge}, {Yang}, \& {Li}}]{jiang2012111}
{Jiang}, D., {Han}, Z., {Ge}, H., {Yang}, L., \& {Li}, L. 2012, \mnras, 421, 2769, \dodoi{10.1111/j.1365-2966.2011.20323.x}

\bibitem[{{Jiang} {et~al.}(2010){Jiang}, {Han}, {Wang}, {Jiang}, \& {Li}}]{jiang2010}
{Jiang}, D., {Han}, Z., {Wang}, J., {Jiang}, T., \& {Li}, L. 2010, \mnras, 405, 2485, \dodoi{10.1111/j.1365-2966.2010.16615.x}

\bibitem[{{Jones} {et~al.}(2020){Jones}, {Conroy}, {Horvat}, {Giammarco}, {Kochoska}, {Pablo}, {Brown}, {Sowicka}, \& {Pr{\v{s}}a}}]{Jones2020}
{Jones}, D., {Conroy}, K.~E., {Horvat}, M., {et~al.} 2020, \apjs, 247, 63, \dodoi{10.3847/1538-4365/ab7927}

\bibitem[{{Knote} {et~al.}(2022){Knote}, {Caballero-Nieves}, {Gokhale}, {Johnston}, \& {Perlman}}]{occ1}
{Knote}, M.~F., {Caballero-Nieves}, S.~M., {Gokhale}, V., {Johnston}, K.~B., \& {Perlman}, E.~S. 2022, \apjs, 262, 10, \dodoi{10.3847/1538-4365/ac770f}

\bibitem[{{Koleva} {et~al.}(2009){Koleva}, {Prugniel}, {Bouchard}, \& {Wu}}]{ULySS}
{Koleva}, M., {Prugniel}, P., {Bouchard}, A., \& {Wu}, Y. 2009, \aap, 501, 1269, \dodoi{10.1051/0004-6361/200811467}

\bibitem[{{Kwee}(1958)}]{k1958}
{Kwee}, K.~K. 1958, \bain, 14, 131

\bibitem[{{Kwee} \& {van Woerden}(1956)}]{kw1956}
{Kwee}, K.~K., \& {van Woerden}, H. 1956, \bain, 12, 327

\bibitem[{{Lanza} \& {Rodon{\`o}}(2002)}]{L2002}
{Lanza}, A.~F., \& {Rodon{\`o}}, M. 2002, Astronomische Nachrichten, 323, 424, \dodoi{10.1002/1521-3994(200208)323:3/4<424::AID-ASNA424>3.0.CO;2-1}

\bibitem[{{Li} {et~al.}(2016){Li}, {Gao}, {Hu}, {Guo}, {Jiang}, \& {Chen}}]{li_o-c_2016}
{Li}, K., {Gao}, D.~Y., {Hu}, S.~M., {et~al.} 2016, \apss, 361, 63, \dodoi{10.1007/s10509-016-2649-8}

\bibitem[{{Li} {et~al.}(2022){Li}, {Gao}, {Liu}, {Gao}, {Li}, {Chen}, \& {Sun}}]{li2022fo}
{Li}, K., {Gao}, X., {Liu}, X.-Y., {et~al.} 2022, \aj, 164, 202, \dodoi{10.3847/1538-3881/ac8ff2}

\bibitem[{Li {et~al.}(2020)Li, Kim, Xia, Michel, Hu, Gao, Guo, \& Chen}]{Li_2020}
Li, K., Kim, C.-H., Xia, Q.-Q., {et~al.} 2020, The Astronomical Journal, 159, 189, \dodoi{10.3847/1538-3881/ab7cda}

\bibitem[{{Li} {et~al.}(2014){Li}, {Qian}, {Hu}, \& {He}}]{li_o-c_2014}
{Li}, K., {Qian}, S.~B., {Hu}, S.~M., \& {He}, J.~J. 2014, \aj, 147, 98, \dodoi{10.1088/0004-6256/147/5/98}

\bibitem[{{Li} {et~al.}(2021){Li}, {Xia}, {Kim}, {Hu}, {Guo}, {Jeong}, {Chen}, \& {Gao}}]{li2021a}
{Li}, K., {Xia}, Q.-Q., {Kim}, C.-H., {et~al.} 2021, \apj, 922, 122, \dodoi{10.3847/1538-4357/ac242f}

\bibitem[{{Li} {et~al.}(2019){Li}, {Xia}, {Michel}, {Hu}, {Guo}, {Gao}, {Chen}, \& {Gao}}]{li}
{Li}, K., {Xia}, Q.-Q., {Michel}, R., {et~al.} 2019, \mnras, 485, 4588, \dodoi{10.1093/mnras/stz715}

\bibitem[{{Li} \& {Zhang}(2006)}]{liandzhang}
{Li}, L., \& {Zhang}, F. 2006, \mnras, 369, 2001, \dodoi{10.1111/j.1365-2966.2006.10462.x}

\bibitem[{{Li} {et~al.}(2024){Li}, {Zhu}, {Ding}, {Xu}, {Zheng}, {Qiu}, \& {Liu}}]{lixuzhi}
{Li}, X.-Z., {Zhu}, Q.-F., {Ding}, X., {et~al.} 2024, \apjs, 271, 32, \dodoi{10.3847/1538-4365/ad226a}

\bibitem[{{Liu}(2021)}]{liuliang}
{Liu}, L. 2021, \pasp, 133, 084202, \dodoi{10.1088/1538-3873/ac1ac1}

\bibitem[{{Liu} \& {Yang}(2003)}]{liu}
{Liu}, Q.-Y., \& {Yang}, Y.-L. 2003, \cjaa, 3, 142, \dodoi{10.1088/1009-9271/3/2/142}

\bibitem[{Liu \& Yang(2003)}]{Liu_2003}
Liu, Q.-Y., \& Yang, Y.-L. 2003, Chinese Journal of Astronomy and Astrophysics, 3, 142, \dodoi{10.1088/1009-9271/3/2/142}

\bibitem[{{Liu} {et~al.}(2023){Liu}, {Li}, {Michel}, {Gao}, {Gao}, {Liu}, {Yin}, {Wang}, \& {Sun}}]{liuxinyi}
{Liu}, X.-Y., {Li}, K., {Michel}, R., {et~al.} 2023, \mnras, 519, 5760, \dodoi{10.1093/mnras/stad026}

\bibitem[{{Lucy}(1967)}]{Lucy1967}
{Lucy}, L.~B. 1967, \zap, 65, 89

\bibitem[{{Masci} {et~al.}(2019){Masci}, {Laher}, {Rusholme}, {Shupe}, {Groom}, {Surace}, {Jackson}, {Monkewitz}, {Beck}, {Flynn}, {Terek}, {Landry}, {Hacopians}, {Desai}, {Howell}, {Brooke}, {Imel}, {Wachter}, {Ye}, {Lin}, {Cenko}, {Cunningham}, {Rebbapragada}, {Bue}, {Miller}, {Mahabal}, {Bellm}, {Patterson}, {Juri{\'c}}, {Golkhou}, {Ofek}, {Walters}, {Graham}, {Kasliwal}, {Dekany}, {Kupfer}, {Burdge}, {Cannella}, {Barlow}, {Van Sistine}, {Giomi}, {Fremling}, {Blagorodnova}, {Levitan}, {Riddle}, {Smith}, {Helou}, {Prince}, \& {Kulkarni}}]{ztf2}
{Masci}, F.~J., {Laher}, R.~R., {Rusholme}, B., {et~al.} 2019, \pasp, 131, 018003, \dodoi{10.1088/1538-3873/aae8ac}

\bibitem[{{Michaels}(2018)}]{LAST2}
{Michaels}, E.~J. 2018, \jaavso, 46, 27

\bibitem[{{O'Connell}(1951)}]{Connell}
{O'Connell}, D.~J.~K. 1951, Publications of the Riverview College Observatory, 2, 85

\bibitem[{{Pecaut} \& {Mamajek}(2013)}]{tab5}
{Pecaut}, M.~J., \& {Mamajek}, E.~E. 2013, \apjs, 208, 9, \dodoi{10.1088/0067-0049/208/1/9}

\bibitem[{{Pribulla} {et~al.}(2003){Pribulla}, {Kreiner}, \& {Tremko}}]{Apeople}
{Pribulla}, T., {Kreiner}, J.~M., \& {Tremko}, J. 2003, Contributions of the Astronomical Observatory Skalnate Pleso, 33, 38

\bibitem[{{Pr{\v{s}}a} \& {Zwitter}(2005)}]{Prša2005}
{Pr{\v{s}}a}, A., \& {Zwitter}, T. 2005, \apj, 628, 426, \dodoi{10.1086/430591}

\bibitem[{{Pr{\v{s}}a} {et~al.}(2016){Pr{\v{s}}a}, {Conroy}, {Horvat}, {Pablo}, {Kochoska}, {Bloemen}, {Giammarco}, {Hambleton}, \& {Degroote}}]{Prša2016}
{Pr{\v{s}}a}, A., {Conroy}, K.~E., {Horvat}, M., {et~al.} 2016, \apjs, 227, 29, \dodoi{10.3847/1538-4365/227/2/29}

\bibitem[{{Qian} {et~al.}(2017){Qian}, {He}, {Zhang}, {Zhu}, {Shi}, {Zhao}, \& {Zhou}}]{qian2017}
{Qian}, S.-B., {He}, J.-J., {Zhang}, J., {et~al.} 2017, Research in Astronomy and Astrophysics, 17, 087, \dodoi{10.1088/1674-4527/17/8/87}

\bibitem[{{Qian} {et~al.}(2005){Qian}, {Zhu}, {Soonthornthum}, {Yuan}, {Yang}, \& {He}}]{qian2005b}
{Qian}, S.~B., {Zhu}, L.~Y., {Soonthornthum}, B., {et~al.} 2005, \aj, 130, 1206, \dodoi{10.1086/432544}

\bibitem[{{Qian} {et~al.}(2014){Qian}, {Wang}, {Zhu}, {Snoonthornthum}, {Wang}, {Zhao}, {Zhou}, {Liao}, \& {Liu}}]{qian}
{Qian}, S.~B., {Wang}, J.~J., {Zhu}, L.~Y., {et~al.} 2014, \apjs, 212, 4, \dodoi{10.1088/0067-0049/212/1/4}

\bibitem[{{Rasio}(1995)}]{rasio}
{Rasio}, F.~A. 1995, \apjl, 444, L41, \dodoi{10.1086/187855}

\bibitem[{{Ricker} {et~al.}(2015){Ricker}, {Winn}, {Vanderspek}, {Latham}, {Bakos}, {Bean}, {Berta-Thompson}, {Brown}, {Buchhave}, {Butler}, {Butler}, {Chaplin}, {Charbonneau}, {Christensen-Dalsgaard}, {Clampin}, {Deming}, {Doty}, {De Lee}, {Dressing}, {Dunham}, {Endl}, {Fressin}, {Ge}, {Henning}, {Holman}, {Howard}, {Ida}, {Jenkins}, {Jernigan}, {Johnson}, {Kaltenegger}, {Kawai}, {Kjeldsen}, {Laughlin}, {Levine}, {Lin}, {Lissauer}, {MacQueen}, {Marcy}, {McCullough}, {Morton}, {Narita}, {Paegert}, {Palle}, {Pepe}, {Pepper}, {Quirrenbach}, {Rinehart}, {Sasselov}, {Sato}, {Seager}, {Sozzetti}, {Stassun}, {Sullivan}, {Szentgyorgyi}, {Torres}, {Udry}, \& {Villasenor}}]{tess}
{Ricker}, G.~R., {Winn}, J.~N., {Vanderspek}, R., {et~al.} 2015, Journal of Astronomical Telescopes, Instruments, and Systems, 1, 014003, \dodoi{10.1117/1.JATIS.1.1.014003}

\bibitem[{{Rovithis-Livaniou} {et~al.}(2000){Rovithis-Livaniou}, {Kranidiotis}, {Rovithis}, \& {Athanassiades}}]{R2000}
{Rovithis-Livaniou}, H., {Kranidiotis}, A.~N., {Rovithis}, P., \& {Athanassiades}, G. 2000, \aap, 354, 904

\bibitem[{{Ruci{\'n}ski}(1973)}]{Rucinski1969}
{Ruci{\'n}ski}, S.~M. 1973, \actaa, 23, 79

\bibitem[{{Rucinski}(1992)}]{xin1996}
{Rucinski}, S.~M. 1992, \aj, 103, 960, \dodoi{10.1086/116118}

\bibitem[{{Shappee} {et~al.}(2014){Shappee}, {Prieto}, {Grupe}, {Kochanek}, {Stanek}, {De Rosa}, {Mathur}, {Zu}, {Peterson}, {Pogge}, {Komossa}, {Im}, {Jencson}, {Holoien}, {Basu}, {Beacom}, {Szczygie{\l}}, {Brimacombe}, {Adams}, {Campillay}, {Choi}, {Contreras}, {Dietrich}, {Dubberley}, {Elphick}, {Foale}, {Giustini}, {Gonzalez}, {Hawkins}, {Howell}, {Hsiao}, {Koss}, {Leighly}, {Morrell}, {Mudd}, {Mullins}, {Nugent}, {Parrent}, {Phillips}, {Pojmanski}, {Rosing}, {Ross}, {Sand}, {Terndrup}, {Valenti}, {Walker}, \& {Yoon}}]{shappee2014}
{Shappee}, B.~J., {Prieto}, J.~L., {Grupe}, D., {et~al.} 2014, \apj, 788, 48, \dodoi{10.1088/0004-637X/788/1/48}

\bibitem[{{Shaw}(1994)}]{shaw1994}
{Shaw}, J.~S. 1994, \memsai, 65, 95

\bibitem[{{Sriram} {et~al.}(2016){Sriram}, {Malu}, {Choi}, \& {Vivekananda Rao}}]{sriram2016}
{Sriram}, K., {Malu}, S., {Choi}, C.~S., \& {Vivekananda Rao}, P. 2016, \aj, 151, 69, \dodoi{10.3847/0004-6256/151/3/69}

\bibitem[{{Stepien}(2006)}]{step2006}
{Stepien}, K. 2006, \actaa, 56, 347, \dodoi{10.48550/arXiv.astro-ph/0701529}

\bibitem[{{Sun} {et~al.}(2020){Sun}, {Chen}, {Deng}, \& {de Grijs}}]{sunwei}
{Sun}, W., {Chen}, X., {Deng}, L., \& {de Grijs}, R. 2020, \apjs, 247, 50, \dodoi{10.3847/1538-4365/ab7894}

\bibitem[{{Terrell} \& {Wilson}(2005)}]{Bpeople}
{Terrell}, D., \& {Wilson}, R.~E. 2005, \apss, 296, 221, \dodoi{10.1007/s10509-005-4449-4}

\bibitem[{{Tylenda} {et~al.}(2011){Tylenda}, {Hajduk}, {Kami{\'n}ski}, {Udalski}, {Soszy{\'n}ski}, {Szyma{\'n}ski}, {Kubiak}, {Pietrzy{\'n}ski}, {Poleski}, {Wyrzykowski}, \& {Ulaczyk}}]{mercy2011}
{Tylenda}, R., {Hajduk}, M., {Kami{\'n}ski}, T., {et~al.} 2011, \aap, 528, A114, \dodoi{10.1051/0004-6361/201016221}

\bibitem[{{Wadhwa} {et~al.}(2021){Wadhwa}, {De Horta}, {Filipovi{\'c}}, {Tothill}, {Arbutina}, {Petrovi{\'c}}, \& {Djura{\v{s}}evi{\'c}}}]{Wadhwa2021}
{Wadhwa}, S.~S., {De Horta}, A., {Filipovi{\'c}}, M.~D., {et~al.} 2021, \mnras, 501, 229, \dodoi{10.1093/mnras/staa3637}

\bibitem[{{Wilsey} \& {Beaky}(2009)}]{WB2009}
{Wilsey}, N.~J., \& {Beaky}, M.~M. 2009, Society for Astronomical Sciences Annual Symposium, 28, 107

\bibitem[{{Wilson}(1979)}]{w1979}
{Wilson}, R.~E. 1979, \apj, 234, 1054, \dodoi{10.1086/157588}

\bibitem[{{Wilson}(1990)}]{w1990}
---. 1990, \apj, 356, 613, \dodoi{10.1086/168867}

\bibitem[{{Wilson}(1994)}]{w1994}
---. 1994, \pasp, 106, 921, \dodoi{10.1086/133464}

\bibitem[{{Wilson} \& {Devinney}(1971)}]{w1971}
{Wilson}, R.~E., \& {Devinney}, E.~J. 1971, \apj, 166, 605, \dodoi{10.1086/150986}

\bibitem[{{Y{\i}ld{\i}z}(2014)}]{yd2014}
{Y{\i}ld{\i}z}, M. 2014, \mnras, 437, 185, \dodoi{10.1093/mnras/stt1874}

\bibitem[{{Yildiz} \& {Do{\u{g}}an}(2013)}]{yd2013}
{Yildiz}, M., \& {Do{\u{g}}an}, T. 2013, \mnras, 430, 2029, \dodoi{10.1093/mnras/stt028}

\bibitem[{{Zhang} {et~al.}(2023){Zhang}, {Chen}, {Fu}, \& {Li}}]{zhangxiaobin2023}
{Zhang}, X., {Chen}, X., {Fu}, J., \& {Li}, Y. 2023, \mnras, 523, 1394, \dodoi{10.1093/mnras/stad1465}

\bibitem[{{Zhou} {et~al.}(2016){Zhou}, {Qian}, {Zhang}, {Jiang}, {Zhang}, \& {Kreiner}}]{2016ApJ_new}
{Zhou}, X., {Qian}, S.~B., {Zhang}, J., {et~al.} 2016, \apj, 817, 133, \dodoi{10.3847/0004-637X/817/2/133}

\end{thebibliography}
%% This command is needed to show the entire author+affiliation list when
%% the collaboration and author truncation commands are used.  It has to
%% go at the end of the manuscript.
%\allauthors

%% Include this line if you are using the \added, \replaced, \deleted
%% commands to see a summary list of all changes at the end of the article.
%\listofchanges

\end{document}